\documentclass[prb,twocolumn,showpacs,superscriptaddress]{revtex4-1}

\usepackage{graphicx}
\usepackage{amsmath,amssymb,bm}

\usepackage{color}

\begin{document}


\title{
Field-Enhanced Kondo Correlations in a Half-Filling Nanotube Dot:
\\
Evolution of an SU($N$) Fermi-Liquid Fixed Point 
}


\author{Yoshimichi Teratani}
\affiliation{%
Department of Physics, Osaka City University, Sumiyoshi-ku, 
Osaka 558-8585, Japan
}

\author{Rui Sakano}
\affiliation{
The Institute for Solid State Physics, 
the University of Tokyo, Kashiwa, Chiba 277-8581, Japan
}

\author{Ryo Fujiwara}  
\author{Tokuro Hata }
\author{\\ Tomonori Arakawa}
\affiliation{
Department of Physics, Osaka University, Toyonaka, 
Osaka 560-0043, Japan
}

\author{Meydi Ferrier}
\affiliation{
Department of Physics, Osaka University, Toyonaka, 
Osaka 560-0043, Japan
}
\affiliation{
Laboratoire de Physique des Solides, CNRS, 
Universit\'{e} Paris-Sud, 
Universit\'{e} Paris Saclay, 
91405 Orsay Cedex, 
France
}

\author{Kensuke Kobayashi}
\affiliation{
Department of Physics, Osaka University, Toyonaka, 
Osaka 560-0043, Japan
}

\author{Akira Oguri}
\affiliation{%
Department of Physics, Osaka City University, Sumiyoshi-ku, 
Osaka 558-8585, Japan
}


\date{\today}


\begin{abstract}
Carbon nanotube quantum dot has four-fold degenerate one-particle levels,  
 which bring a variety to the Kondo effects taking place 
in a wide tunable-parameter space. 
We theoretically study an emergent SU($2$) symmetry 
that is suggested by recent magneto-transport measurements,  
carried out near two electrons filling. 
It does not couple with the magnetic field, 
and emerges in the case where 
the spin and orbital Zeeman splittings cancel each other out 
in two of the one-particle levels among four. 
This situation seems to be realized in the recent experiment. 
Using the Wilson numerical renormalization group,  
we show that a crossover from the SU($4$) to SU($2$) 
Fermi-liquid behavior occurs 
as magnetic field increases at two impurity-electrons filling. 
We also find that the quasiparticles are significantly renormalized 
as the remaining two one-particle levels 
move away from the Fermi level and are frozen at high magnetic fields.
 Furthermore, we consider how the singlet ground state 
evolves during such a crossover. 
Specifically, 
we reexamine the SU($N$) Kondo singlet for $M$ impurity-electrons filling 
in the limit of strong exchange interactions.
We find that the nondegenerate Fermi-liquid fixed point 
of Nozi\`{e}res and Blandin can be described 
as a bosonic Perron-Frobenius vector for $M$ composite pairs, 
each of which consists of one impurity-electron and one conduction-{\it hole}. 
This interpretation in terms of the Perron-Frobenius theorem 
can also be extended to the Fermi-liquid fixed-point without the SU($N$) symmetry.  
\end{abstract}



\maketitle

\section{Introduction}
\label{sec:introduction} 

Combination of the spin and orbital degrees of freedom causes  
an interesting variety in the Kondo effects\cite{HewsonBook} 
in quantum dots. 
It arises in various ways, depending on  
the electron fillings and external fields. 
For instance,  
the Kondo states which involve the orbital components 
such as the ones of the SU($4$) and the spin-triplet Kondo effects  
have been observed  as well as the spin-based SU($2$) 
Kondo state.\cite{Sasaki2000,Goldhaber-Gordon2014,Franceschi2005,Makarovski,Izumida1998,Broda2003,Choi2005,Lim,Busser2007,Anders2008,GalpinLoganAnders2010,MantelliMocaZarandGrifoni,Grifoni_PRB2015,NishikawaHewson2016} 
Furthermore, recent ultra-sensitive current and current-noise measurements 
have precisely identified the Fermi-liquid states  
with the SU($2$) and SU($4$) symmetries.\cite{Ferrier2016}

In this paper, we focus on electron-correlation effects  
in a carbon nanotube (CNT) quantum dot\cite{RMP-Kouwenhoven,IzumidaCNT}
 with two electrons filling.
It is inspired by 
recent magneto-transport experiment,   
which observes an unexpected evolution of the Kondo plateau 
that can be regarded as an indication of 
a crossover from the SU($4$) to SU($2$) Fermi-liquid state. 
\footnote{
M.\ Ferrier, T.\ Arakawa, T.\ Hata, R.\ Fujiwara, 
R.\ Delagrange, Y.\ Teratani, R.\ Sakano, A.\ Oguri, and K.\ Kobayashi, 
unpublished.} 
As magnetic field increases, 
the Kondo plateau near half-filling reduces 
the height from $4e^2/h$ to $2e^2/h$ keeping the  
flat structure unchanged.
This implies that two one-particle levels among the four still  
remain unlifted near the Fermi level in the magnetic field. 
This is possible if the magnetic field $\vec{B}$ 
is applied in such a way that the spin and orbital Zeeman effects 
cancel each other out.

We study the Kondo effect  
taking place in this ideal case 
where the twofold degenerate levels remain near the Fermi level.   
Using the numerical renormalization group (NRG),\cite{WilsonRMP,KWW1} 
we find that the Kondo correlations 
are enhanced as magnetic field increases.
This is due to the fact that 
the number of active one-particle levels decreases 
from four to two, and it makes quantum fluctuations large. 
Then, we also take into account the perturbations 
 that lift the double degeneracy;
specifically  the valley mixing, the spin-orbit interaction, 
and an energy-difference between the spin and orbital 
Zeeman splittings.
We find that the crossover can be seen   
if the energy gap which is induced by these perturbations 
is smaller than the Kondo energy scale. 
For a realistic parameter set deduced 
from the recent experiments,\cite{Note1} 
it is satisfied up to $B \lesssim 5$ T.

One of the other most interesting features of the nanotube dots is that  
different class of the SU($N$) Kondo effects 
occur depending on the number of electrons $M$ 
occupying the impurity levels,
and such a variation 
has been observed in recent measurements.\cite{Ferrier2016} 
In this paper, we also discuss how the ground state evolves 
as the number electrons $M$ localized in the dot varies. 
The low-lying energy states 
of the SU($N$) Kondo systems    
show the Fermi-liquid behavior for any $M$ 
as Nozi\`{e}res and Blandin mentioned in their 
well-known paper.\cite{NozieresBlandin} 
However, it seems not to be widely recognized 
how the nondegenerate Fermi-liquid ground state  
is constructed, for arbitrary occupation number 
of impurity electrons $M$, with conduction electrons.
Nozi\`{e}res and Blandin considered the limit of a strong exchange 
interaction $J_K \to \infty$, 
at which major effects of electron correlations are determined by   
the two local sites consisting of the Kondo impurity and one adjacent site 
from the Wilson chain\cite{WilsonRMP,KWW1} for the conduction band. 
This limit provides an important information  
to classify the fixed points of the renormalization group 
 because the effective Kondo-exchange coupling significantly 
increases at low energies. 
It was very briefly suggested 
that the ground state is a non-degenerate singlet 
which is constructed with $M$ impurity-electrons 
and $N-M$ conduction electrons 
in the adjacent site, and this state describes 
a Fermi-liquid fixed point.\cite{NozieresBlandin} 
An explicit derivation of the energy spectrum  
in this limit, $J_K \to \infty$,
has been given later by Pacollet {\it et al\/}, 
using an SU($N$) group-representation theory.\cite{ParcolletPRB}

In this paper, we also  present an alternative interpretation   
based on a {\it hole\/} picture introduced for the conduction electrons. 
The impurity-electrons and {\it conduction-holes\/} 
with the same {\it flavor\/} strongly bind with each other 
to form  $M$ composite pairs. 
It gives a natural description of the nondegenerate Fermi-liquid 
fixed point 
as a bosonic Perron-Frobenius vector, 
which is a robust and a unique nodeless ground state  
of $M$ composite hard-core bosons.
This description does 
 {\it not\/} require the SU($N$) symmetry, 
and thus can be extended 
to some cases {\it without\/} this symmetry.

This paper is organized as follows. 
We first of all present an interpretation of the Fermi-liquid fixed point 
for general impurity-electrons filling $M$ 
in terms of a bosonic Perron-Frobenius vector 
with and without SU($N$) symmetry in the first half of the paper.
Then, after these general discussions about the SU($N$) Kondo effect,
we consider the field-induced crossover from the SU($4$) to SU($2$) 
Fermi-liquid  behavior observed in recent experiments 
of a CNT quantum dot\cite{Note1} in the second half.

 In Sec.\ \ref{sec:SU_N_Kondo}, 
we consider the ground state of 
 SU($N$) Anderson and Kondo impurity models 
to describe a relation between the Fermi-liquid fixed point 
of  Nozi\`{e}res and Blandin 
and a {\it totally antisymmetric representation\/} of the SU($N$).  
In Sec.\ \ref{sec:Hard_core_boson}, 
we show that the ground state in the limit of strong Kondo-exchange coupling  
can be described by a Perron-Frobenius eigenvector 
for the composite hard-core bosons. 
%
Then, in Sec.\ \ref{sec:CFT_Fermi_liquid}, 
we introduce a microscopic Hamiltonian $\bm{H}_{d}^{0}$  
to determine one-particle energy levels of  CNT quantum dots, 
and define a set of renormalized parameters for quasiparticles   
to describe the Fermi-liquid behavior at low energies.
In Sec.\ \ref{sec:cnt_dot}, 
we show that the field-induced crossover observed in a CNT dot 
can be explained as a result of a matching of the spin 
and orbital Zeeman splittings. 
 This matching yields an emergent SU($2$) symmetry 
which does not couple to the magnetic field.
We present NRG results for the linear-response conductance  
and the renormalized parameters for quasiparticles, 
obtained for the case with this emergent SU($2$) symmetry 
and also for a realistic case where this symmetry becomes only approximate.
Summary is given in Sec.\ \ref{sec:summary}.

\section{SU($N$) Kondo effect for $M$ impurity-electrons}
\label{sec:SU_N_Kondo}

We describe the Anderson impurity model 
for carbon nanotube quantum dots in this section.
One of the most significant features of the CNT dot is that  
different class of the SU($N$) Kondo effects 
can occur depending on the number of electrons $M$ 
occupying the impurity levels.
In this section, we introduce the model to describe 
CNT dots, and describe how the singlet ground state evolves  
as the impurity occupation number $M$ varies   
in the strong-coupling limit, $J_K \to \infty$, 
of the SU($N$) Kondo model.

\subsection{$N$-orbital Anderson impurity model for quantum dots}
\label{sec:Anderson_model}

Carbon nanotube quantum dots connected to two leads  
can be described by an $N$ ($=4$) orbital Anderson impurity model:  
$\mathcal{H} = \mathcal{H}_{d}^{} +\mathcal{H}_{c}+\,\mathcal{H}_{T}$,  
\begin{align}
&\mathcal{H}_{d}^{} \,= \, 
\sum_{m=1}^N
 \epsilon_m^{}  
d_{m}^{\dagger} d_{m}^{} 
\,+\, 
 U \sum_{m< m'} n_{d,m} n_{d,m'} \;,
\label{eq:Hd}
\\
&\mathcal{H}_{c}\,=\, 
 \sum_{\nu=L,R}
\sum_{m=1}^N
\int_{-D}^D \! d\varepsilon \, 
\varepsilon\, 
\Bigl (
c_{\nu,\varepsilon m}^{\dagger} 
c_{\nu,\varepsilon m}^{} 
\,- n_c^{0}(\varepsilon)  \Bigr),
\label{eq:HC}
 \\
&\mathcal{H}_{T}^{} =
\sum_{\nu=L,R}\sum_{m=1}^N v_{\nu}^{}
\left( 
\psi_{\nu,m}^{\dagger} d_{m}^{} + 
d_{m}^{\dagger} \psi_{\nu,m}^{} 
\right) , 
\label{eq:HT}
\\
&\psi_{\nu,m} 
\,\equiv\,
 \int_{-D}^D \! d\varepsilon \,\sqrt{\rho_c^{}} \, 
c_{\nu,\varepsilon m}^{} , 
 \qquad n_{d,m}^{} \equiv  d_m^\dagger d_m^{} . 
\end{align}
Here, 
$d_{m}^{\dagger}$ creates an electron with energy $\epsilon_m$ 
in the $m$-th one-particle level ($m\,=\,1,2,\ldots,N$)  
of the dot. 
We also call $m$ the \lq\lq {\it flavor}'' in the following.
In the present study we assume that 
the Coulomb interaction between whole the electrons occupying 
the dot can be characterized by a single parameter $U$.  
The conduction electrons are described by the operator 
 $c_{\nu,\varepsilon m}^{\dagger}$ 
for the lead on the left and right ($\nu=L,R$). 
It  is  normalized as  
$\{ c_{\nu,\varepsilon m}^{}\,,\, c_{\nu',\varepsilon' 
m'}^{\dagger} \} = \delta(\varepsilon-\varepsilon')\, 
\delta_{\nu\nu'}\delta_{mm'} $.  
The Fermi level is situated at the center $\varepsilon_F^{}=0$ 
of the conduction band  with  the width $2D$.  
For subtracting a constant energy of the conduction electrons 
filling the noninteracting band, 
we introduce $n_c^{0}(\varepsilon)\equiv \Theta(-\varepsilon)$ 
with  $\Theta(\varepsilon)$ the step function. 
The tunneling matrix element $v_{\nu}^{}$ 
in Eq.\ \eqref{eq:HT} is assumed that 
it preserves the orbital index $m$. 
The resonance energy scale is denoted as 
 $\Delta_{\nu}^{} \equiv \pi\rho_{c}v_{\nu}^{2}$ 
with $\rho_c^{}=1/(2D)$,  
and $\Delta \equiv \Delta_L^{} + \Delta_R^{}$.   
Among whole the conduction electron degrees of freedom,
only the following linear combination corresponding to the bonding component  
is coupled to the impurity levels  corresponding to 
the dot. Therefore, 
 the tunneling Hamiltonian can be expressed such that 
\begin{align}
& 
 \mathcal{H}_{T}^{} =
 \! \sum_{m=1}^N 
 v \left( 
 d_{m}^{\dagger} a_{m}^{} +
 a_{m}^{\dagger} d_{m}^{} 
 \right) , \qquad 
v \equiv \sqrt{v_L^2 +v_R^2}, 
\\
& a_m \equiv   \sum_{\nu=L,R} \! \frac{v_{\nu}^{}}{v}\, \psi_{\nu,m}, 
\qquad 
c_{\epsilon m} \equiv 
 \sum_{\nu=L,R} \! \frac{v_{\nu}^{}}{v}\,c_{\nu, \epsilon m} .
\label{eq:HT2}
\end{align}

For carbon nanotube quantum dots, 
the one-particle levels with the energy $\epsilon_m$ for $m=1,2,3,4$ 
consist of  the spin ($\uparrow$, \!$\downarrow$) 
and valley ($\mathrm{K}$, \!\! $\mathrm{K'}$) degrees of freedom. 
We will explicitly determine $\epsilon_m$ 
in Sec.\ \ref{sec:CFT_Fermi_liquid},  
using a microscopic one-particle Hamiltonian $\bm{H}_{d}^{0}$   
which takes into account the spin-orbit interaction, 
the valley mixing, and  spin and orbital Zeeman couplings. 
In Eq.\ \eqref{eq:Hd}, we have taken the intra- and inter-valley 
Coulomb repulsions to be identical, and have also neglected Hund's rule 
coupling $J_H$. 
This is consistent with recent measurements 
for the SU($4$) Kondo effect\cite{Ferrier2016,Note1} 
and with previous nanotube data.\cite{Makarovski}
Some corrections due to $J_H$ have been found for 
the ridges other than the SU($4$) one,   
or in other experiments.\cite{MoriyamaPRL} 
That is, however, beyond the scope of this paper 
and will be discussed elsewhere. 
The {\it flavor\/} \lq\lq$m$'' conserving tunneling, 
described by Eqs.\ \eqref{eq:HT} and \eqref{eq:HT2}, 
can physically be realized 
for the leads which are formed in the same nanotube. 

%

The multi-orbital Anderson impurity model $\mathcal{H}$, 
defined in Eqs.\ \eqref{eq:Hd}--\eqref{eq:HT}, 
has an SU($N$) symmetry in the case of  
$\epsilon_m \equiv \varepsilon_d$ for $m=1,2,\ldots,N$, where 
all the one-particle energies of impurity levels are identical. 
It is a rotation symmetry in an $N$ dimensional orbital, 
or {\it flavor\/}, space.  
The total Hamiltonian $\mathcal{H}$ becomes invariant to 
transformations by arbitrary unitary matrix $\mathcal{U}_{mm'}$,
\begin{align} 
d_m' =\sum_{m'=1}^M\mathcal{U}_{mm'} d_{m'}^{}, \qquad 
c_{\nu,\epsilon m}' =\sum_{m'=1}^N\mathcal{U}_{mm'} c_{\nu,\epsilon m'}. 
\end{align} 
The SU($N$) symmetric Anderson model has intensively been studied, 
particularly in the limit of large Coulomb interaction $U\to \infty$ 
where 
the average number of impurity-electrons 
$M = \sum_m \langle d_{m}^{\dagger}d_{m}^{}\rangle$ 
takes a value in the range  $M\leq 1$.\cite{KuramotoKojima1984,Coleman1984,ReadNewns83a,Schlottmann1983,BickersNCA} 
%
Whether or not the system has the SU($N$) symmetry 
does {\it not\/} depend on the impurity level-position of $\varepsilon_d$, 
and  thus different class of the SU($N$) Kondo effects can occur 
depending on the number of electrons occupying the dot levels $M$, 
which varies with the parameters $\varepsilon_d$, $U$, and $\Delta$.
For instance, $N=4$ for CNT dots, and the Fermi-liquid behavior 
has been observed for $M=1,2$, and $3$ varying 
the gate voltage which corresponds to $\varepsilon_d$.\cite{Ferrier2016}

At half-filling $M=N/2$, 
which is achieved for the level-position  $\varepsilon_d = -(N-1)U/2$,   
 the Hamiltonian also has an electron-hole symmetry as well as the SU($N$).  
For this case,  perturbation expansion with respect to 
the Coulomb interaction $U$ has been examined,  
extending the calculations 
of Yamada-Yosida for $N=2$\cite{YamadaYosida1,YamadaYosida2} 
to general $N$.\cite{SakanoFujiiOguri} 
The wavefunction renormalization factor $Z$ 
and the vertex correction $\Gamma_{mm';m'm}^{}(0,0;0,0)$ 
for $m \neq m'$ 
have been calculated up to order $U^3$ and $U^4$, 
respectively,
\begin{align}
& 
\!\!\!\!
\frac{1}{Z}
\ = \ 
1+ \left( 3-\frac{\pi^2}{4}\right)(N-1)\, u^2 
\nonumber \\
 & \quad 
-\left( \frac{21}{2} \zeta(3) - 7 - \frac{\pi^2}{2} \right)(N-1)(N-2)\, u^3 
+ \cdots 
, 
\label{eq:enhancementfactor-ex} \\
& \!\!\!\! 
\frac{1}{\pi \Delta}
\Gamma_{mm';m'm}^{}(0,0;0,0)
\  = \ 
\nonumber 
\\ 
& 
\quad u 
- (N-2) \,u^2
+ \left[ N^2-\left(\frac{\pi ^2}{2}-1\right) N
+9 -\frac{\pi^2}{2}\right] u^3 \nonumber \\
& \quad - (N-2) \Biggl[\,  N^2 
- \left(12+ \frac{7}{4} \pi^2 -21 \zeta(3)  \right) N 
\nonumber \\
&
\qquad \qquad \qquad \ 
- 17 - \frac{71}{12} \pi^2  + \frac{133}{2} \zeta (3) \,\Biggr]\, u^4
+ \cdots 
,
\label{eq:forvertex-ex}
\end{align}
where $u \equiv U/(\pi\Delta)$,  
and $\zeta(x)$ is the Riemann zeta function. 
For $N>2$, both $1/Z$ and $\Gamma_{mm';m'm}^{}(0,0;0,0)$ 
become {\it not} even nor odd function of $U$. 
These results explicitly show that 
the power series expansion in $U$ works 
at least for small $U$, 
or small rescaled-coupling  $g \equiv (N-1)u$,\cite{OguriSakanoFujii,ao2012}
 because the coefficients are finite. 
Therefore, the ground state can evolve from the non-interacting one  
through the adiabatic switching-on of $U$ at half-filling. 
The NRG calculations for the SU($4$) 
Anderson model, carried out in a wide range of the 
 impurity-electron 
filling  $0<M<4$ and the Coulomb interaction $U$ or $g\equiv (N-1)u$,
also clearly indicate the  
Fermi-liquid behavior.\cite{Anders2008,GalpinLoganAnders2010,NishikawaCrowHewson2,ao2012}
We will describe the renormalized parameters for the CNT dots in more detail 
in Sec.\ \ref{sec:CFT_Fermi_liquid}.

\subsection{Fermi-liquid fixed point for $M$ impurity-electrons 
}

In order to gain an insight into how the ground state 
varies with $N$ and $M$, we next consider 
 a strong-coupling limit, $J_K \to \infty$, of 
the SU($N$) Kondo model.
In this case,  
the impurity electrons and 
 the {\it adjacent\/} conduction electrons 
which directly couple to the impurity electrons via $J_K$, 
are decoupled from the rest of the conduction-electron 
degrees of freedom. 
Thus, the Hamiltonian for the impurity and {\it adjacent\/} 
conduction electrons can be described 
in an $N^2$-dimensional Hilbert space, 
and can be diagonalized. 
For finite $J_K$, there are quantitative corrections 
due to the  rest of the conduction-electron degrees of freedom. 
Nevertheless, the effective Kondo coupling $\widetilde{J}_K$ significantly 
increases at low energies, as it can be expected from  
the poor-man's scaling theory\cite{PoormansSC,NozieresBlandin}
(see appendix \ref{sec:Poorman_SUn}).
Therefore, qualitatively, the fixed points of the renormalization group  
can be classified according to eigenvectors of 
the two-site model describing the $J_K \to \infty$ limit.

Nozi\`{e}res and Blandin gave a brief important 
statement in the footnote 9 of 
their well-known paper, without providing details.\cite{NozieresBlandin} 
It perfectly describes the ground-state wavefunction 
in the strong exchange-interaction limit: 
for general $N$ and $M$ the ground state is a nondegenerate singlet 
 consisting of $M$ impurity-electrons and    
 $N - M$ adjacent conduction electrons $a_m$. 
These $N$ electrons are distributed evenly  
into the one-particle levels 
with {\it different\/} orbital index \lq\lq$m$'',  
and this state 
 describes a fixed point with the usual Fermi-liquid behavior.  
An explicit proof has been provided later   
by Parcollet {\it et al\/}, 
applying a representation theory to the SU($N$) Kondo model.\cite{ParcolletPRB}
Note that the statement of Nozi\`{e}res and Blandin 
is based on the Coqblin-Schrieffer form of 
the exchange interaction,\cite{CoqblinSchrieffer}   
which can also be written in the SU($N$) Kondo form.
We will discuss this singlet state more precisely in this section.

For certain special finite values of the exchange coupling,  
Affleck has shown that, at half-filling $M=N/2$ for even $N$,  
the SU($N$) Kondo impurity can be absorbed into 
that of the orbital,  or {\it flavor\/}, 
sector of the conduction electron degrees of freedom, 
using the non-Abelian bosonization approach.\cite{AffleckNPhys} 
For this special case, 
the Hamiltonian can be diagonalized using the Kac-Moody algebra, 
and  it shows that the excitation spectrum are described by 
the quasiparticle excitations of the local Fermi liquid.
Low-energy Fermi-liquid properties of the 
SU($N$) Kondo model have been also been studied  
away from half-filling,\cite{MoraSUnKondoI,MoraSUnKondoII}
extending Nozi\`{e}res's description of 
the local Fermi liquid.\cite{NozieresFermiLiquid}

\subsubsection{ 
Coqblin-Schrieffer model vs.\  SU($N$)-Kondo model}
\label{subsec:CoqblinSchrieffer_SUn_Kondo_A}

To describe the singlet state of Nozi\`{e}res and Blandin in more detail,
we consider the case  $\Delta \ll U$ 
 where the hybridization energy 
is much smaller than the Coulomb interaction. 
Since the eigenvalue of $\mathcal{H}_{d}^{}$ 
is $E_M \equiv \varepsilon_d M + U M(M-1)/2$, 
the impurity contains $M$ electrons 
 in the atomic limit $v\to 0$  for 
\begin{align}
-M U < \varepsilon_d < -(M-1)\, U \;.
\label{eq:M_electron_step}
\end{align}
For large Coulomb interactions $U \gg \Delta$, 
the effective Hamiltonian for the subspace 
with fixed $M$ impurity-electrons 
can be obtained through the second-order perturbation 
in $\mathcal{H}_{T}^{}$,\cite{CoqblinSchrieffer}
\begin{align}
& 
\!\!\!  
\mathcal{H}_\mathrm{eff}^{}
\,\equiv  \    
\mathcal{H}_{T}^{}\, 
\frac{1}{E_M-(\mathcal{H}_{d}^{}+\mathcal{H}_{c}^{} )}
\, \mathcal{H}_\mathrm{T}^{} 
\ \simeq  \ \, 
\mathcal{H}_{K}   + \mathcal{H}_\mathrm{ps}  ,
\label{eq:Coqblin-Schrieffer_def} 
\\
& 
\!\!\!
\mathcal{H}_{K} \equiv     
\frac{J_K}{2}
\sum_{mm'} 
\left( 
 a_{m}^{\dagger} a_{m'}^{} 
d_{m'}^{\dagger} d_m^{} 
- \frac{1}{N}   
 a_{m}^{\dagger} a_m^{}
d_{m'}^{\dagger} d_{m'}^{}
 \right)   ,
\label{eq:Coqblin-Schrieffer} 
\\
%
& \!\!\!
J_K\equiv \, 2\left(
 \frac{v^2}{E_{M-1}-E_{M}} 
 + 
 \frac{v^2}{E_{M+1}-E_{M}} 
\right) 
\ > 0 
,  
\label{eq:JK}
\\
& \!\!\!
\mathcal{H}_\mathrm{ps} \equiv  \,  
V_\mathrm{ps}\, 
\sum_{m} a_{m}^{\dagger} a_{m}^{} \;,
\label{eq:V_potential}
\\ 
& 
\!\!\! 
V_\mathrm{ps} \equiv \,   
\frac{v^2}{E_{M-1}-E_{M}} 
\frac{M}{N}
-\frac{v^2}{E_{M+1}-E_{M}} 
\Bigl(1-\frac{M}{N}\Bigr)
.
\label{eq:V_potential_coeff}
\end{align}
The exchange coupling is positive $J_K >0$,   
for $\varepsilon_d$ given in Eq.\ \eqref{eq:M_electron_step}.
We focus on the exchange interaction $\mathcal{H}_{K}$ 
and will omit  the potential 
scattering term $\mathcal{H}_\mathrm{ps}$ 
in the following.

Equation \eqref{eq:Coqblin-Schrieffer} is the usual 
Coqblin-Schrieffer form of the exchange interaction,
 applicable to general $M$. 
It  can also be  written 
in the SU($N$) Kondo form,\cite{AffleckNPhys,ParcolletPRB}
using an identity shown in appendix \ref{susec:CS_to_SUnKondo}, 
\begin{align}
 \mathcal{H}_{K} 
\, = \, J_K\, \left(\bm{a}^\dagger \bm{T}^{\mu} \bm{a}\right) \,
  \left(\bm{d}^\dagger \bm{T}^{\mu} \bm{d} \right) .
\label{eq:SUn_Kondo_original} 
\end{align}
Here, $\bm{a}^\dagger \equiv ( a_{1}^\dagger,\ldots , a_{N}^\dagger )$
is a row vector of the operators.
We use the Einstein convention for Greek superscripts, 
namely the repeated ones are summed.  
The SU($N$) generators $\bm{T}^{\mu}$ for $\mu=1,2,\ldots N^2-1$  
are traceless $N \times N$ Hermitian matrices 
of the {\it fundamental\/} representation,  
satisfying the commutation relations,
\begin{align}
&  
\left[ \,
\bm{T}^{\mu} 
,\,
\bm{T}^{\nu} \,\right] \,=\, i f^{\mu\nu\lambda}\, \bm{T}^{\lambda} , 
%
 \label{eq:Commutation_fundamental} 
\end{align}
where $f^{\mu\nu\lambda}$ is the structure factors. 
Explicit expressions for $\bm{T}^{\mu}$ are given in Eqs.\  
\eqref{eq:SUn_Generator_explicit_1}--\eqref{eq:SUn_Generator_explicit_3}.
The local \lq\lq spin''  $\bm{d}^\dagger \bm{T}^{\mu} \bm{d}$ 
for $M$ impurity-electrons 
can be expressed in terms of the matrices,   
as shown in appendix \ref{subsec:M_electron_mat}:  
\begin{align}
 \mathcal{H}_{K}^{(M)} 
\, = \, J_K\, \left(\bm{a}^\dagger \bm{T}^{\mu} \bm{a}\right) \,
  \mathcal{S}_{r_M^{}}^{\mu} .
\label{eq:SUn_Kondo_spin} 
\end{align}
Here, a set of the matrices $\mathcal{S}_{r_M^{}}^{\mu}$ 
consists a $\binom{N}{M}$ dimensional representation of the SU($N$),
which we denote $r_M^{}$, and satisfies the same commutation relations 
as Eq.\ \eqref{eq:Commutation_fundamental},   
\begin{align}
 \left[ \,
\mathcal{S}_{r_M^{}}^{\mu} \,,\, \mathcal{S}_{r_M^{}}^{\nu} \,\right] \,=\, 
i f^{\mu\nu\lambda} \mathcal{S}_{r_M^{}}^{\lambda} \;.
 \label{eq:Commutation_rM} 
\end{align}

The Kondo form of the exchange interaction, 
Eq.\ \eqref{eq:SUn_Kondo_original} or Eq.\ \eqref{eq:SUn_Kondo_spin}, 
 makes the symmetric property of the Hamiltonian explicit. 
For instance, with this Kondo Hamiltonian, 
the one-loop scaling equation can be calculated simply 
following along the same line  carried out for the SU($2$) case, 
replacing the spin matrices for the conduction 
and impurity electrons by the SU($N$) ones $\bm{T}_{}^{\mu}$ 
and $\mathcal{S}_{r_M^{}}^{\mu}$, 
 respectively (see appendix \ref{sec:Poorman_SUn}).
To the one-loop order, 
the scaling equation does not depend on $M$ 
as shown in Eq.\ \eqref{eq:scaling_SUn},
but a factor that is  proportional to $N$ gives 
the Kondo temperature of the form  
$T_K = D  \exp\bigl[ -\frac{2}{N \rho_c J_{K}}\bigr]$.

One thing we would like to emphasize in this section is that 
 a {\it hole picture\/}, which we introduce for the conduction electrons  
such that 
\begin{align}
b_m^{} \equiv  a_m^{\dagger}, \qquad \qquad 
h_{\epsilon m}^{} \equiv c_{\epsilon m}^{\dagger},
\label{eq:b_operator}
\end{align}
keeping the impurity electrons $d_{m}$ unchanged,   
becomes a natural description of the singlet ground state for $N>2$.
With these conduction holes, 
the Hamiltonian 
Eq.\ \eqref{eq:SUn_Kondo_original} takes the form, 
\begin{align}
\mathcal{H}_K^{} \,=  
& \ J_K \Bigl( 
\bm{b}^\dagger \overline{\bm{T}}^{\mu}   \bm{b} \Bigr)
    \Bigl(\bm{d}^\dagger \,\bm{T}^{\mu}\, \bm{d} \Bigr)  \;.
\label{eq:SUn_Kondo_conjugate}
\end{align}
The matrices 
 $\overline{\bm{T}}^{\mu} \equiv \left(- \bm{T}^{\mu} \right)^*$ 
satisfy the same commutation relations 
as Eq.\ \eqref{eq:Commutation_fundamental}, 
namely the {\it hole picture\/} correspond to 
the {\it conjugate representation\/} of the SU($N$).
We note that the {\it hole picture\/} 
of this form is suitable for $M \leq N/2$. 
It should be modified in an opposite way for $M > N/2$,   
applying the electron-hole transformation to the impurity-electrons 
keeping conduction electrons unchanged.
In the following, we assume $0<M\leq N/2$   
since the other case corresponds to the charge conjugate  
defined with respect to the whole electrons.

\subsubsection{Singlet state in $J_K \to \infty$ limit 
for $M$ impurity-electrons }

\label{subsubsec:electron_picture_TAR}

As the renormalized exchange coupling $\widetilde{J}_K$, 
defined in Eq.\ \eqref{eq:scaling_SUn}, grows large 
at low energies,  
behavior in the strong coupling limit $J_K\to \infty$   
determines the fixed-point of the renormalization group 
in the first approximation.\cite{WilsonRMP}  
In this limit, the wavefunction is determined by 
diagonalizing $\mathcal{H}_K$ consisting of 
 the impurity electrons and the adjacent conduction electrons $a_m$,   
neglecting $\mathcal{H}_c$ the bulk part of the conduction electrons.

The number of the adjacent conduction holes,  
 $N_b=\sum_m b_m^{\dagger }b_m^{}$, is conserved in this limit.
Therefore, 
the wavefunction for given $M$  and  $N_b$ 
can be expanded using a direct-product basis set 
which consists of $\binom{N}{M} \otimes \overline{ \binom{N}{N_b} }$ states,  
where the bar on the top of the binomial coefficient for the {\it holes\/} 
is a label assigned for 
the {\it conjugate representation\/}.\cite{GeorgiLieAlg,PeskinSchroeder}   
 The Hilbert space 
can be decomposed into a direct sum of irreducible representations,
in a similar way such that the product states 
of 2 spins are decomposed 
into $2\otimes 2 = 1 \oplus 3$ in the SU($2$) case. 
The decomposition for the 2-site SU($N$) Kondo model  
has been carried out by Parcollet {\it et al\/} 
in the electron picture.\cite{ParcolletPRB} 
They showed that 
a 
 {\it one-dimensional\/} representation 
emerges at $N_b=M$ as a Young tableau of a single 
column with the {\it greatest possible number of boxes\/} $N$,  
 and it becomes the ground state for $J_K \to \infty$. 
This state corresponds 
to a {\it totally antisymmetric representation\/} (TAR),  
and in our {\it hole picture\/} it can be interpreted as 
\begin{align}
\binom{N}{M} \otimes \overline{ \binom{N}{M} }
\, =\  1 \, \oplus \, \cdots \, .
\label{eq:quark_antiquark}
\end{align}
For example, in the SU($4$) case, 
the product states for $N_b=M$ can be decomposed into 
$4\otimes\overline{4} = 1\oplus 15$ for $M=1$, and 
$6\otimes 6 = 1\oplus 15 \oplus 20$ for $M=2$.
Note that, at half-filling $M=N/2$ for even $N$,   
the conjugate representation for the holes 
$\bigl(-\mathcal{S}_{r_M^{}}^{\mu}\bigr)^*$ 
becomes equivalent to $r_M^{}$ which is 
for the electrons $\mathcal{S}_{r_M^{}}^{\mu}$.

This {\it nondegenerate singlet\/} state 
emerges in the case where the total number of electrons  
becomes $N=M+N_a$ with $N_a\equiv N-N_b$ the number of adjacent 
conduction electrons,  and can be  explicitly expressed 
in the form   
 \begin{align}
& 
\!\!\!\!
\left| \mathrm{TAR} \right\rangle_M  
 \equiv    
\frac{1}{\sqrt{\binom{N}{M}}\ (N-M)!M!} 
\nonumber \\
&    
\!\!\!
\times 
 \sum_{\{j\}} \mathrm{sgn}(\{j\})\,  
 d_{j_1}^{\dagger} 
 d_{j_2}^{\dagger} \cdots 
 d_{j_M}^{\dagger}
 a_{j_{M+1}}^{\dagger} \cdots 
 a_{j_{N-1}}^{\dagger} 
 a_{j_{N}}^{\dagger}
 | 0 \rangle .
 \label{eq:TAR}
 \end{align}
 Here,  
 $\{j\}=\{j_1,j_2,\ldots,j_N\}$ is a set 
 of $N$ integers $1, 2,\ldots ,N$, and  
 the summation extends over all the $N!$ permutations 
with the sign factor $\mathrm{sgn}(\{j\})=+1$ or $-1$ 
being taken according to whether $\{j\}$ is even or odd.
This wavefunction can also be 
 rewritten in a totally symmetric form  
in terms of the {\it conduction holes\/} 
through a simple rearrangement of the conduction-electron operators, 
\begin{align}
& 
\!\!\!\!\!\!\!
\left| \mathrm{TAR} \right\rangle_M  
 \equiv    
\nonumber \\
&    
\frac{1}{\sqrt{\binom{N}{M}}} 
 \sum_{j_1<j_2<\cdots<j_M} 
 d_{j_1}^{\dagger} 
 a_{j_{1}}^{} \,
 d_{j_2}^{\dagger} 
 a_{j_{2}}^{} \,\cdots \,
 d_{j_M}^{\dagger}
 a_{j_{M}}^{}
 | \widetilde{0} \rangle , 
%
 \label{eq:TAR_boson}
 \end{align}
where $| \widetilde{0} \rangle  \equiv 
 a_{1}^{\dagger} a_{2}^{\dagger}\cdots  a_{N}^{\dagger} | 0 \rangle$.
It can also be interpreted as a bosonic wavefunction 
for the particle-hole pairs $d_{j}^{\dagger} a_{j}^{}$.
The TAR truly corresponds to the singlet ground state 
that Nozi\`{e}res and Blandin mentioned.\cite{NozieresBlandin,ParcolletPRB}
For finite $J_K$, this state evolves to a Kondo singlet, 
 involving  low-energy conduction electrons 
far away from the impurity site. Successive inclusion 
of such  conduction electrons can be carried out with the NRG, 
and the low-lying excitations show the SU($N$) 
Fermi-liquid behavior.

\section{
Fermi-liquid fixed point as a bosonic Perron-Frobenius vector}
\label{sec:Hard_core_boson}

We describe more precisely the interpretation of 
the TAR in terms of a {\it hard-core boson\/},  
which is a composite particle 
consisting of one impurity-electron and one adjacent conduction-hole 
with the same {\it flavor\/}  \lq\lq$m$'' 
appearing  in Eq.\ \eqref{eq:TAR_boson}. 
With a basis set of the  hard-core bosons,
all the off-diagonal elements of the Hamiltonian 
becomes {\it non-positive\/}, 
for which the Perron-Frobenius theorem is applicable. 
It clearly explains why the TAR becomes   
the {\it unique\/} lowest-energy state 
that can be written as a {\it nodeless bosonic\/} wavefunction  
in the form of Eq.\  \eqref{eq:TAR_boson}.   
This description does {\it not\/} require the SU($N$) symmetry, 
and the Perron-Frobenius ground state 
remains robust against perturbations which break the SU($N$) symmetry 
as long as the off-diagonal elements of the Hamiltonian 
can be kept {\it non-positive\/}.

\subsection{Hard-core bosons in the SU($N$) symmetric case}
\label{subsec:boson_SUn}

The discussions given in Sec.\ \ref{subsubsec:electron_picture_TAR}
 were based on a group-representation description of the SU($N$). 
In order to gain physical insights into the nondegenerate state, 
we now go back to the Coqblin-Shrieffer form,  
but with the {\it hole picture\/} 
using Eq.\ \eqref{eq:b_operator},
\begin{align}
\mathcal{H}_{K} 
=   
\frac{J_K}{2}
\left( 
-
\sum_{mm'} 
d_{m'}^{\dagger}  b_{m'}^{\dagger} b_{m}^{}  d_m^{} 
+ \frac{1}{N}   
\sum_{mm'} b_{m}^{\dagger} b_m^{} 
d_{m'}^{\dagger} d_{m'}^{}
 \right) .
\label{eq:CF_hard_core_boson}
\end{align}
For $J_K \to \infty$, 
the second term in the bracket becomes a constant $N_b M/N$,  
taking the summation over $m$ and $m'$.
The first term can be interpreted as 
a {\it tunneling\/} Hamiltonian for    
the composite hard-core bosons, 
$Q_{m}\equiv b_{m}^{} d_m^{}$  and $Q_{m}^2=0$, 
 which consists of one impurity-electron and one conduction-hole 
of the same flavor \lq\lq$m$''. 
This Hamiltonian is also equivalent to 
a reduced BCS model, 
for which exact solution was obtained by Richardson\cite{Richardson67} 
 in the context of nuclear physics,  
and it was also applied to ultra-small-grain superconductors 
by von Delft and Braun.\cite{vonDelft2000} 
We apply the same approach to 
the ground state of Eq.\ \eqref{eq:CF_hard_core_boson} 
for $M$ impurity-electrons filling.

In our case, the hard-core bosons can {\it hop\/} around, 
via the matrix element $J_K$, onto the vacant orbitals,  
which are {\it not\/} occupied by other hard-core bosons 
nor the immobile unpaired objects such as  
a singly-occupied impurity-electron and 
a singly-occupied conduction-hole. 
The number of the composite pairs $N_\mathrm{pair}$ is conserved. 
Similarly, the number of unpaired impurity-electrons 
 $N_{\mathcal{O}d}=M-N_\mathrm{pair}$ 
and that of the unpaired conduction-holes  
$N_{\mathcal{O}h}=N_b-N_\mathrm{pair}$ are also conserved. 
Thus, the number of the {\it unblocked\/} orbitals 
which are {\it not\/} occupied by these unpaired objects 
becomes  $N_\mathcal{UB}^{} \equiv N-N_{\mathcal{O}d}-N_{\mathcal{O}h}
 =N-M-N_b+2N_\mathrm{pair}$.  
Symbolically, $\mathcal{O}_d$ and  $\mathcal{O}_h$ 
denote a set of orbitals occupied 
by the unpaired impurity-electrons and unpaired conduction-holes, 
respectively,  
while $\mathcal{UB}$ denotes 
a set of {\it unblocked\/} orbitals.\cite{vonDelft2000} 
Taking $N_\mathrm{pair}$ as a quantum number, 
the eigenstates can be expanded in the form 
\begin{align}
&
\!\!\!\!
\left| N_\mathrm{pair}; \left\{m_h\right\}, \left\{m_d\right\} \right\rangle 
\equiv  
\! \prod_{m_h\in \mathcal{O}_h}
\!\!  b^{\dagger}_{m_h}  
\! \prod_{m_d\in \mathcal{O}_d}
\!\!  d^{\dagger}_{m_d} 
\left |\Psi_\mathrm{pair} \right \rangle, 
\label{eq:basis_HCboson_1}
\\
& 
\!\!\!\!
\left |\Psi_\mathrm{pair} \right \rangle = 
\nonumber \\
& 
\sum_{j_1<\ldots<j_{N_\mathrm{pair}} \in \mathcal{UB}}
\psi_{\mathrm{pair}}(j_1,\ldots,j_{N_\mathrm{pair}})\,
Q_{j_1}^{\dagger}   
Q_{j_2}^{\dagger} \cdots   
Q_{j_{N_\mathrm{pair}}}^{\dagger} 
\!  
| \widetilde{0}\rangle  
.
\label{eq:basis_HCboson_2}
\end{align}
The level indices $\{m_h\}=\{m_{h,1},\ldots,m_{h,N_{\mathcal{O}_h}}\}$
and $\{m_d\}=\{m_{d,1},\ldots,m_{h,N_{\mathcal{O}_d}}\}$  
for the unpaired objects can also be regarded as quantum numbers. 
The vacuum for the conduction holes is defined such that
 $b_{j}^{} | \widetilde{0}\rangle=0$ and 
 $d_{j}^{} | \widetilde{0}\rangle=0$.
The configuration of 
these unpaired objects does not affect the energy while it causes 
the degeneracy $g_{M,N_b,N_\mathrm{pair}}^{}$ of the eigenstates, 
\begin{align}
g_{M,N_b,N_\mathrm{pair}}^{} \,=\, 
\frac{N!}{\left(M-N_\mathrm{pair}\right)! \left(N_b-N_\mathrm{pair}\right)!\,
N_\mathcal{UB}^{}!} \;.
\end{align}
In Eq.\ \eqref{eq:basis_HCboson_2}, the summation for each of $j$'s
runs over 
{\it unblocked\/} orbitals with 
a requirement $j_1<\ldots<j_{N_\mathrm{pair}}$. 
The pair wavefunction 
$\psi_{\mathrm{pair}}(j_1,\ldots,j_{N_\mathrm{pair}})$ 
is determined by the Sch\"{o}dinger equation   
$
\mathcal{H}_{K} 
\left |\Psi_\mathrm{pair}\right \rangle =   \epsilon  
\left |\Psi_\mathrm{pair}\right \rangle 
$, which describes the self-avoiding motion of the $N_\mathrm{pair}$ 
hard-core bosons onto the $N_\mathcal{UB}^{}$ vacant orbitals. 
With this basis set, the Hamiltonian for the pairs is written as  
a $\binom{N_\mathcal{UB}^{}}{N_\mathrm{pair}}$ dimensional matrix.
It has  $\left(N_\mathcal{UB}^{}-N_\mathrm{pair}\right)N_\mathrm{pair}$ 
 non-zero off-diagonal elements in each row, and in each column, 
as the total number of the allowed {\it hopping\/} processes is given by  
 product of the number of vacant orbitals and the number of pairs. 
Furthermore, all such non-zero off-diagonal elements 
take the same negative value $-J_K/2<0$ 
owing to the bosonic commutation relation between the two different pairs. 
To those matrices of this form, 
the Perron-Frobenius theorem is applicable  
since the {\it connectivity\/} condition necessary for this theorem   
also holds for the {\it pair hopping\/} 
in this $\binom{N_\mathcal{UB}^{}}{N_\mathrm{pair}}$ dimensional space. 
The Perron-Frobenius theorem states 
that the lowest-energy state becomes nondegenerate 
and has a nodeless eigenvector,\cite{Tasaki}
which in our case is  
a uniform $\binom{N_\mathcal{UB}^{}}{N_\mathrm{pair}}$ dimensional vector. 
Thus, the lowest energy state in each subspace labeled 
by the quantum number $(M,N_b,N_\mathrm{pair})$ 
 can be explicitly written in the form, 
\begin{align}
&\psi_{\mathrm{pair}}^\mathrm{min}(j_1,\ldots,j_{N_\mathrm{pair}}) 
\,=\,  
\frac{1}{\sqrt{
\binom{N_\mathcal{UB}^{}}{N_\mathrm{pair}}
}}\; , 
\label{eq:psi_M_Nb}
\\
&\epsilon_{M,N_b,N_\mathrm{pair}}^\mathrm{min} 
=\frac{J_K}{2}  \left[ 
-\left(N_\mathcal{UB}^{}  -N_\mathrm{pair}  +1\right) 
N_\mathrm{pair}+ \frac{N_bM}{N} 
\right] .
\label{eq:energy_M_Nb}
\end{align}
It takes the absolute minimum at $N_b=N_\mathrm{pair}=M$,   
where 
all the orbitals are {\it unblocked\/} $N_\mathcal{UB}^{}=N$.
Thus, the ground state is nondegenerate and 
 is identical to the TAR, given in
Eqs.\  \eqref{eq:TAR} and \eqref{eq:TAR_boson},
\begin{align}
\epsilon_{M}^\mathrm{GS} 
 =& \ - J_K  \, \frac{M(N-M)(N+1)}{2N}  
\label{eq:energy_hard_core_boson} ,
\\
\left |\Psi \right \rangle_{M}^\mathrm{GS} 
= & \ 
\frac{1}{\sqrt{\binom{N}{M}}\ M!}\,
\left[\, 
\sum_{m=1}^N  d_{m}^{\dagger} b_{m}^{\dagger}
\, \right]^{M}
| \widetilde{0}\rangle 
\;.
\label{eq:wavefunction_hard_core_boson} 
\end{align}
The energy can also be expressed in terms of the Casimir operator, 
 defined in appendix \ref{subsec:M_electron_mat}, as  
 $\epsilon_{M}^\mathrm{GS}= -J_K  C_2(r_M^{})$. It 
 reproduces the result of Pacollet{\it et al\/},\cite{ParcolletPRB} 
which is obtained using an addition rule 
for the SU($N$) {\it flavors\/}
of the impurity and conduction electrons.\cite{GeorgiLieAlg,PeskinSchroeder}

\begin{widetext}

\begin{figure}[t]
\begin{center}
 \begin{minipage}{0.84\linewidth}
\includegraphics[width=\linewidth]{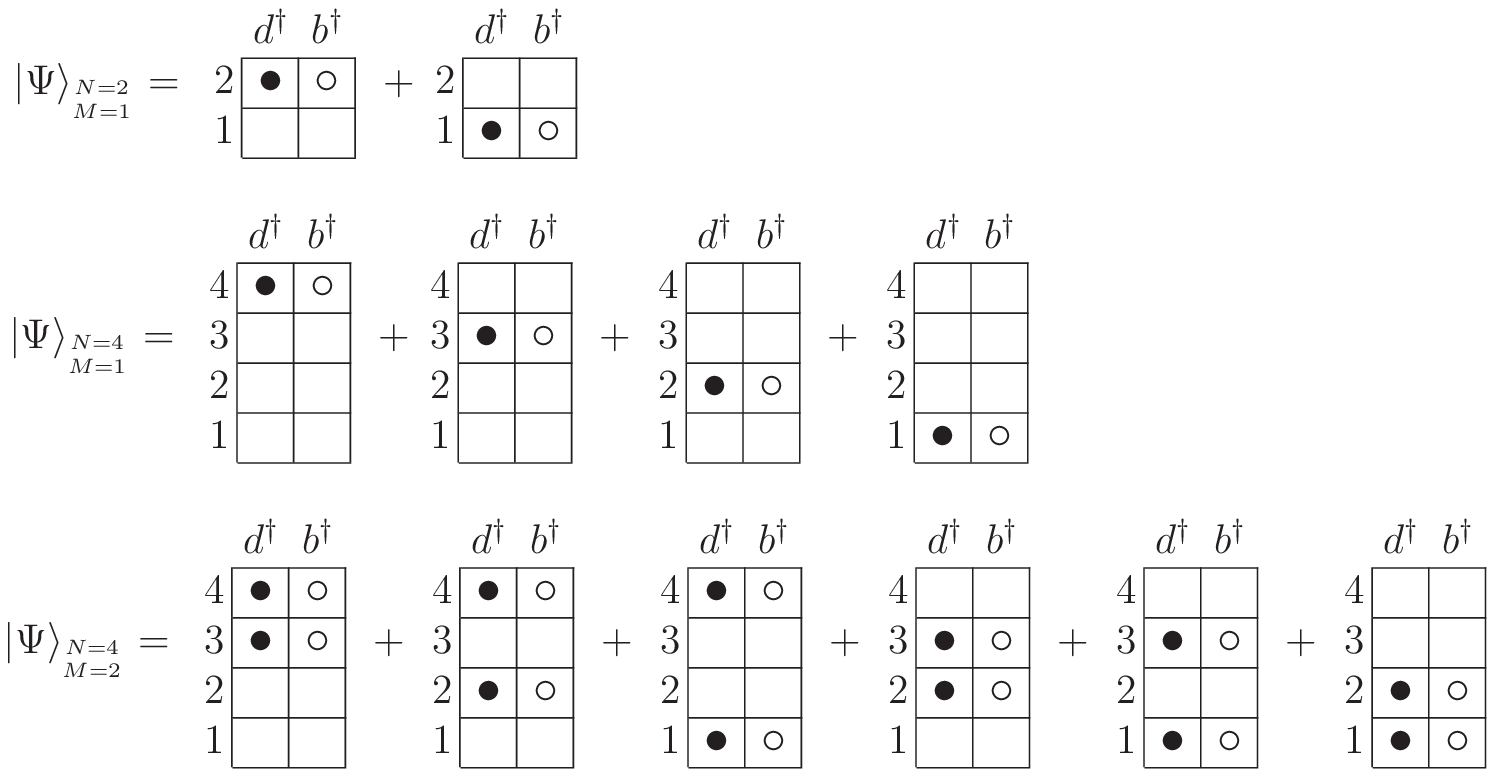}
 \end{minipage}
\end{center}
\vspace{-0.3cm}
\caption{
Schematic picture of the nondegenerate singlet ground state 
of the SU($4$) Kondo model with  $M=1$ and $M=2$ impurity-electrons  
 in the limit of $J_K \to \infty$. 
The wavefunction for the SU($2$) case with $M=1$ is 
also illustrated on the top for comparison.
Each row represents {\it flavor\/}  \lq\lq$m$'' of the levels $m=1,2,3,4$.
The first ($d^\dagger$) and second ($b^\dagger$) columns represent 
orbitals in the impurity and the adjacent conduction site, respectively, 
with ($\bullet$) the impurity-electrons and ($\circ$) the conduction-holes. 
The explicit expressions for these states are given in 
Eqs.\ \eqref{eq:TAR_boson_N4M1} and \eqref{eq:TAR_boson_N4M2}, respectively.
}
\label{fig:PF_vector}
\end{figure}

\end{widetext}

The wavefunction  $\left |\Psi \right \rangle_{M}^\mathrm{GS}$ 
describes how the pairs distribute in the {\it flavor\/} space. 
%
For example, in the SU($2$) case, it takes the form
\begin{align}
\left |\Psi \right \rangle_{\! N=2 \atop \! M=1}^\mathrm{} 
= & \  
\frac{1}{\sqrt{2}}
\left( 
d_{1}^{\dagger}b_{1}^{\dagger} 
+ d_{2}^{\dagger} b_{2}^{\dagger} 
\right)
| \widetilde{0} \rangle 
\label{eq:TAR_boson_N2M1}
\\
= & \  
\frac{1}{\sqrt{2}}
\left( 
d_{1}^{\dagger}a_{2}^{\dagger} 
- d_{2}^{\dagger} a_{1}^{\dagger} 
\right)
| 0 \rangle , 
\nonumber
\end{align}
since $| \widetilde{0} \rangle = a_{1}^{\dagger}a_{2}^{\dagger} | 0 \rangle$ 
for $N=2$.
Our main interest is in the SU($4$) symmetric CNT quantum dots, 
for which the ground state for $M=1$ and that for $M=2$ 
in the $J_K \to \infty$ limit are given, respectively, by
\begin{align}
\left |\Psi \right \rangle_{\! N=4 \atop \! M=1}^\mathrm{} 
\,= &  \  
\frac{1}{2}
\left( 
  d_{1}^{\dagger} b_{1}^{\dagger} 
+ d_{2}^{\dagger} b_{2}^{\dagger} 
+ d_{3}^{\dagger} b_{3}^{\dagger} 
+ d_{4}^{\dagger} b_{4}^{\dagger} 
\right)
| \widetilde{0}\rangle  ,
\label{eq:TAR_boson_N4M1}
\\
\left |\Psi \right \rangle_{\! N=4 \atop \!  M=2}^\mathrm{} 
= &  \  
\frac{1}{\sqrt{6}}
\Bigl( 
d_{1}^{\dagger} b_{1}^{\dagger} \,
d_{2}^{\dagger} b_{2}^{\dagger} 
+
d_{1}^{\dagger} b_{1}^{\dagger} \,
d_{3}^{\dagger} b_{3}^{\dagger} 
+
d_{1}^{\dagger} b_{1}^{\dagger} \,
d_{4}^{\dagger} b_{4}^{\dagger} 
\nonumber \\
&   
+
d_{2}^{\dagger} b_{2}^{\dagger} \,
d_{3}^{\dagger} b_{3}^{\dagger} 
+
d_{2}^{\dagger} b_{2}^{\dagger} \,
d_{4}^{\dagger} b_{4}^{\dagger} 
+ 
d_{3}^{\dagger} b_{3}^{\dagger} \,
d_{4}^{\dagger} b_{4}^{\dagger} 
\Big)
| \widetilde{0}\rangle  .
\label{eq:TAR_boson_N4M2}
\end{align}
In Fig.\ \ref{fig:PF_vector}, 
the configuration of the electrons and holes in the ground state 
 of these cases are schematically illustrated.
Each of $M$ composite electron-hole pairs is in the same row 
labeled by the {\it flavor\/} \lq\lq$m$''. 
These pairs are uniformly distributed along the vertical direction 
in the $N$ different rows, 
constructing a nondegenerate wavefunction 
as a  nodeless linear combination 
in which all the coefficients are identical.
This agrees with the statement of 
 Nozi\`{e}res and Blandin stated given in the footnote 9 
of their paper,\cite{NozieresBlandin} 
namely the singlet ground state for $M$ impurity electrons 
is constructed with $N-M$ adjacent conduction electrons. 

The charge distribution of this form 
can also be understood from an Anderson-impurity point of view. 
Suppose the situation where $U=0$ and $\varepsilon_d \ll - \Delta$. 
The noninteracting ground state is nondegenerate, 
and the impurity site is almost full filled by $N$ electrons. 
If the Coulomb repulsion $U$ is switched on gradually keeping 
the level position $\varepsilon_d$ unchanged, 
the number of impurity-electrons will decrease 
as the ground state evolves continuously. 
At a certain value of $U$, 
the average number of impurity-electrons becomes $M$ and the 
phase shift becomes  $\delta=\pi M/N$ as 
totally $N-M$ electrons have already moved towards the conduction band.

For finite exchange interaction $J_K$,  
the wavefunction $\left |\Psi \right \rangle_{M}^\mathrm{GS}$  
evolves to the Kondo singlet state which shows 
the SU($N$) Fermi-liquid behavior. 
There is also an interesting analogy with the SU($N$) gauge theory, 
in which the TAR corresponding to Fig.\ \ref{fig:PF_vector}  
describes a $M$-meson state 
that is constructed from the quarks ($\bullet$) 
 and anti-quarks ($\circ$).\cite{GeorgiLieAlg,PeskinSchroeder}

\subsection{
Singlet ground state without SU($N$) symmetry}
\label{subsec:boson_general}

We next discuss an evolution of the nondegenerate ground state 
of Nozi\`{e}res and Blandin in the case without the SU($N$) symmetry. 
In the hard-core boson interpretation mentioned in the above, 
the singlet state corresponds to the Perron-Frobenius eigenvector,  
which is a linear-combination of all the basis vectors 
with strictly {\it positive\/} coefficients.\cite{Tasaki} 
Specifically, in the SU($N$) symmetric case, 
all the coefficients are identical, or {\it uniform\/}, 
as seen in Eq.\ \eqref{eq:psi_M_Nb}.
For some perturbations which breaks the SU($N$) symmetry,  
the Perron-Frobenius eigenvector can continuously evolve  
to a vector with {\it non-uniform\/}  coefficients 
preserving the {\it nodeless\/} structure.

In order to given an explicit example, 
we examine the case where each of the one-particle 
energies $\epsilon_m$ takes a certain value bounded 
in the range $\delta \epsilon$ 
near the middle of the $M$-electron region 
defined in  Eq.\ \eqref{eq:M_electron_step},
\begin{align}
\epsilon_{m} =-\left( M - \frac{1}{2} \right) U + \delta \epsilon_m, 
\quad \    -\frac{\delta \epsilon}{2} < 
\delta \epsilon_m <\frac{\delta \epsilon}{2}.
\end{align}
Specifically, we only take into account an extended 
exchange-interaction term derived for this case, 
neglecting the potential scattering term. 
Details for this anisotropic exchange coupling, $\mathcal{H}_K^\mathrm{ais}$, 
are described in appendix \ref{sec:anisotropic_exchange}. 

The fixed-point Hamiltonian in the strong exchange interaction limit 
can be deduced from $\mathcal{H}_K^\mathrm{ais}+\mathcal{H}_d$,  
which can be expressed in the {\it hole-picture\/}, as 
\begin{align}
&
\!\!\!\!\!\!\!\!\!
\mathcal{H}_{K}^\mathrm{ais} 
+\mathcal{H}_{d} 
\, = 
\nonumber \\
&  \quad   \sum_{m\neq m'}
\sum_{\{\alpha\}}
-\frac{J_{m'm}^{\{\alpha\}}}{2}\, 
d_{m'}^{\dagger} 
 b_{m'}^{\dagger} 
b_{m}^{} 
d_m^{} 
\left| \{\alpha\} \right\rangle_M^{}  \!
{_M^{}}\!\left\langle \{\alpha\} \right|  
\nonumber \\
& 
+  \sum_{m} \sum_{\{ \alpha \}}
\frac{J_{mm}^{\{\alpha\}}}{2}\, 
 \left(
1 - 
b_{m}^{\dagger} b_{m}^{} 
\right)
d_{m}^{\dagger} d_{m}^{} 
\left| \{\alpha\} \right\rangle_M^{}  \!
{_M^{}}\!\left\langle \{\alpha\} \right|  
\nonumber \\
& -\, 
\frac{\overline{J}M(N-N_b) }{2N}
\ + 
\sum_{\{ \alpha \}}
E_{M}^{\{\alpha\}}
\left| \{\alpha\} \right\rangle_M^{}  \!
{_M^{}}\!\left\langle \{\alpha\} \right|   
\label{eq:CF_hard_core_boson_aniso_2_2site}
\end{align}
Here, $J_{m'm}^{\{\alpha\}}$  and $\overline{J}$ are defined 
in Eqs.\ \eqref{eq:J_aniso_appendix} and \eqref{eq:J_aniso_mean_appendix},
 respectively.
The exchange coupling  $J_{m'm}^{\{\alpha\}}$ 
becomes anisotropic and depends on  
the energy of initial and final $M$ impurity-electrons states.  
The couplings are  positive, $J_{m'm}^{\{\alpha\}} >0$ and $\overline{J}>0$,
for all $m$, $m'$, and $\{\alpha\}$ for large Coulomb interactions 
$U/2 \gg (M + 1/2 ) \delta \epsilon$, 
as shown in Eq.\ \eqref{eq:asym_conditions}.
The diagonal elements, which correspond to 
the last terms in the right-hand side 
of Eqs.\ \eqref{eq:CF_hard_core_boson_aniso_2_2site}, 
depend also on the distribution of 
unpaired impurity-electrons ${\{m_d\}}$ 
as well as the energy of the initial state $E_{M}^{\{\alpha\}}$.  
These are the main differences from the SU($N$) symmetric case.

The off-diagonal elements of $J_{m'm}^{\{\alpha\}}$ 
can also be regarded as the {\it hopping\/} matrix elements 
for the hard-core bosons 
$Q_{m}^{\dagger} \equiv   d_m^{\dagger} b_{m}^{\dagger}$
 as those in the SU($N$) case.
The unpaired particles cannot {\it hop\/} around the 
{\it flavor\/} space also for this anisotropic case. 
Therefore, the number of unpaired impurity-electrons $N_{\mathcal{O}_d}$ 
and that of  unpaired conduction-holes $N_{\mathcal{O}_h}$ are conserved,  
taking values in the range: $N_{\mathcal{O}_d}=0,1,\ldots,M$, and 
$N_{\mathcal{O}_h}=0,1,\ldots,N-M$. Similarly, 
the numbers 
 $N_b=M-N_{\mathcal{O}_d}+N_{\mathcal{O}_h}$, 
$N_\mathrm{pair}=M-N_{\mathcal{O}_d}$, 
and 
$N_\mathcal{UB}^{}=N-N_{\mathcal{O}_d}-N_{\mathcal{O}_h}$   
are also conserved. 

Therefore, the eigenstates of 
$\mathcal{H}_{K}^\mathrm{ais} +\mathcal{H}_{d}$ 
can also be expanded using  
the same basis set as  
Eqs.\ \eqref{eq:basis_HCboson_1} and  \eqref{eq:basis_HCboson_2}:
\begin{align}
&
\!\!\!\!\!\!\!\!\!\!\!\!\!\!\!\!\!\!\!\!\!\!\!
\bigl| \left\{m_\mathrm{p}\right\}; 
\left\{m_h\right\}, \left\{m_d\right\} \bigr\rangle 
\nonumber \\
& \equiv  
\! \prod_{m_h\in \mathcal{O}_h}
\!\!  b^{\dagger}_{m_h}  
\! \prod_{m_d\in \mathcal{O}_d}
\!  d^{\dagger}_{m_d} 
 \prod_{m_\mathrm{p}\in \,\mathrm{pairs}}
\!\! Q_{m_\mathrm{p}}^{\dagger}
\,  |\widetilde{0} \rangle  
\label{eq:basis_aniso_1}
\\
& =  \  
 \prod_{i=1}^{N_b} \,
 b^{\dagger}_{\beta_i}  \, 
 \prod_{j=1}^{M} \, 
 d^{\dagger}_{\alpha_j}  \, |\widetilde{0} \rangle\; . 
\label{eq:basis_aniso_2}
\end{align}
Here, 
$\{\alpha_1,\alpha_2,\ldots,\alpha_M\}$ and  
$\{\beta_1,\beta_2,\ldots,\beta_{N_b}\}$ 
represent a set of occupied impurity-electron levels and that 
of occupied conduction-hole levels, respectively, 
which we symbolically write  $\{\alpha\}=\{m_d\}+\{m_\mathrm{p}\}$ and  
$\{\beta\}=\{m_h\}+\{m_\mathrm{p}\}$ 
as sums of the unpaired and paired elements. 
The coordinates of the unpaired objects 
 $\left\{m_h\right\}$ and $\left\{m_d\right\}$ 
can be regarded as quantum numbers whereas 
the coordinates of the pairs 
$\left\{m_\mathrm{p}\right\}=\{ m_{\mathrm{p},1}^{},
\ldots,m_{\mathrm{p,}N_\mathrm{pair}}^{}\}$  
constitute a  $\binom{N_\mathcal{UB}^{}}{N_\mathrm{pair}}$ 
 dimensional subspace for the pair wavefunction 
 $\psi_{\mathrm{pair}}^{}(m_{\mathrm{p},1}^{},
\ldots,m_{\mathrm{p,}N_\mathrm{pair}}^{})$.


In the representation using the basis set Eq.\ \eqref{eq:basis_aniso_1}, 
the off-diagonal elements of the Hamiltonian matrix 
become {\it negative} or zero: 
there emerge  $\left(N_\mathcal{UB}^{}-N_\mathrm{pair}\right)N_\mathrm{pair}$ 
 {\it negative\/} off-diagonal elements $- J_{m'm}^{\{\alpha\}}/2$ 
in each column, and their Hermitian-conjugate elements emerge in each row. 
Therefore,  similarly to the SU($N$) symmetric case,  
the lowest energy state in each subspace 
is given by the Perron-Frobenius vector, 
and  all the coefficients 
which correspond to the pair wavefunction 
become {\it positive\/}    
$\psi_{\mathrm{pair}}^{}(m_1^{},\ldots,m_{N_\mathrm{pair}}^{})>0$. 
However, the coefficients are {\it not\/} uniform in the anisotropic case. 

The ground state corresponds to 
 the Perron-Frobenius vector 
for the subspace of  $N_{\mathcal{O}_h}=N_{\mathcal{O}_d}=0$, 
where all the impurity-electrons and conduction-holes form the pairs,  
for $U/2 \gg (M +1/2 ) \delta \epsilon$  
as discussed in appendix \ref{subsec:ground-state_anisotropic}.
Thus, the ground state  
for Eq.\ \eqref{eq:CF_hard_core_boson_aniso_2_2site} 
is also a nondegenerate singlet which 
is constructed with the $M$ hard-core bosons,
\begin{align}
\left |\Psi \right \rangle_{M}^\mathrm{GS} 
\,=
 {\sum_{\{m\}}}'
\, 
\psi_{\mathrm{pair}}^\mathrm{GS}(m_1,\ldots,m_{M})\,
Q_{m_1}^{\dagger}   
Q_{m_2}^{\dagger} \cdots   
Q_{m_{M}}^{\dagger} 
\!  
| \widetilde{0} \rangle \! .
\label{eq:basis_HCboson_3}
\end{align}
 Here, 
each component of $\{m\} = \{m_1,m_2,\ldots,m_M\}$
is summed over $1$ to $N$ with the constraint: $m_1<\ldots<m_{M}$.
Alternatively,  it can also be symmetrized 
 using the properties of the hard-core bosons: 
$Q_{m}^2=0$ and $[Q_{m}^{}\,,\,Q_{m'}^{\dagger}]=0$ for $m\neq m'$.
The pair wavefunction is nodeless 
$\psi_{\mathrm{pair}}^\mathrm{GS}(m_1,\ldots,m_{M})>0$ 
but is {\it not\/} a uniform function 
in the case without the SU($N$) symmetry.
It evolves further for finite exchange couplings $J_{mm}^{\{\alpha\}}$,  
and will describe the fixed point with the Fermi-liquid behavior.


\section{Fermi-liquid theory for CNT dots}

\label{sec:CFT_Fermi_liquid}

\subsection{One-particle Hamiltonian for CNT dots}

The one-particle energy levels of the  carbon nanotube 
quantum dots with the energy  $\epsilon_m$ ($m=1,2,3,4$) 
consist of  the spin ($\uparrow$, \!$\downarrow$) 
and valley  ($\mathrm{K}$, \!\! $\mathrm{K'}$) degrees of freedom. 
Owing to the cylindrical geometry of the CNT, 
the valley degrees of freedom capture 
a magnetic moment along the direction of the CNT axis,
and it couples to an external magnetic field parallel 
to the axis.\cite{RMP-Kouwenhoven,MantelliMocaZarandGrifoni,Grifoni_PRB2015} 
Furthermore, the four levels are coupled each other through 
the spin-orbit interaction $\Delta_\mathrm{SO}$ 
and valley mixing $\Delta_\mathrm{KK'}^{}$ term.  
The one-particle energy $\epsilon_m$ is determined 
as an eigenvalue of the following $4\times 4$ matrix $\bm{H}_d^{0}$, 
which includes these couplings.\cite{RMP-Kouwenhoven,MantelliMocaZarandGrifoni,Grifoni_PRB2015} 
Using a basis set consisting of 
the spin ($\uparrow$, \!$\downarrow$) 
and  the valley pseudo spin ($\mathrm{K}$, $\mathrm{K'}$),  
which can be described by the operators 
 $\bm{\psi}^{\dagger}_{d} 
 \equiv  (
 \psi^{\dagger}_{\mathrm{K}\uparrow}, 
 \psi^{\dagger}_{\mathrm{K}\downarrow}, 
 \psi^{\dagger}_{\mathrm{K}'\uparrow}, 
 \psi^{\dagger}_{\mathrm{K}'\downarrow} 
 )$, 
the one-particle part of the CNT-dot Hamiltonian 
can be  written in the form,
\begin{align}
&\mathcal{H}_d^{0} \equiv  \, 
\bm{\psi}^{\dagger}_{d} \,
\bm{H}_d^0 \, 
\bm{\psi}^{}_{d}  \, = \, \sum_{m} 
\epsilon_m d_{m}^{\dagger} d_{m}^{} \;,  
\label{eq:Hd0_micro}
\\
& \bm{H}_d^0 
=  
\, \varepsilon_d \bm{1}_\mathrm{s} \!\otimes\! \bm{1}_\mathrm{orb}
 + \frac{\Delta_\mathrm{KK'}}{2} 
\bm{1}_\mathrm{s} \!\otimes\! \bm{\tau}^x \! 
+ \frac{\Delta_\mathrm{SO}}{2}  \bm{\sigma}^z \!\otimes\! \bm{\tau}^z \! 
 - \overrightarrow{\bm{\mathcal{M}}} \cdot \vec{b},
\label{eq:H_d0_mat_1}
\\
& \overrightarrow{\bm{\mathcal{M}}} \, \equiv  \, 
-\frac{1}{2} \, g_\mathrm{s} \, \vec{\bm{\sigma}} 
\!\otimes\! \bm{1}_\mathrm{orb}
- g_\mathrm{orb} \,
\bm{1}_\mathrm{s} \!\otimes\! \bm{\tau}^z  
 \,\vec{e}_{z}^{} 
.
\label{eq:dot_magnetization}
\end{align}
Here, $\bm{\sigma}^j$ and  $\bm{\tau}^j$ 
for $j=x,y,z$ are the Pauli matrices 
for the spin and the valley pseudo-spin spaces, respectively.  
The Zeeman splitting is determined by the 
total magnetization  
$\overrightarrow{\bm{\mathcal{M}}}$, for which  
 $g_\mathrm{s}= 2$ and  $g_\mathrm{orb}$ are
the g-factors for the spin and valley magnetic moments, 
respectively, $\vec{b}\equiv \mu_B \vec{B}$ is 
the external magnetic field with $\mu_B$ the Bohr magneton, 
  and   
 $\vec{e}_{z}^{}$  
is a unit vector along the nanotube axis. 
In Eq.\ \eqref{eq:H_d0_mat_1}, 
 the SU($4$) invariant component $\varepsilon_d$, 
which can be tuned through the applied gate voltage, 
is also included in the diagonal part
with the unit matrices $\bm{1}_\mathrm{s}$ and $\bm{1}_\mathrm{orb}$.

The  unitary transform  $\bm{\mathcal{U}}_d^{}$ to diagonalize $\bm{H}_d^0$, 
the eigenvalue $\epsilon_m$ 
and the eigenvector $\bm{u}_m$ are defined such that 
\begin{align}
\bm{H}_d^0\, \bm{u}_m =& \  \epsilon_m \,\bm{u}_m \;,
\qquad \quad \ \bm{u}_m^\dagger \cdot \bm{u}_{m'}^{} \,= \, \delta_{mm'} 
\;, 
\label{eq:one_particle_level_CNT_eigen0}
\\
\bm{d}^{\dagger}_{} \,\equiv & \  \bm{\psi}^{\dagger}_{d} 
\,\bm{\mathcal{U}}_d^{} , 
\qquad \qquad 
\bm{\mathcal{U}}_d^{}
=(\bm{u}_{1}^{},\bm{u}_{2}^{},\bm{u}_{3}^{},\bm{u}_{4}^{}) \;. 
\end{align}
Here, $
\bm{d}^{\dagger}_{} 
=  (
d^{\dagger}_{1}, 
d^{\dagger}_{2}, 
d^{\dagger}_{2}, 
d^{\dagger}_{4} 
)$, and we will assign the label $m=1,2,3,4$  
such that $\epsilon_1\leq \epsilon_2\leq \epsilon_3\leq \epsilon_4$ 
unless otherwise noted.
We also note that 
the matrix $\bm{H}_d^0$ can also be expressed 
in the following $4\times 4$ matrix form: 
for  $\vec{b} = b_{\parallel}^{} \vec{e}_{z}^{} 
+ b_{\perp}^{} \vec{e}_{x}^{}$ with 
 $b_{\parallel}^{} \equiv  b \cos{\Theta}$,   
 $b_{\perp}^{} \equiv  b \sin{\Theta}$, and  
 $\Theta$ the angle of the magnetic field relative to the nanotube axis,  
\begin{widetext}
%
\begin{align}
\bm{H}_d^0 \,= 
\left [ \,
\begin{matrix} 
\varepsilon_d +\frac{\Delta_\mathrm{SO}}{2} 
+ ( g_\mathrm{orb} + \frac{g_\mathrm{s}}{2} ) \,b_{\parallel}^{}
& \frac{g_\mathrm{s}}{2}\, b_{\perp}^{} & \frac{\Delta_\mathrm{KK'}}{2}& 0  
\cr 
 \frac{g_\mathrm{s}}{2} \,b_{\perp}^{} 
& \varepsilon_d -\frac{\Delta_\mathrm{SO}}{2}  
+( g_\mathrm{orb} - \frac{g_\mathrm{s}}{2}) \, b_{\parallel}^{}
& 0 & \frac{\Delta_\mathrm{KK'}}{2} 
\cr 
  \frac{\Delta_\mathrm{KK'}}{2} & 0 & 
\varepsilon_d - \frac{\Delta_\mathrm{SO}}{2} 
-( g_\mathrm{orb} - \frac{g_\mathrm{s}}{2} ) \,b_{\parallel}^{}
& \frac{g_\mathrm{s}}{2} \,b_{\perp}^{}  
\cr 
 0 & \frac{\Delta_\mathrm{KK'}}{2}  & \frac{g_\mathrm{s}}{2} \,b_{\perp}^{} &
\varepsilon_d + \frac{\Delta_\mathrm{SO}}{2}  
-( g_\mathrm{orb} + \frac{g_\mathrm{s}}{2}) \,b_{\parallel}^{}
\cr 
\end{matrix}          
\, \right ]  .
\label{eq:H_d0_mat_wide}
\end{align}
\end{widetext}

\subsection{Renormalized parameters for quasiparticles}

Low-energy Fermi-liquid behavior of 
the Anderson impurity, $\mathcal{H}$, can be explored  
using the the Matsubara imaginary-frequency Green's function,\cite{NozieresFermiLiquid,YamadaYosida1,YamadaYosida2,ShibaKorringa,Yoshimori,HewsonRPT2001}
\begin{align}
G^{}_{m}(i\omega)
 \, \equiv & \,   - \int_0^{1/T} \!\!  d\tau \, 
e^{i \omega \tau} 
\left\langle T_\tau \, d_m^{}(\tau)\,d_m^{\dagger}
\right\rangle  
\nonumber \\
\,= & \ 
\frac{1}
{i\omega-\epsilon_m^{}-\Sigma^{}_{m}(i\omega)+i\Delta\,\mathrm{sgn}\, \omega}
\;.
 \label{eq:GM}
\end{align}
Here, 
$\langle \mathcal{O} \rangle \equiv 
\mathrm{Tr}\,[\mathcal{O}\,e^{-\mathcal{H}/T} ]/\mathrm{Tr}\,[e^{-\mathcal{H}/T}]$ denotes the thermal average. 
In the following, we consider the zero-temperature limit $T \to 0$, 
where the Matsubara frequency $\omega$ can be treated 
as a continuous variable.
Behavior of the self energy $\Sigma^{}_{m}(i\omega)$  
near the Fermi level 
 $\omega \simeq 0$ determines the characteristics of the  quasiparticles,  
\begin{align}
G^{}_{m}(i\omega)\, \simeq 
\, 
\frac{Z_{m}}
{i\omega-\widetilde{\epsilon}_{m}+i\widetilde{\Delta}_{m}\,\mathrm{sgn}\,\omega} \;. 
\label{eq:GR_FERMI}
\end{align}
The renormalized resonance width 
 $\widetilde{\Delta}_{m} \equiv Z_{m} \Delta$ 
and the peak position 
$\widetilde{\epsilon}_{m} \equiv Z_{m} 
\left [\epsilon_m^{}+\Sigma_{m}(0)\right]$ 
of the local level are parameterized by  
the wavefunction renormalization factor  
\begin{align}
 \frac{1}{Z_{m}} \,\equiv \,  
1- \left. \frac{\partial\Sigma_{m}^{}(i\omega)}{\partial\, i\omega}
\right|_{\omega=0} \;.
\label{eq:Z_factor}
\end{align}
The width of the level $m$ closest to the Fermi level  
determines the Kondo energy scale $T_K \sim \widetilde{\Delta}_m$.

The phase shift, which is defined as the argument of     
$G^{}_{m}(i0^+)=-|G^{}_{m}(i0^+)| e^{i \delta_m}$
in the complex plane, plays an very important role on  
the ground-state properties.
At zero temperature, 
it determines the occupation number of the local level 
 through the Friedel sum rule 
\begin{align}
\left\langle n_{d,m} \right\rangle =\frac{\delta_{m}}{\pi} \;, 
\qquad
\cot \delta_m =\, \frac{\epsilon_m +\Sigma_{m}(0)}{\Delta} \;.
\label{eq:Friedel}
\end{align}
Furthermore, 
the transmission probability $\mathcal{T}_{m}^{}$ 
through the $m$-th dot level can also be expressed 
in terms of the phase shift, 
or the density of states $\rho_{d,m}^{}(0)$ at the Fermi level,
\begin{align}
\mathcal{T}_{m}^{} = & \ 
\frac{4\Delta_{L}\Delta_{R}}{(\Delta_{L}+\Delta_{R})^{2}}\,
\sin^{2}\delta_{m} , 
\label{eq:transmission}
\\
\rho_{d,m}^{}(0) \equiv& \,   
- \frac{1}{\pi} \mbox{Im}\, G_m(i0^+) 
 = \frac{\sin^2 \delta_m^{}}{\pi \Delta}.
\label{eq:DOS_int}
\end{align}
With this $\mathcal{T}_{m}^{}$   
and  $\mathcal{S}_m \equiv 
 \mathcal{T}_{m}^{} \left(1-\mathcal{T}_{m}^{}\right)$,  
the linear-response conductance and 
 noise for the current flowing 
 between the two leads, $L$ and $R$, 
can be expressed in a Landauer form 
\begin{align}
\mathcal{G}  = \frac{e^2}{h}
\sum_{m=1}^{N} \mathcal{T}_{m}^{} \; , 
\qquad  
\mathcal{S} = \frac{2e^3}{h}
\sum_{m=1}^{N} 
\mathcal{S}_m \;.
\label{eq:Conductance_T0}
\end{align}

The residual interaction   $\widetilde{U}_{mm'}^{}$ 
between the quasiparticles is  another important 
 local-Fermi-liquid parameter.   
It is defined in terms of  the vertex correction    
$\Gamma_{m,m';m',m}(i\omega,i\omega';i\omega',i\omega)$ 
 at zero frequencies, 
\begin{align}
\widetilde{U}_{mm'}^{} \,\equiv\,   Z_m Z_{m'} \Gamma_{m,m';m',m}(0,0;0,0)\;.  
\end{align}
Note that  $\Gamma_{mm:mm}(0 , 0; 0, 0)=0$ due to 
the Pauli principle. 
These residual interactions also become 
level dependent in the case where the SU($4$) symmetry is broken. 
We introduce a dimensionless parameter  
 $R_{mm'}^{}$ as an analogue of the Wilson ratio in the symmetric case   
\begin{align}
R_{mm'}^{} \,\equiv & \ \,   
1+\sqrt{\widetilde{\rho}_{d,m}^{}(0)\widetilde{\rho}_{d,m'}^{}(0)} 
\ \widetilde{U}_{mm'}^{}  . 
\label{eq:wilson_ratio}
\end{align}
Here, $\widetilde{\rho}_{d,m}^{}(0)$ is the renormalized 
density of states for the quasiparticles, 
\begin{align}
\widetilde{\rho}_{d,m}^{}(0) \,\equiv & \   
\frac{\rho_{d,m}^{}(0)}{Z_m}  
\;. 
\end{align}
The quasiparticle density of states is also  
one of the important parameters. For instance,  
contributions of the quantum-dot part on the $T$-linear specific heat 
$C_\mathrm{dot}^{} = \gamma \,T $ 
can be written as,\cite{Yoshimori} 
\begin{align}
\gamma \,=\, \frac{\pi^2}{3} \sum_m  
\widetilde{\rho}_{d,m}^{}(0) \;. 
\end{align}
Corrections due to the 
 residual interaction   $\widetilde{U}_{mm'}^{}$ appears 
for higher-order correlation functions. 
For example, the charge susceptibility for  
impurity electrons can be written as 
$\chi_c^{} \equiv \sum_m \chi_{c,m}^{}$, 
\begin{align}
\chi_{c,m}^{} \,\equiv& \  
- \frac{\partial \left\langle n_{d,m}\right\rangle}{\partial \varepsilon_d}  
\ = \  -\sum_{m'}
\frac{\partial \epsilon_{m'}}{\partial \varepsilon_d}
\frac{\partial  \left\langle n_{d,m}\right\rangle }{\partial \epsilon_{m'}}  
\nonumber \\
=& \ 
\widetilde{\rho}_{d,m}^{}(0) \left[ 
\, 1 - \sum_{m' (\neq m)}
 \widetilde{U}_{mm'} \widetilde{\rho}_{d,m'}^{}(0) 
\, \right] . 
\label{eq:chi_c}
\end{align}
Note that $\partial \epsilon_{m'}/\partial \varepsilon_d =1$ 
by definition 
Eq.\ \eqref{eq:one_particle_level_CNT_eigen0},
the last line of Eq.\ \eqref{eq:chi_c} 
 follows from the {\it Fermi-liquid relations}:
\begin{align}
-\,\frac{\partial \left\langle n_{d,m}\right\rangle }{\partial \epsilon_{m'}}  
\,= \,
\widetilde{\rho}_{d,m}^{}(0)
\left[ \delta_{mm'} - 
\widetilde{U}_{mm'}\,\widetilde{\rho}_{d,m'}^{}(0)\,
\right] ,
\label{eq:two_Fermi_liquid_relations}
\end{align}
where $\widetilde{U}_{mm}^{}=0$ by definition. 
These relations correspond to the Ward identities,\cite{Yoshimori,aoJPSJ2001}
described in the appendix \ref{sec:Ward_identities}. 
The last line of Eq.\ \eqref{eq:chi_c} can also be interpreted physically 
such that the factor  $\widetilde{\rho}_{d,m}^{}(0) $ in front 
represents contributions of the free renormalized quasiparticles 
while the bracket represents a relative dimensionless value,  
which is reduced from the free-quasiparticle value by 
the residual interactions $\widetilde{U}_{mm'}$.

In order to write the magnetic susceptibilities in a similar form, 
we need the matrix elements of 
magnetization matrix $\overrightarrow{\bm{\mathcal{M}}}$ 
with respect to one-particle eigenvector $\bm{u}_m^{}$, 
which can be expressed 
in the following forms using the Feynman theorem,  
\begin{align}
\vec{F}_{m}
\equiv\,
\bm{u}_m^{\dagger} \,
\overrightarrow{\bm{\mathcal{M}}} 
\,\bm{u}_m^{}  
\, = \,  - \,\bm{u}_m^{\dagger} 
\frac{\partial \bm{H}_d^0}{\partial \vec{b}} 
 \,\bm{u}_m^{}  
\,=\, - \,\frac{\partial \epsilon_{m}}{\partial \vec{b}} .
\end{align}
The ground-state average of the magnetization 
 $\overrightarrow{\mathcal{M}}$ 
can be written  in terms of  these matrix elements,  
\begin{align}
\overrightarrow{\mathcal{M}} \,\equiv \,
\left \langle 
 \bm{\psi}^{\dagger}_{d} \, \overrightarrow{\bm{\mathcal{M}}} 
\,
 \bm{\psi}^{}_{d} 
\right \rangle
\,=\, 
\sum_m 
\vec{F}_{m} 
\left \langle 
n_{d,m}
\right \rangle . 
\label{eq:magnetization}
\end{align}
The magnetic susceptibility,
$\chi_{\mathcal{M}}^{\mu\nu} \equiv 
\partial \mathcal{M}^{\mu}/\partial b_\nu$, 
can be expressed in the form, 
\begin{align}
\chi_{\mathcal{M}}^{\mu\nu}
 \, = & \    
\sum_m \frac{\partial F_{m}^{\mu} 
}{\partial b_\nu}
\,\left \langle 
n_{d,m}
\right \rangle 
+ \sum_{mm'} 
F_{m}^{\mu}\,
\frac{\partial \epsilon_{m'}}{\partial b_\nu}
\frac{\partial  \left \langle n_{d,m}\right \rangle 
}{\partial \epsilon_{m'}}
\nonumber \\
\,=& \   
\sum_m \frac{\partial F_{m}^{\mu} 
}{\partial b_\nu}
\,\left \langle 
n_{d,m}
\right \rangle 
+ \sum_{m} 
F_{m}^{\mu}\,
F_{m}^{\nu}\,\widetilde{\rho}_{d,m}^{}(0) 
\nonumber \\
& \ - \sum_{m\neq m'}
F_{m}^{\mu}\,
F_{m'}^{\nu}\,
\widetilde{\rho}_{d,m}^{}(0)\,
\widetilde{\rho}_{d,m'}^{}(0) \,\widetilde{U}_{mm'}.
\end{align}
Here, the last term in the right-hand side represents 
the contributions of the residual interaction, 
or the vertex corrections.

Specifically for $\Delta_\mathrm{SO}=\Delta_\mathrm{KK'}=0$, 
 the spin component of the magnetization 
becomes parallel to the field  
 $\vec{e}_\Theta^{} = \cos \Theta \,\vec{e}_{z}^{} 
+ \sin \Theta \,\vec{e}_{x}^{}$ while the orbital component is   
always  along the nanotube axes. 
Therefore,
in this case Eq.\ \eqref{eq:magnetization} can be written in the form,    
\begin{align}
\overrightarrow{\mathcal{M}} \,= &  \  
\mathcal{M}_\mathrm{orb}  \, \vec{e}_z 
+  \mathcal{M}_\mathrm{s} \, \vec{e}_\Theta^{}\;,
\label{eq:magnetization_example1}
\\
\mathcal{M}_\mathrm{orb} 
\,=& \ 
g_\mathrm{orb}
\Bigl[\,
\left \langle n_{d,1} \right \rangle 
-\left \langle n_{d,4} \right \rangle 
+\left \langle n_{d,2} \right \rangle 
-\left \langle n_{d,3} \right \rangle 
\,\Bigr], 
\label{eq:magnetization_example2}
\\
\mathcal{M}_\mathrm{s} 
\,= & \ 
\frac{g_\mathrm{s}}{2}\,
\Bigl[\,
\left \langle n_{d,1} \right \rangle 
-\left \langle n_{d,4} \right \rangle 
-\left \langle n_{d,2} \right \rangle 
+\left \langle n_{d,3} \right \rangle 
\,\Bigr].
\label{eq:magnetization_example3}
\end{align}
Here, the label $m=1,2,3$ and $4$  
are assigned to 
$|\mathrm{K'}\! \downarrow_{\vec{b}}\rangle$,    
$|\mathrm{K'}\! \uparrow_{\vec{b}}\rangle$,  
$|\mathrm{K}\! \downarrow_{\vec{b}}\rangle$ and  
$|\mathrm{K}\! \uparrow_{\vec{b}}\rangle$, respectively,   
with   
$\uparrow_{\vec{b}}$  and $\downarrow_{\vec{b}}$    
the spin defined with respect to the direction 
along the field $\vec{b}$ for $\Theta < \pi/2$.  
The corresponding one-particle energies are given by 
$\epsilon_1 = 
\varepsilon_d - (g_\mathrm{orb }\cos\Theta +g_\mathrm{s}/2) b$, 
$\,\epsilon_2 =  
\varepsilon_d - (g_\mathrm{orb }\cos\Theta -g_\mathrm{s}/2) b$, 
$\,\epsilon_3 =  
\varepsilon_d + (g_\mathrm{orb }\cos\Theta -g_\mathrm{s}/2) b$, 
and 
$\epsilon_4 =  
\varepsilon_d + (g_\mathrm{orb }\cos\Theta +g_\mathrm{s}/2) b$.


\section{Field-induced crossover in a half-filling CNT dot}
\label{sec:cnt_dot}

One of the most interesting experimental findings of 
carbon nanotube quantum dots 
is that the SU($4$) Kondo effects for different 
impurity-occupation numbers  $M=1$, $2$, and $3$ 
can successively occur as the dot level $\epsilon_m$ is varied 
by tuning the gate voltages,\cite{Ferrier2016} as mentioned. 
We next consider a crossover from the SU($4$) to SU($2$) Fermi-liquid state 
occurring near half-filling,  
where two electrons are occupied in the local levels of the quantum dot.

It is a different class of the SU($4$) to SU($2$) crossover 
from those considered previously 
for the CNT quantum dots,\cite{Choi2005,Lim,Anders2008,GalpinLoganAnders2010,MantelliMocaZarandGrifoni}
and is inspired by recent magneto-transport experiment    
which observes an unexpected evolution of the Kondo plateau.\cite{Note1} 
As magnetic field increases, 
the Kondo plateau near half-filling reduces 
the height from $4e^2/h$ to $2e^2/h$ keeping the  
flat structure unchanged.
This implies that two one-particle levels among the four still  
remain unlifted near the Fermi level in the magnetic field. 
This is possible if the  magnetic field $\vec{B}$ 
is applied in such a way that the spin and orbital Zeeman effects 
cancel each other out. 
In order to explain these experimental findings, 
we propose a model on the basis of the Anderson impurity $\mathcal{H}$ 
given in Eqs.\ \eqref{eq:Hd}--\eqref{eq:HT} 
with the one-particle part defined in 
Eqs.\ \eqref{eq:Hd0_micro}--\eqref{eq:dot_magnetization}, 
and calculate magneto conductance and Fermi-liquid parameters using the NRG.


\subsection{Matching of spin and orbital Zeeman splittings}

We introduce a model in which the double degeneracy remains unlifted 
near half-filling in a finite magnetic field $b$, 
setting the parameters such that  
$\Delta_\mathrm{SO}=\Delta_\mathrm{KK'}=0$  
together with a condition 
\begin{align}
g_\mathrm{orb} \cos \Theta \,=\, \frac{g_\mathrm{s}}{2}\;, 
\qquad \qquad  g_\mathrm{s} = 2.
\label{eq:zeeman_matching}
\end{align}
This is {\it not\/} rare for CNT dots as 
the orbital magnetic moment 
can take some values around $g_\mathrm{orb}\sim 10$.\cite{RMP-Kouwenhoven}
In this case, 
the orbital Zeeman splitting $\pm (g_\mathrm{orb} \cos \Theta) b$ 
matches the spin Zeeman splitting  $\pm (g_\mathrm{s}/2) b$, 
so that the one-particle levels become 
\begin{align}
\epsilon_1= \varepsilon_d -2\, b, \quad \ \,
\epsilon_2=\epsilon_3= \varepsilon_d, \quad \ \,
\epsilon_4= \varepsilon_d+2\,b .   
\label{eq:caseA}
\end{align}
%
 The two levels  in the middle, $m=2$ and $3$, 
lost the coupling to the magnetic field as 
 the spin and orbital Zeeman effects cancel out, 
and thus the energies $\epsilon_2$ and $\epsilon_3$ 
 become independent of $b$. 
The other two levels, $\epsilon_1$ and $\epsilon_4$,  
move away from $\varepsilon_d$  as $b$ increases. 
The total Hamiltonian $\mathcal{H}$ has a symmetry of 
U(1)$_{m=1}$$\times$SU($2$)$_{m=2,3}$$\times$U(1)$_{m=4}$ 
 for finite magnetic fields.
The two degenerate states, $m=2$ and $3$, 
have an SU($2$) symmetry while each of the the other two,   
 $m=1$ and $4$, only has the $U(1)$ symmetry corresponding 
to the charge conservation of the electrons 
 carrying the {\it flavor\/} \lq\lq$m$''.
This SU($2$) symmetric part shows a Kondo effect which 
 evolves from the SU($4$) symmetric two-electron 
 Kondo singlet state  as magnetic field increases. 
Furthermore, for the one-particle levels $\epsilon_m$ given 
in Eq.\ \eqref{eq:caseA},  the Hamiltonian $\mathcal{H}$ has 
an extended particle-hole symmetry 
which is accompanied by an inversion of the {\it flavor\/} \lq\lq$m$'':
\begin{align} 
d_1^\dagger \Rightarrow h_4^{}, \quad  
d_2^\dagger \Rightarrow h_3^{}, \quad  
d_3^\dagger \Rightarrow h_2^{}, \quad 
d_4^\dagger \Rightarrow h_1^{},  
\label{eq:p_h_symmetry}
\end{align} 
and 
 $c_{\nu,\varepsilon m}^{\dagger} 
\Rightarrow -f_{\nu,-\varepsilon m'}^{}$ for 
$(m,m')$ = (1,4), (2,3), (3,2), (4,1), 
 where $h_m$ and $f_{\nu,-\varepsilon m'}$
 are fermion operators describing the holes.

In the real CNT dot used for 
recent magneto-transport measurements,\cite{Ferrier2016,Note1}
the Coulomb interaction  
 $U \approx 6$ meV  
and the hybridization energy 
 $\Delta \equiv \Delta_L + \Delta_R \approx 0.9$ meV 
with $\Delta_L \approx\Delta_R$ 
dominate the other energy scales.      
The valley mixing and spin-orbit interaction are smaller 
than these two  $\Delta_\mathrm{KK'} \sim \Delta_\mathrm{SO} \sim 
0.2$ meV. 
The orbital Zeeman coupling is 
estimated to be $g_\mathrm{orb} \cos \Theta \approx 0.7$, 
which is still not far from the matching value $1.0$.
Nevertheless, in order to clarify how the deviations 
from the case described by Eq.\ \eqref{eq:caseA} 
affect this crossover, we also examine the realistic case   
using  $\epsilon_m$'s determined through $\bm{H}_d^0$ 
 with the parameters  deduced from the experiment:
\begin{align}
\Delta_\mathrm{KK'} = \Delta_\mathrm{SO} = 0.07 \pi \Delta, 
\qquad   
g_\mathrm{orb} \cos \Theta = 0.7.  
\label{eq:caseB}
\end{align}
In this case, the extended particle-hole symmetry does {\it not\/} hold. 
Furthermore, the Hamiltonian $\mathcal{H}$ no longer has 
the SU($2$)$_{m=2,3}$ symmetry, and it is lowered to 
the U($1$)$_{m=2}$$\times$U($1$)$_{m=3}$ 
corresponding to 
charge conservation in each of these two channels $m=2$ and $3$.

We have carried out NRG calculations,   
taking the discretization parameter to be $\Lambda = 6.0$.\cite{KWW1}
We have kept typically the lowest 3000 eigenstates in each NRG step 
using the U(1)$\times$U(1)$\times$U(1)$\times$U(1) symmetry. 
The renormalized parameters have been deduced from 
flow of the low-energy eigenvalues near 
the fixed point of the NRG.\cite{WilsonRMP,KWW1,HewsonOguriMeyer,aoJPSJ2005NRG}
Note that the Coulomb interaction in the above-mentioned two cases 
are scaled as $U=2\pi \Delta$ with $\Delta  =0.9$ meV, 
and a magnetic field of order $b=0.1\pi\Delta$ 
corresponds to $B=4.9$ T in a real scale. 
The tunneling couplings can be well approximated  
by a symmetric one $\Delta_L=\Delta_R$, which 
 simplifies the transsimsion probability and the current noise 
as $\mathcal{T}_m = \pi \Delta \rho_{d,m}^{}(0)  = \sin^2 \delta_m$ 
and $\mathcal{S}_m \equiv
\mathcal{T}_{m}^{} \left(1-\mathcal{T}_{m}^{}\right) = (\sin^2 2\delta_m)/4$, 
respectively.


\begin{figure}[t]
\begin{center}
\rule{0,2cm}{0cm}
\begin{minipage}{0.68\linewidth}
  \includegraphics[width=\linewidth]{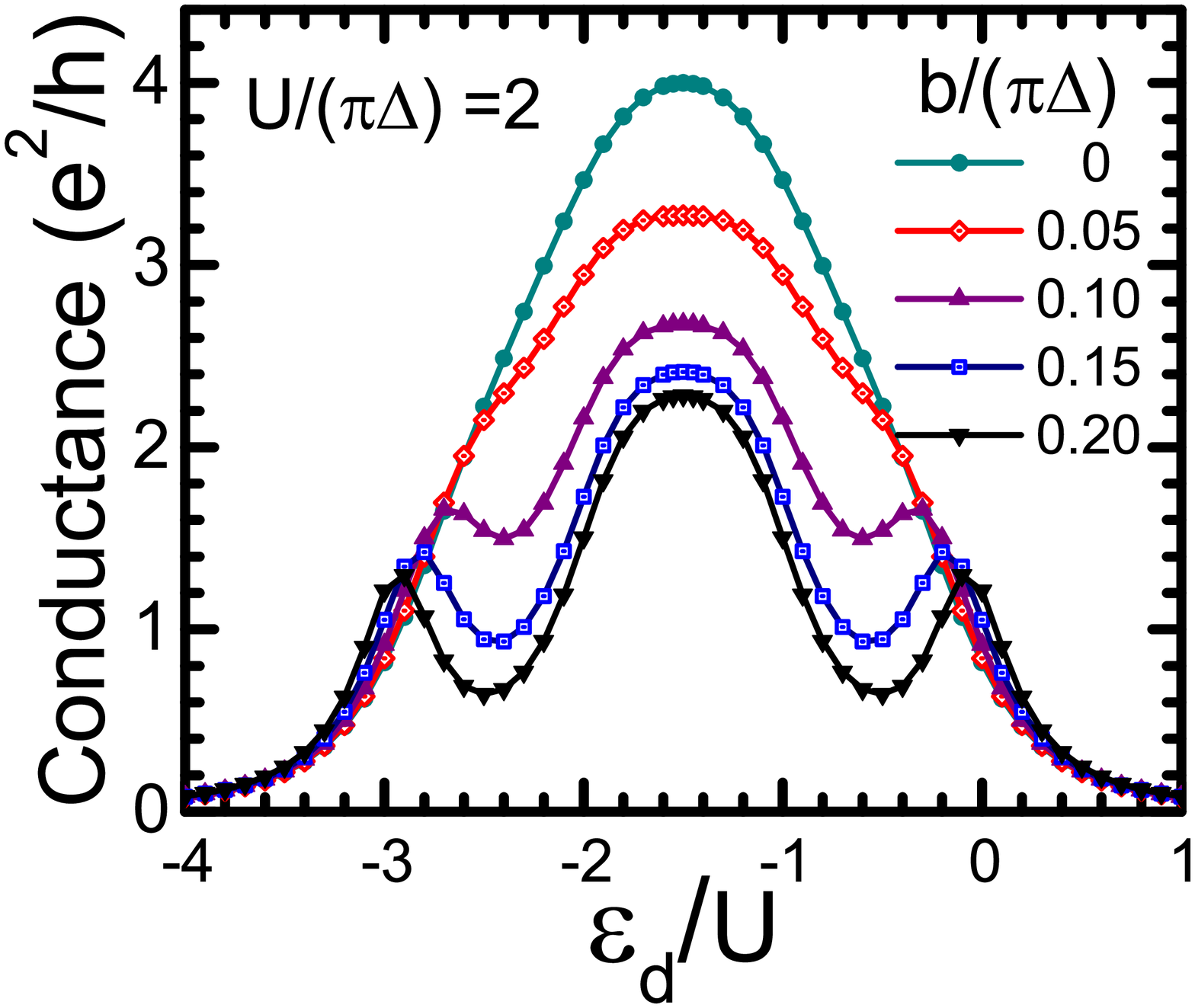}
\end{minipage}
\\
\begin{minipage}{0.72\linewidth}
  \includegraphics[width=\linewidth]{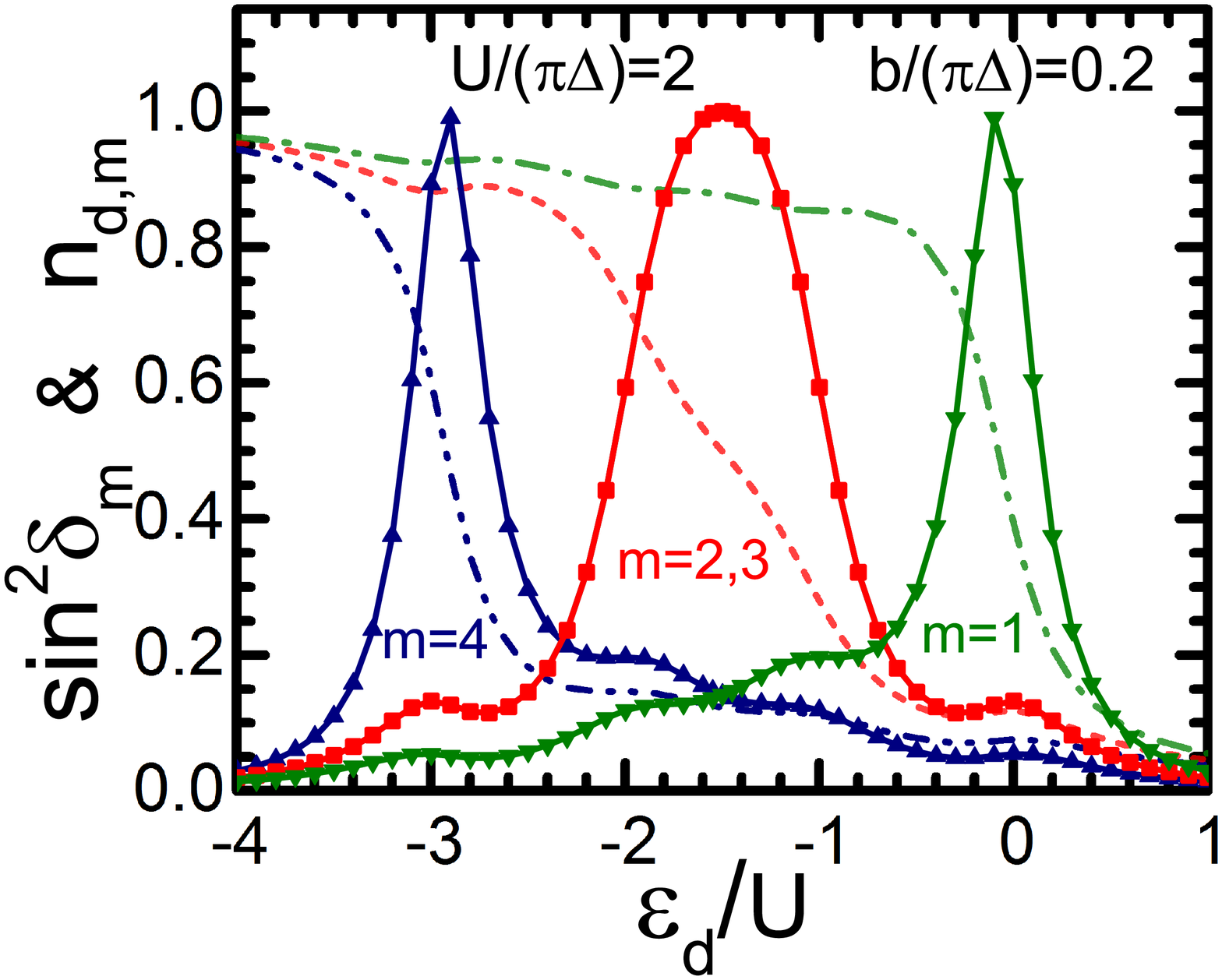}
\end{minipage}
\end{center}
 \caption{
(Color online) 
NRG results are plotted vs $\varepsilon_d$ 
for $\epsilon_m$ given in  
Eq.\ \eqref{eq:caseA} and $U/(\pi\Delta)=2.0$.
Upper panel: conductance 
at magnetic fields of 
$b/(\pi\Delta) = 0,\, 0.05,\, 0.1,\, 0.15$, and $0.2$.  
Lower panel: $\sin^2 \delta_m$ (solid line)  
and  $\langle n_{d,m}^{} \rangle$ (dashed line)
 at $b/(\pi\Delta)=0.2$ which corresponds to a real field 
of $B =9.8$ T for $\Delta=0.9$ meV.\cite{Ferrier2016} 
}
 \label{fig:Conductance_A}
\end{figure}



\begin{figure}[t]
\begin{center}

\begin{minipage}{0.69\linewidth}
\includegraphics[width=\linewidth]{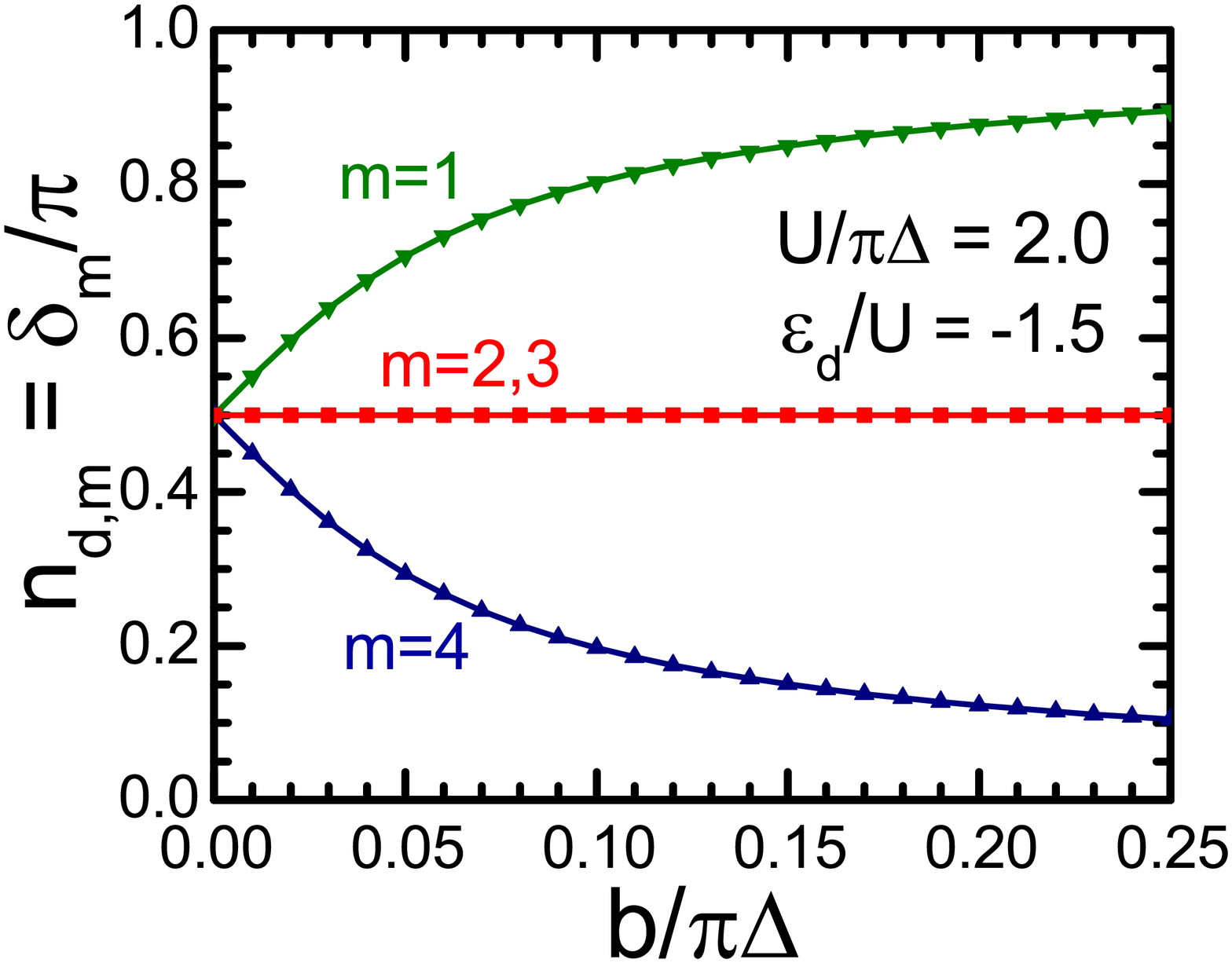}
\end{minipage}
\\
\begin{minipage}{0.7\linewidth}
\includegraphics[width=\linewidth]{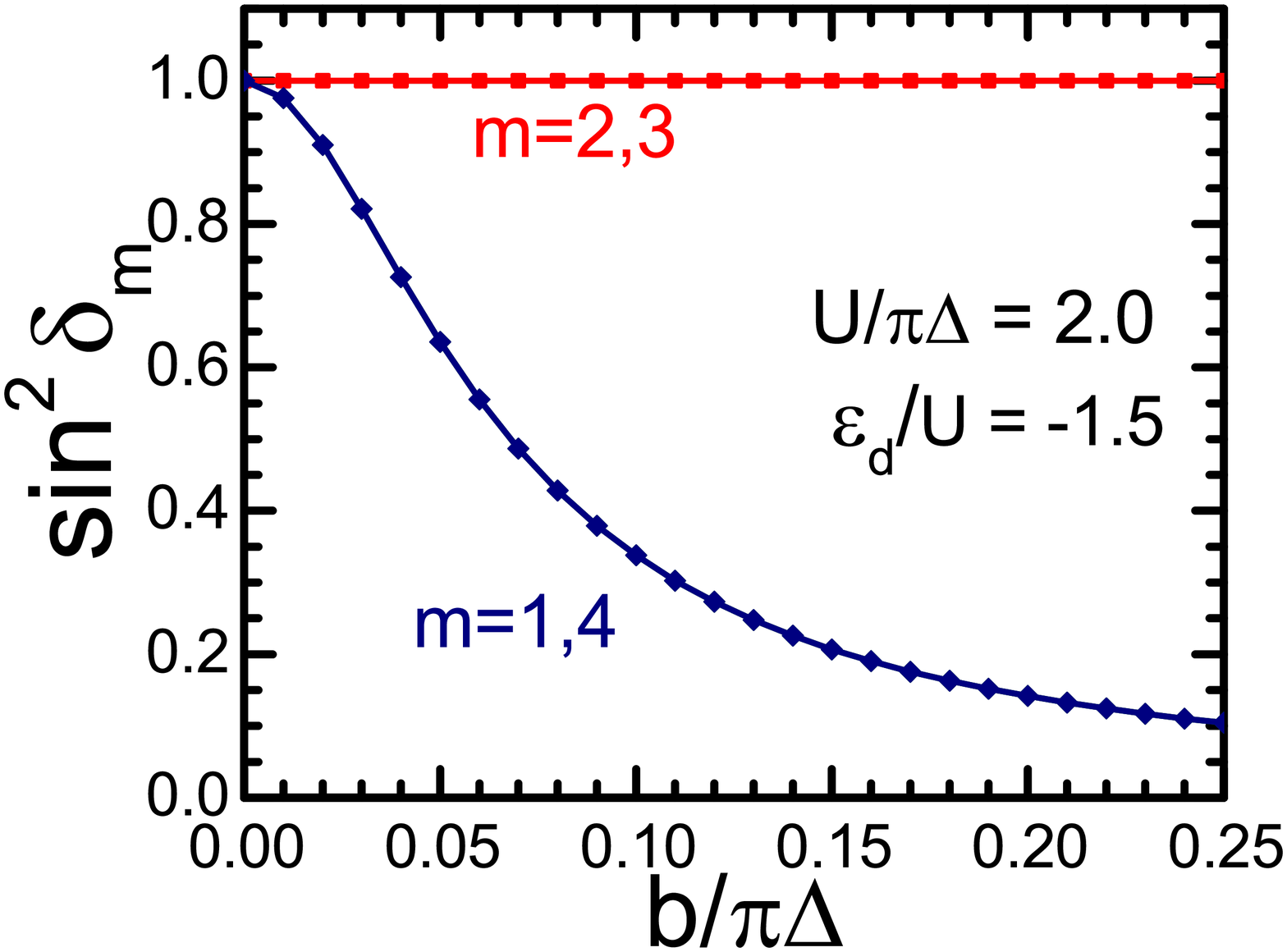}
\end{minipage}
\\
\rule{0,15cm}{0cm}
\begin{minipage}{0.68\linewidth}
\includegraphics[width=\linewidth]{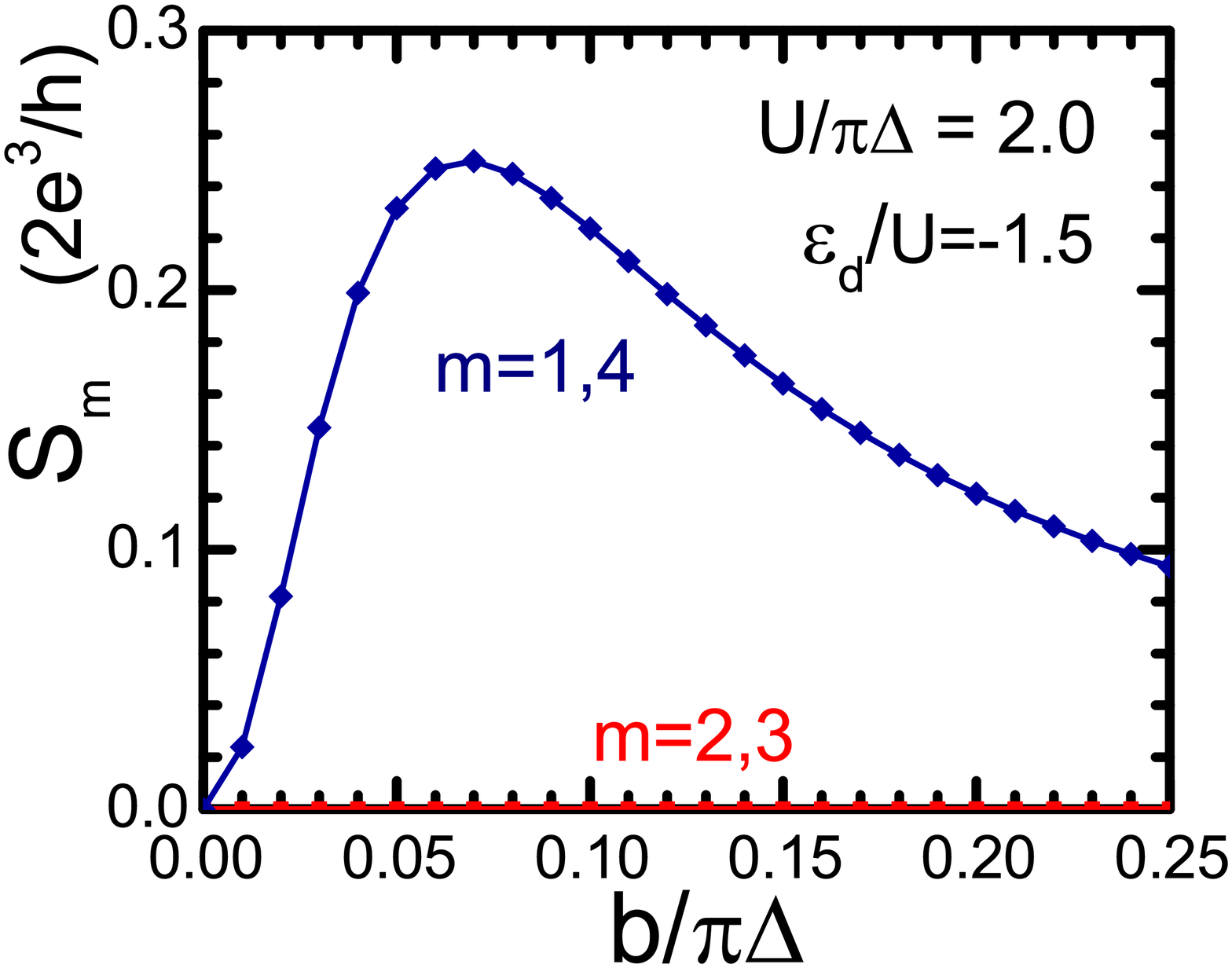}
\end{minipage}
\end{center}
 \caption{ 
(Color online) 
 $\langle n_{d,m}^{} \rangle =\delta_m/\pi$, 
$\sin^2 \delta_m$, and current noise $\mathcal{S}_m$        
are 
plotted vs $b$ at half-filling $\varepsilon_d/U = -1.5$ 
 and $U/(\pi\Delta)=2.0$ 
for $\epsilon_m$ given in Eq.\ \eqref{eq:caseA}.
}
 \label{fig:nd_cond_half_A}
\end{figure}



\begin{figure}[t]
\begin{center}
\rule{0,2cm}{0cm}
\begin{minipage}{0.7\linewidth}
  \includegraphics[width=\linewidth]{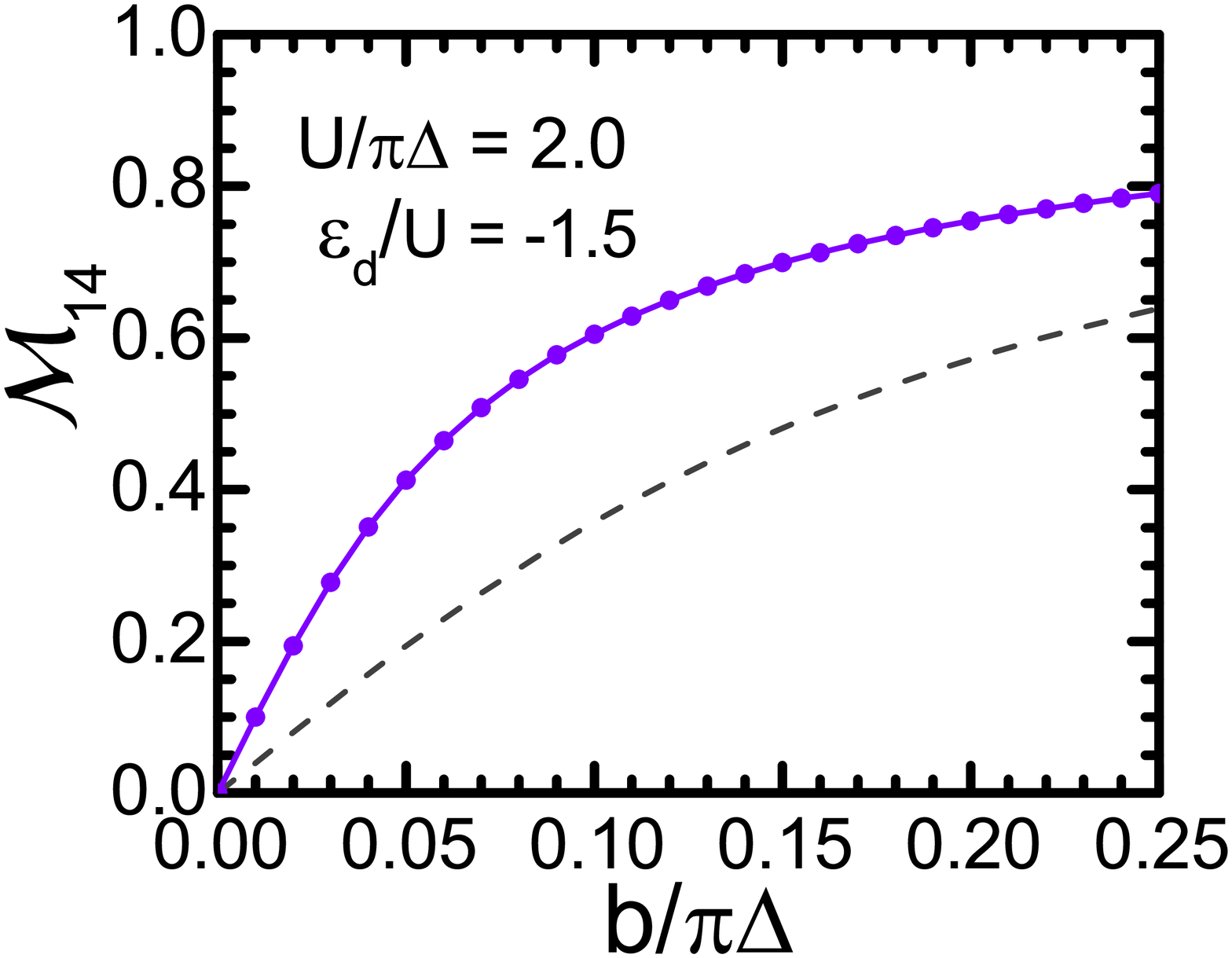}
\end{minipage}
\end{center}
 \caption{
(Color online) 
Magnetization $\mathcal{M}_{14}$ 
defined in Eq.\ \eqref{eq:Magnetization_14} is plotted vs $b$ 
at half-rilling  $\varepsilon_d/U = -1.5$  
and $U/(\pi\Delta)=2.0$ 
for $\epsilon_m$ given in Eq.\ \eqref{eq:caseA}.  
The dashed line shows the noninteracting results 
obtained for $U=\varepsilon_d=0$.
}
 \label{fig:magnetization_14}
\end{figure}

\subsection{U(1)$_{m=1}$$\times$SU($2$)$_{m=2,3}$$\times$U(1)$_{m=4}$ symmetric case }
\label{subsec:results_A}

The conductance $\mathcal{G}$  obtained 
for energy levels $\epsilon_m$ given in Eq.\ \eqref{eq:caseA} 
is plotted vs $\varepsilon_d$ 
in the upper panel of Fig.\ \ref{fig:Conductance_A}  
for several values of $b$. 
The lower panel shows  $\sin^2 \delta_m$ and  $\langle n_m \rangle$ 
for $m=1,2,3,4$ at $b/(\pi\Delta) = 0.2$. 
We can see that the conductance has a broad peak 
near half-filling $\varepsilon_d/U \simeq -1.5$.  
The system has the SU($4$) symmetry at zero field $b=0$, 
and the conductance peak reaches the unitary-limit value $4e^2/h$.
Because the interaction $U/(\pi \Delta) =2.0$ is still not very large, 
the conductance peak is not completely flat and the shoulders 
near $1/4$ and $3/4$ fillings are less pronounced.
Such a flat structure will become clearer 
if the Coulomb interaction $U$ is much larger. 
Nevertheless,
other Fermi-liquid parameters are already renormalized 
significantly,\cite{NishikawaCrowHewson2,ao2012} 
as we will describe later.

 As magnetic field $b$ increases, the broad conductance peak decreases   
and in the limit of $b \to \infty$ it approaches 
the SU($2$) unitary-limit value $2e^2/h$, 
keeping the typical flat form of the Kondo plateau. 
This is caused by 
the doubly degenerate levels $m=2$ and $3$ remaining at the Fermi level, 
and is consistent with the behavior observed 
in the recent measurements.\cite{Note1}
The contributions of the other two levels, $m=1$ and $4$, 
on the conductance are separately seen 
for large fields $b/(\pi\Delta) \gtrsim 0.1$ as  
the two additional sub peaks at $\varepsilon_d/U \approx 0.0$ and $-3.0$  
 with the height close to $e^2/h$.
We can also see in the lower panel of Fig.\ \ref{fig:Conductance_A},
which shows the results at $b/(\pi\Delta) = 0.2$,  
that two of the channels $m=2$ and $3$ contribute to 
the SU($2$) Kondo plateau near half-filling  
while the other two contribute to the side peaks. 
Furthermore,  $\sin^2 \delta_m$ has a long tail 
in the off-resonance region with a weak step-structure 
that is caused by the inter-channel correlations.     
The steps emerge as the resonance peaks cross the Fermi level. 
The phase shift $\delta_m$ varies 
from $0$ to $\pi$ for $m=1$ and $4$ 
at the crossing point as a single electron enters into the impurity level.
The Kondo half-step $\delta_m =\pi/2$ emerges 
near half-filling  $\varepsilon_d/U \simeq -1.5$ 
for the levels $m=2$ and $3$ in the middle, 
and this half-step will be more pronounced if $U$ is much larger.

The Kondo effect which is caused by the doubly degenerate states,   
$m=2$ and $3$, is most enhanced at half-filling $\varepsilon_d/U = -3/2$.  
We next investigate 
the magnetic field dependence of the Kondo correlations 
in more detail at half-filling.
Figure \ref{fig:nd_cond_half_A} shows 
 $\langle n_m \rangle$,  
 $\sin^2 \delta_m$, and the current noise $\mathcal{S}_m$  
as a function of $b$.
 Even at finite magnetic fields $b$,
the average occupation number of the twofold degenerate states, 
$m=2$ and $3$, is unchanged $\langle n_2 \rangle = \langle n_3 \rangle = 1/2$. 
This is caused by the  matching of spin and orbital Zeeman splittings 
described by Eq.\ \eqref{eq:zeeman_matching}.
As  the phase shifts are locked at $\delta_2=\delta_3 =\pi/2$, 
these two channels give a unitary-limit contribution $2e^2/h$ 
to the total conductance and do not 
induce a current noise $\mathcal{S}_2=\mathcal{S}_3=0$.  
The transmission probability of the other two levels
($\sin^2 \delta_1=\sin^2 \delta_4$) decreases as $b$ increases 
and finally  vanish in the limit of $b\to \infty$.
Correspondingly,  
the current noise for these states 
($\mathcal{S}_1=\mathcal{S}_4$) has a maximum 
at $b/(\pi\Delta) \approx 0.07$ 
where the transmission probability becomes 
$\mathcal{T}_1=\mathcal{T}_4=0.5$, 
namely at the filling of  $\langle n_{d,4}\rangle =1/4$  
and  $\langle n_{d,1}\rangle =3/4$. 
Owing to the  matching condition 
Eq.\ \eqref{eq:zeeman_matching},
 the magnetization $\overrightarrow{\mathcal{M}}$ 
given in Eqs.\ \eqref{eq:magnetization_example1}--\eqref{eq:magnetization_example3} can be expressed in the 
form $\mathcal{M}_\mathrm{orb} = g_\mathrm{orb}\, \mathcal{M}_{14}$ 
and $\mathcal{M}_\mathrm{s} = \frac{g_\mathrm{s}}{2} \mathcal{M}_{14}$ 
with 
\begin{align}
\mathcal{M}_{14}
\,\equiv& \ 
\left \langle n_{d,1} \right \rangle 
-\left \langle n_{d,4} \right \rangle \;.
\label{eq:Magnetization_14}
\end{align}
Both the spin and orbital components of the magnetization are 
determined by the phase shifts of the first and fourth levels: 
 $\delta_1 =\pi(1+\mathcal{M}_{14})/2$ 
and $\delta_4 =\pi(1-\mathcal{M}_{14})/2$.  
We see in Fig.\ \ref{fig:magnetization_14} that  
 $\mathcal{M}_{14}$ is significantly enhanced by the Coulomb interaction $U$, 
and it approaches to the saturation value 1 for large fields. 
In the limit of $b\to \infty$, 
both the charge and magnetic fluctuations 
caused by these two levels are suppressed 
as the occupation numbers tend 
to be full $\langle n_1 \rangle \to 1$ 
and empty $\langle n_4 \rangle \to 0$.

\begin{figure}[t]
\begin{center}
 \begin{minipage}{0.89\linewidth}
\includegraphics[width=\linewidth]{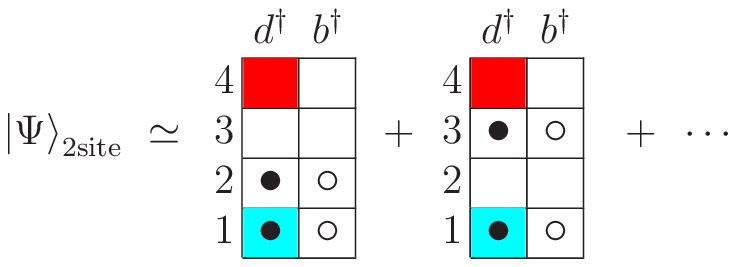}
 \end{minipage}
\end{center}
\vspace{-0.3cm}
\caption{(Color online) 
Schematic picture of the ground-state wave function 
near the CNT dot for large fields and $U\gg\Delta$.
Each row represents one of the levels of $m=1,2,3,4$.
The first ($d^\dagger$) and second ($b^\dagger$) columns 
represent orbitals in the impurity and the adjacent conduction site, 
respectively, with ($\bullet$) the impurity-electrons and ($\circ$) 
the conduction-holes. 
The explicit expression  of the wavefunction is 
given in Eq.\ \eqref{eq:TAR_boson_caseA}. 
The impurity level of $m=1$ ($4$), which is 
illustrated as a light blue (red) box, 
situates deep inside  (far above)  the Fermi level, 
and thus corrections of order $v^2/b$ determine 
the distribution of the {\it conduction-hole\/} next to  
these colored boxes.
}
\label{fig:PF_vector_2}
\end{figure}

Thus, in the limit of large magnetic field $b\to\infty$,
most of the components of the Coulomb interaction defined 
in Eq.\ \eqref{eq:Hd} can be treated with the mean-field theory, 
except for the one between 
the twofold degenerate levels, $m=2$ and $m'=3$.\cite{EdwardsHewsonPandis}  
Thus, the dot part of the Hamiltonian  can be simplified in the form,
\begin{align}
\mathcal{H}_{d}^{}
\xrightarrow{\,b\to \infty\,}&   
\ \ 
U \left[ n_2^{}n_3^{} -\frac{1}{2} 
\left(n_2^{}+n_3^{}\right) \right]
\nonumber \\ 
&  
+ \left(2b + \frac{U}{2} \right) 
\Bigl(  n_4^{} - n_1^{} \Bigr)
+ \mathrm{const.}\; .
\label{eq:Hd_B_inf}
\end{align}
This shows that the degenerate levels remaining at the  Fermi level, 
$m=2$ and $3$, can be 
described by the particle-hole symmetric SU($2$) Anderson model. 
The other two levels, $m=1$ and $4$, are frozen and 
can be separated. 
As discussed in Sec.\ \ref{sec:SU_N_Kondo} and \ref{sec:Hard_core_boson}, 
the ground-state wavefunction for the two-site case gives an insight 
into the Fermi-liquid fixed point. 
For large fields and 
 $U \gg \Delta$, 
dominant components of the ground state are given by 
\begin{align}
\left |\Psi \right \rangle_\mathrm{2site}
\ \propto & \  
\left( 
d_{2}^{\dagger}b_{2}^{\dagger} 
+ 
d_{3}^{\dagger} b_{3}^{\dagger} 
\right) 
d_{1}^{\dagger} b_{1}^{\dagger} 
| \widetilde{0} \rangle  + \cdots
\label{eq:TAR_boson_caseA}
\\
= & \  
\left( 
d_{2}^{\dagger}a_{3}^{\dagger} 
- d_{3}^{\dagger} a_{2}^{\dagger} 
\right)
d_{1}^{\dagger} a_{4}^{\dagger} 
| 0 \rangle   + \cdots \;.
\end{align}
where $| \widetilde{0} \rangle = a_{1}^{\dagger}a_{2}^{\dagger}a_{3}^{\dagger}a_{4}^{\dagger} | 0 \rangle$ and $b_{m}^{\dagger}$ is the 
creation operator for the conduction hole defined 
in Eq.\ \eqref{eq:b_operator}.
 This wavefunction is also illustrated in Fig.\ \ref{fig:PF_vector_2}.
The singlet pair state is constructed by the electrons at $m=2$ and $3$, 
which can evolve to the Fermi-liquid state,  for instance,  
through the successive NRG steps 
that take into account the low-energy conduction-electron degrees of freedom. 
The particle-hole pair is localized  at the bottom  ($m=1$)  
and is absent  at the top ($m=4$).  
The corrections of order $v^2/b$ due to virtual tunneling processes 
 determine the distribution of the conduction electrons at $m=1$ and $m=4$ 
shown in Fig.\ \ref{fig:PF_vector_2}.


\begin{figure}[t]
\begin{center}
 \begin{minipage}{0.69\linewidth}
\includegraphics[width=\linewidth]{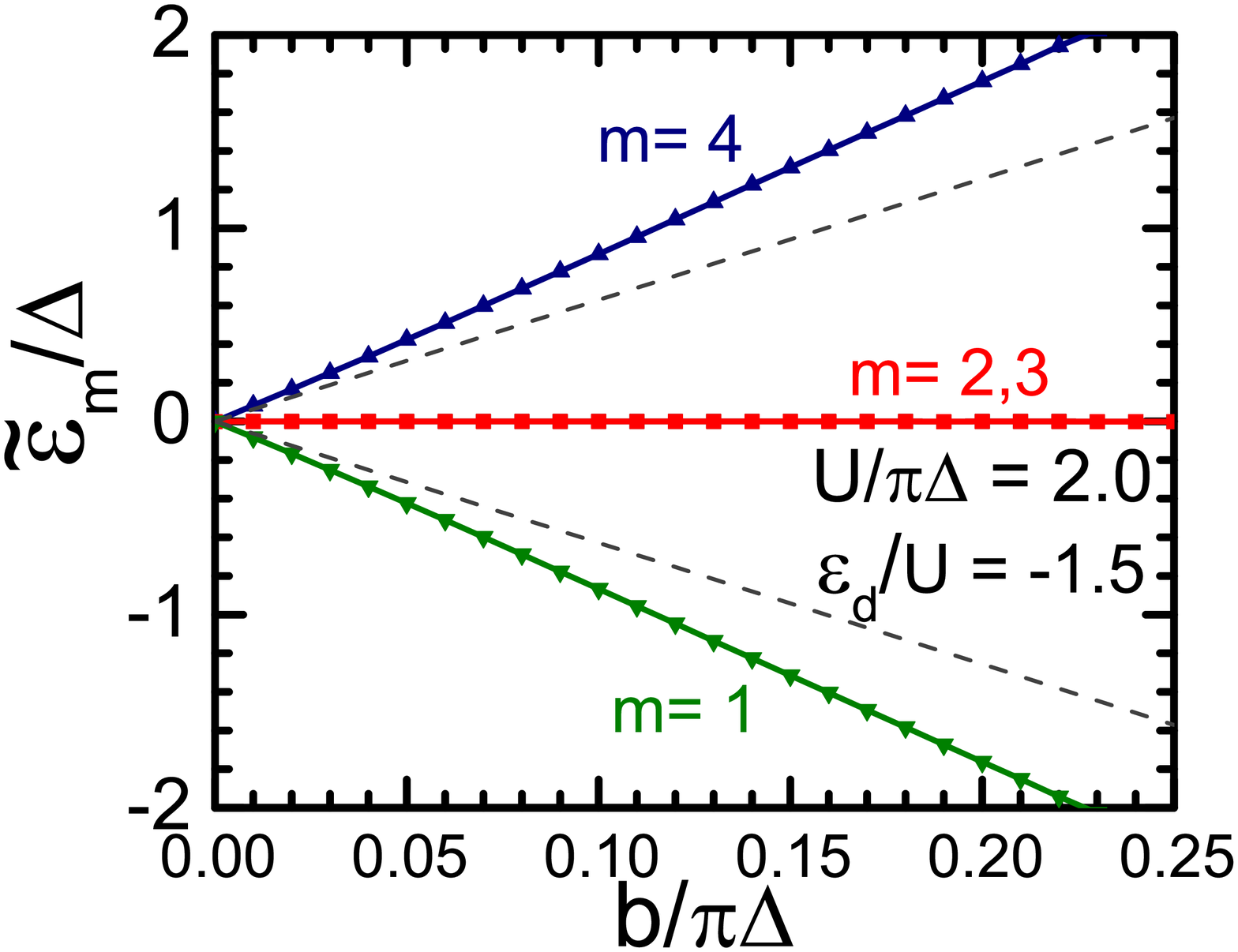}
 \end{minipage}
\\
\begin{minipage}{0.7\linewidth}
\includegraphics[width=\linewidth]{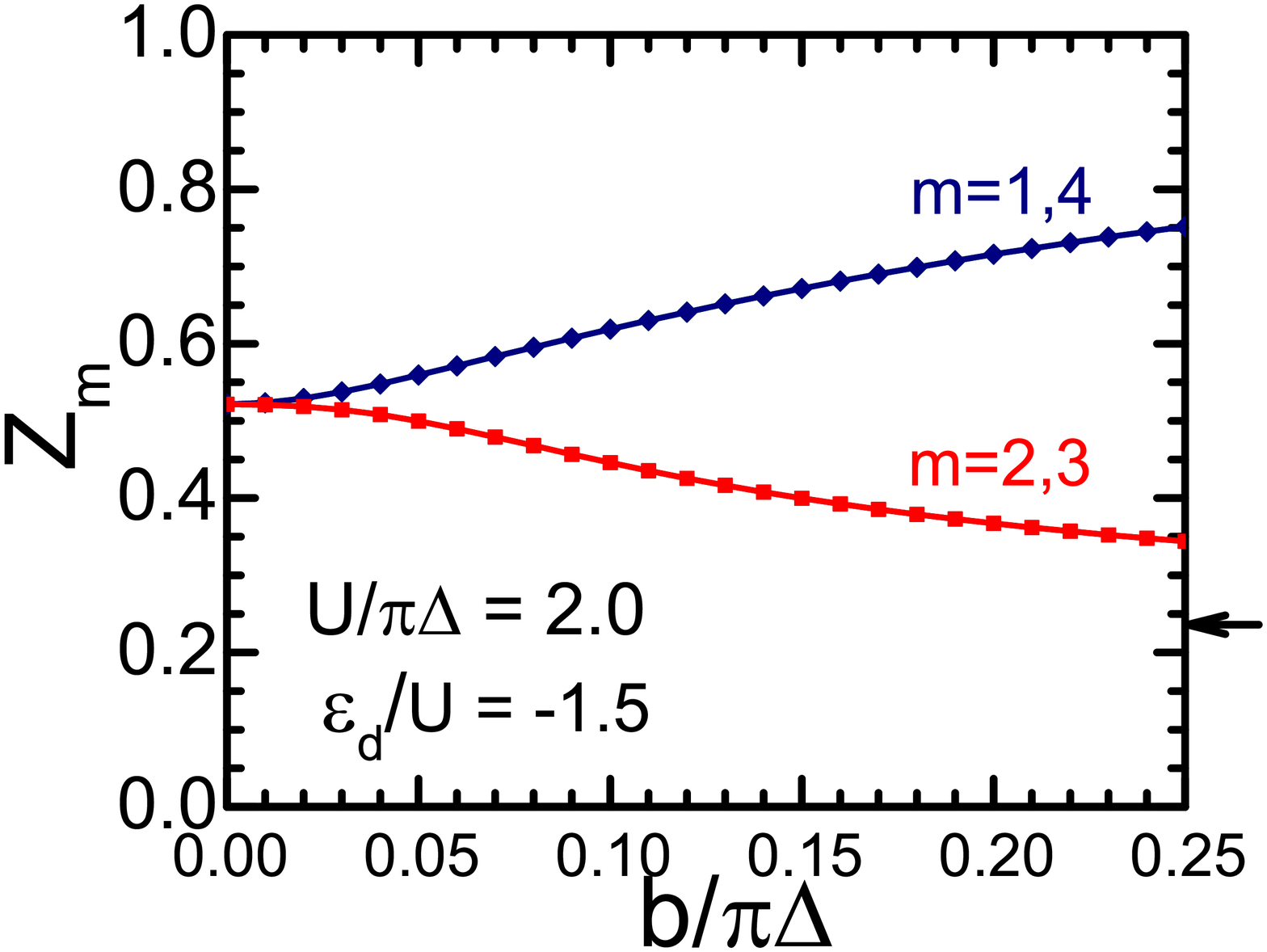}
\end{minipage}
\\
\begin{minipage}{0.71\linewidth}
\includegraphics[width=\linewidth]{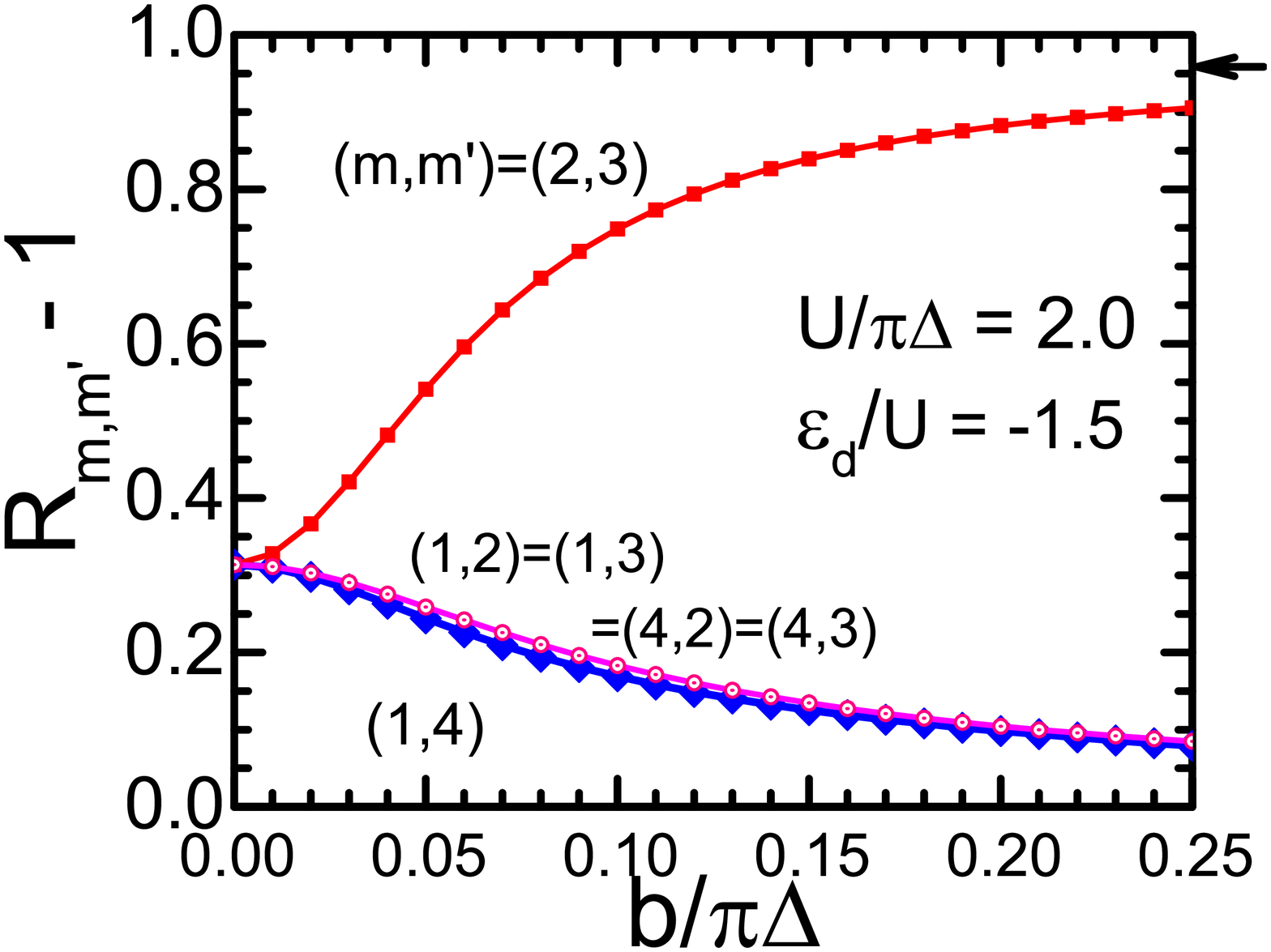}
\end{minipage}
\end{center}
 \caption{ 
(Color online) 
Renormalized parameters  $\widetilde{\epsilon}_m$, 
 $Z_m$  and  $R_{mm'}-1$ plotted vs $b$ at half-rilling 
 $\varepsilon_d/U = -1.5$  
and $U/(\pi\Delta)=2.0$  
for $\epsilon_m$ given in Eq.\ \eqref{eq:caseA}.  
The dashed line in the top panel shows the noninteracting 
level position for $U=0$.  
Arrows in the middle and bottom  panels 
indicate the corresponding values in the SU($2$) limit.
}
 \label{fig:z_uren_half}
\end{figure}


\begin{figure}[h]
\begin{center}
\rule{0,2cm}{0cm}
\begin{minipage}{0.7\linewidth}
  \includegraphics[width=\linewidth]{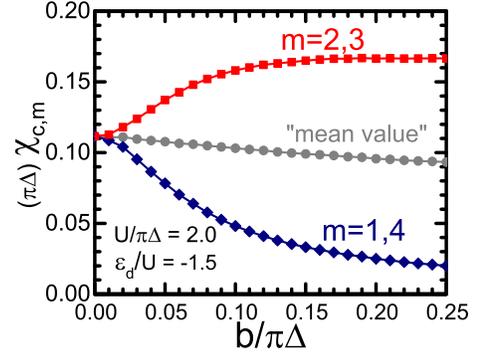}
\end{minipage}
\end{center}
 \caption{
(Color online) 
Charge susceptibility $\chi_{c,m}$ plotted vs $b$ 
for $m=1,2,3,4$ at half-filling,  $\varepsilon_d/U = -1.5$  
and $U/(\pi\Delta)=2.0$ 
for $\epsilon_m$ given in Eq.\ \eqref{eq:caseA}.  
The line in the middle denotes  a normalized total 
susceptibility $(\pi \Delta/4) \sum_m \chi_{c,m}$.
}
 \label{fig:charge_suscep_A}
\end{figure}

How each resonance level shifts as magnetic field  increases 
can be tracked through $\widetilde{\epsilon}_m$ 
shown in the top panel of Fig.\ \ref{fig:z_uren_half}. 
As already deduced from other data,     
the resonance peak
for the doubly degenerate states stays just on the Fermi level  
as  $\widetilde{\epsilon}_2= \widetilde{\epsilon}_3=0$. 
The peaks for the other two, 
$\widetilde{\epsilon}_1$ and $\widetilde{\epsilon}_4$,  
move far away from the Fermi level as $b$ increases. 
The slope of $\widetilde{\epsilon}_1$ and $\widetilde{\epsilon}_4$ against $b$ 
become steeper than those for the noninteracting case.  
This difference causes the enhancement of the magnetization $\mathcal{M}_{14}$ 
described in Fig.\  \ref{fig:magnetization_14}.
For large $b$, these two levels asymptotically approach   
$\pm \left(2b + \frac{U}{2} \right)$, 
given in Eq.\ \eqref{eq:Hd_B_inf}.

The continuous evolution from the Fermi-liquid state 
with the SU($4$) symmetry to the one with the SU($2$) 
appears more sensitively 
in the field-dependence of the renormalization factor $Z_m$  
and the residual interaction $\widetilde{U}_{mm'}$ shown in  
the middle and bottom panels of Fig.\  \ref{fig:z_uren_half}. 
Note that 
 $Z_2=Z_3$ and $Z_1=Z_4$ because of the symmetry 
described in Eq.\  \eqref{eq:p_h_symmetry},  
and there are three independent components for $R_{mm'}$:  
$R_{23}$, $R_{14}$, and $R_{12}=R_{13} = R_{24}=R_{34}$.  
Especially, the coefficients $Z_{2}$ and $R_{23}$ 
for the doubly degenerate states at the Fermi level 
continuously evolve from the SU($4$) value to the SU($2$) value  
as $b$ varies from $0$ to $\infty$.
At zero field, these coefficients take the SU($4$) values: 
 $Z_\mathrm{SU4}^{}=0.52$ and $R_\mathrm{SU4}^{}-1=0.31$ 
for $U/(\pi \Delta)=2.0$.
In the opposite limit $b \to \infty$, 
these two parameters approach those for the SU($2$) Anderson model:  
$Z_{2}\to 0.24$ and $R_{23}-1 \to 0.96$ for the same $U$.
The parameters for the other two levels  
approach the noninteracting  value   
in the limit of large magnetic field $b\to \infty$:  
namely  $Z_{1} \to 1$,  $R_{12} \to 1$,  and $R_{14} \to 1$. 
Note that 
$R_{12}$ is slightly larger than $R_{14}$ for finite $b$ 
as the energy separation 
 $\protect\widetilde{\epsilon}_2-\protect\widetilde{\epsilon}_1$ 
is the half of 
 $\protect\widetilde{\epsilon}_4-\protect\widetilde{\epsilon}_1$. 
All these results shown in the subsection indicate that 
quantum fluctuations and many-body renormalization effects  
 are enhanced as $b$ increases  
because the number of active channels decreases  
as the two levels, $m=1$ and $4$, among the four are frozen.


Figure \ref{fig:charge_suscep_A} shows the charge susceptibility $\chi_{c,m}$, 
 which is obtained using these results of the Wilson ratio with   
Eqs.\ \eqref{eq:wilson_ratio} and \eqref{eq:chi_c}. 
The component for $m=1$ and $4$ ($\chi_{c,1}=\chi_{c,4}$) 
decreases as $b$ increases 
because these two levels are frozen for large magnetic fields. 
In the present case, 
Eq.\ \eqref{eq:chi_c} 
can be rewritten in the following form 
for the component for $m=2$ and $3$ ($\chi_{c,3}=\chi_{c,3}$),  
\begin{align}
\pi \Delta\, \chi_{c,2}^{} 
=& \  
\frac{1}{Z_2}  \Biggl[ 
\  1 -  \left( R_{23}-1 \right) 
 \nonumber \\ 
& \qquad  \quad   
-  2\left( R_{12}-1 \right)  
\sqrt{\frac{Z_2\sin^2 \delta_1}{Z_1\sin^2 \delta_2}} 
\;  \Biggr] . 
\end{align}
Here, the prefactor $1/Z_2$ represents an 
enhancement of the quasiparticle density of states for $m=2$ (or $3$) 
whereas the bracket represents magnitude of the susceptibility 
relative to the one for free renormalized quasiparticles. 
The SU($2$)$_{m=2,3}$ part of 
 $\chi_{c,2}^{}$  becomes larger than 
that for the SU($4$) symmetric case 
because the enhancement due to the quasiparticle density of 
states for $m=2$ and $3$
dominates the reduction due to the residual interactions. 
The total impurity susceptibility $\chi_c=\sum_m \chi_{c,m}$ is 
suppressed as magnetic field increases.

\subsection{Perturbations that break the SU($2$)$_{m=2,3}$}
\label{subsec:results_B}

We next take into account the perturbations that break 
the SU($2$)$_{m=2,3}$ symmetry 
and lift the double degeneracy of the one-particle levels at the Fermi level, 
discussed in the above. 
Typical parameter values for such perturbations in a real CNT dot 
are given in Eq.\ \eqref{eq:caseB}. 
The valley mixing  $\Delta_\mathrm{KK'}^{}$ 
and the spin-orbit interaction  $\Delta_\mathrm{SO}$  
open the gap in the four one-particle levels.
For $g_\mathrm{orb} \cos \Theta \neq g_\mathrm{s}/2$, 
the matching of the spin and orbital Zeeman splittings 
becomes no longer perfect. 
Furthermore, 
an extended particle-hole symmetry such as Eq.\ \eqref{eq:p_h_symmetry} 
does not hold at finite parallel fields $b_{\parallel}^{} \neq 0$ 
in the case where $\Delta_\mathrm{SO} \neq 0$.

At zero-field $b=0$, the eigenvalues of 
$\bm{H}_d^0$ can be explicitly written 
as\cite{MantelliMocaZarandGrifoni,Grifoni_PRB2015} 
\begin{align}
\varepsilon_d 
\pm \frac{1}{2}\sqrt{\Delta_\mathrm{KK'}^2 + \Delta_\mathrm{SO}^2} \;. 
\end{align}
Thus, 
$\sqrt{\Delta_\mathrm{KK'}^2 + \Delta_\mathrm{SO}^2}$ is 
the energy gap between the two different groups of the one-particle levels.
The eigenvectors are doubly degenerate,  
which is caused by an SU($2$) symmetry defined 
with respect to the $\bar{\sigma}= (\Uparrow,\Downarrow)$ component  
of the operator  $g_{\bar{\tau},\bar{\sigma}}^{}$:  
%
\begin{align}
\begin{pmatrix} 
g_{+,\Uparrow}^{}  \cr  
g_{-,\Uparrow}^{}  \cr
\end{pmatrix}  
 \equiv  
\begin{pmatrix} 
\psi^{}_{\mathrm{K}\uparrow}    \cr  
\psi^{}_{\mathrm{K}'\uparrow}   \cr
\end{pmatrix}   
,
\qquad
\begin{pmatrix} 
g_{+,\Downarrow}^{}  \cr  
g_{-,\Downarrow}^{}  \cr
\end{pmatrix}   
 \equiv 
\begin{pmatrix} 
\psi^{}_{\mathrm{K}'\downarrow}  \cr  
\psi^{}_{\mathrm{K}\downarrow}   \cr
\end{pmatrix}   
.
\label{eq:SU2_caseB_b0}
\end{align}
In addition, just at $b=0$, an extended particle-hole symmetry holds 
as a results of an invariance with respect to the transformation, 
\begin{align} 
g_{+,\bar{\sigma}}^{\dagger} \Rightarrow - h_{-,\bar{\sigma}}^{}, \qquad 
g_{-,\bar{\sigma}}^{\dagger} \Rightarrow \ h_{+,\bar{\sigma}}^{}  ,
\label{eq:p_h_symmetry_caseB_b0}
\end{align} 
with the corresponding transforms 
similar to those shown in Eq.\ \eqref{eq:p_h_symmetry}  
for conduction electrons. 


\begin{figure}[t]
\begin{center}
\rule{0,5cm}{0cm}
\begin{minipage}{0.68\linewidth}
  \includegraphics[width=\linewidth]{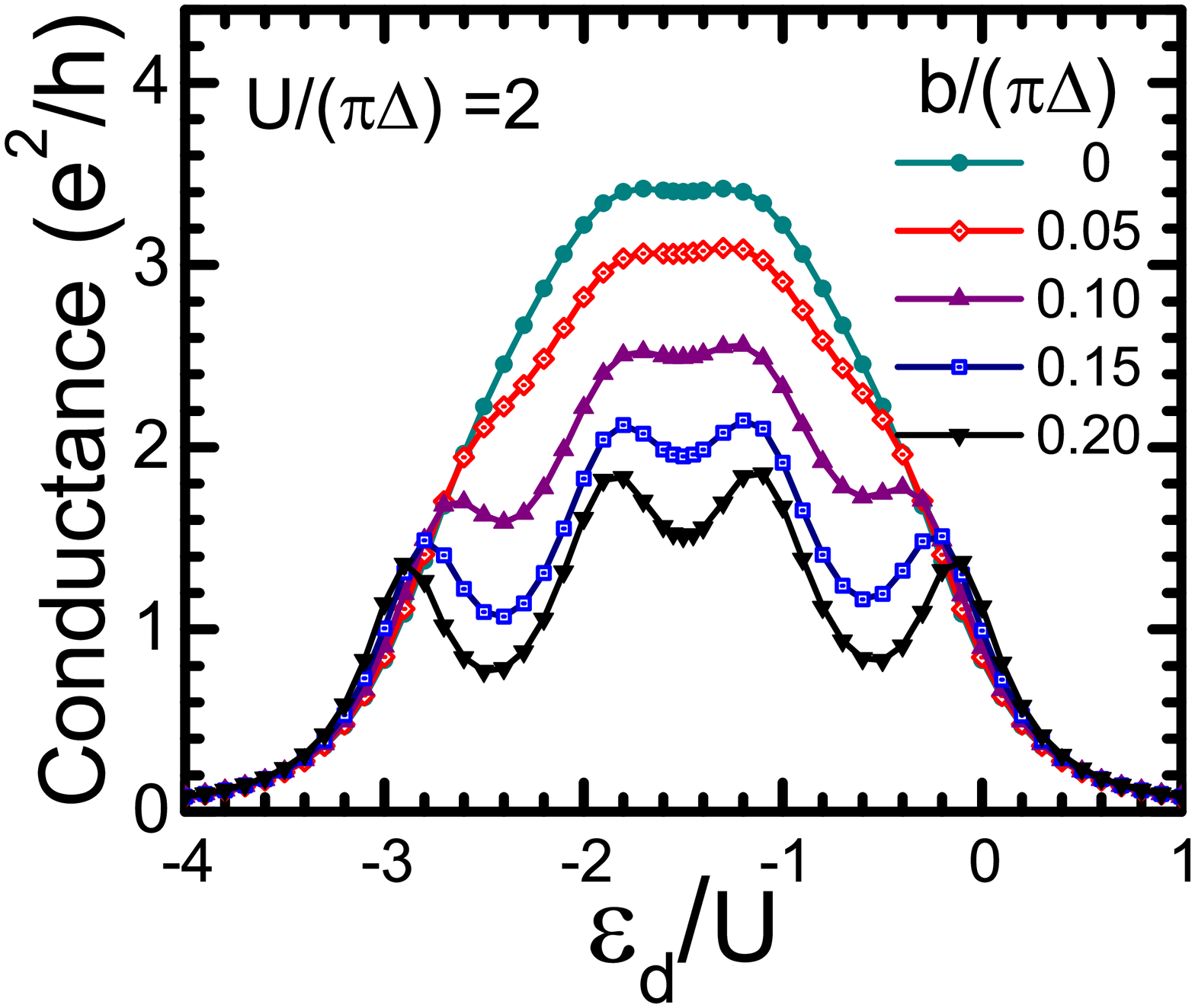}
\end{minipage}
\rule{0.03\linewidth}{0cm}
\begin{minipage}{0.72\linewidth}
  \includegraphics[width=\linewidth]{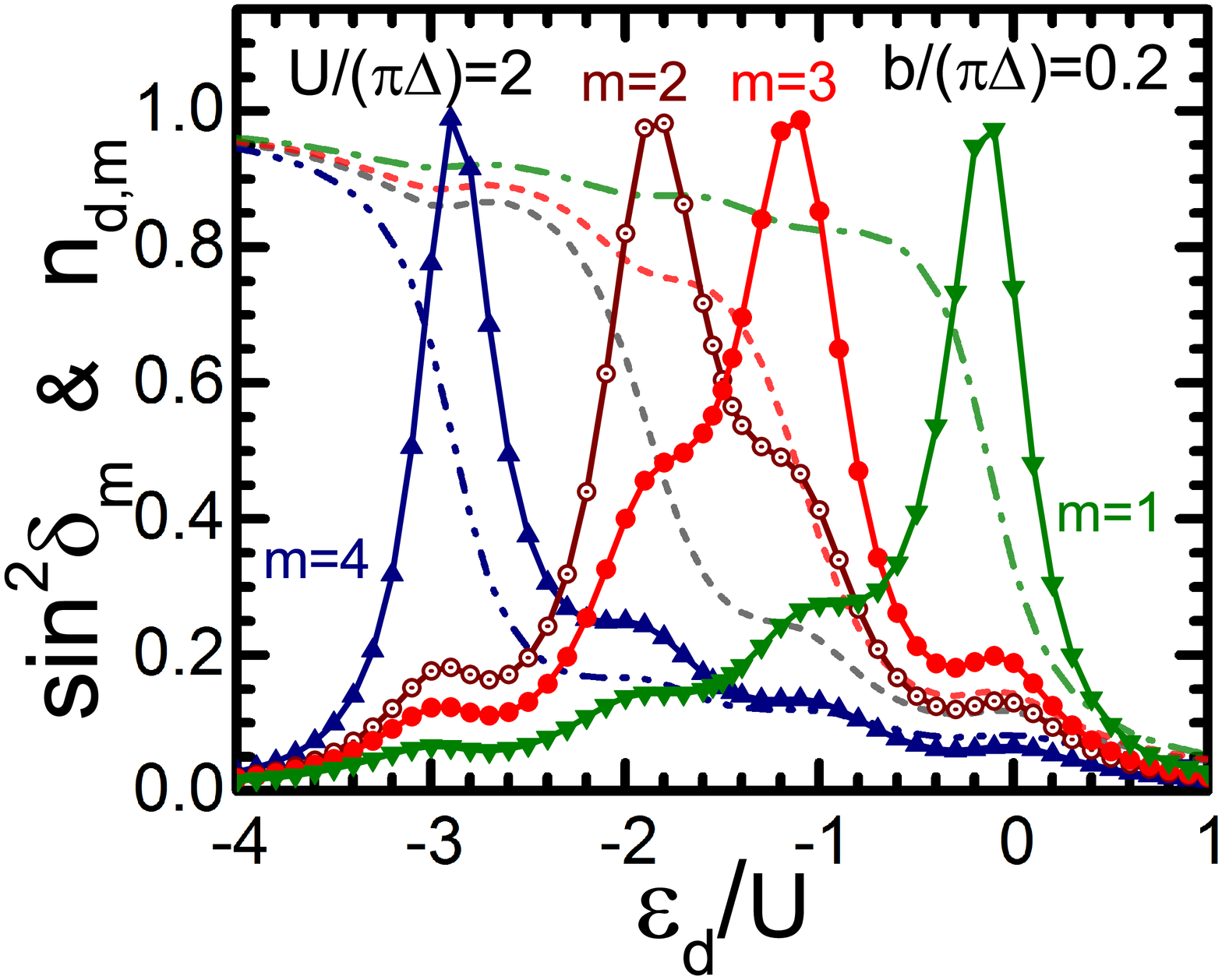}
\end{minipage}
\end{center}
 \caption{ 
(Color online) 
NRG results are plotted vs $\varepsilon_d$ 
for $\epsilon_m$ deduced from Eq.\ \eqref{eq:caseB} 
and $U/(\pi\Delta)=2.0$, 
Upper panel: conductance 
at magnetic fields of 
$b/(\pi\Delta) = 0,\, 0.05,\, 0.1,\, 0.15$, and $0.2$.  
Lower panel: $\sin^2 \delta_m$ (solid line)  
and  $\langle n_{d,m}^{} \rangle$ (dashed line)
 at $b/(\pi\Delta)=0.2$, which corresponds to a real field 
of $B =9.8$ T for $\Delta=0.9$ meV.\cite{Ferrier2016} 
}
 \label{fig:Conductance_B}
\end{figure}


\begin{figure}[t]
\begin{center}
\begin{minipage}{0.69\linewidth}
\includegraphics[width=\linewidth]{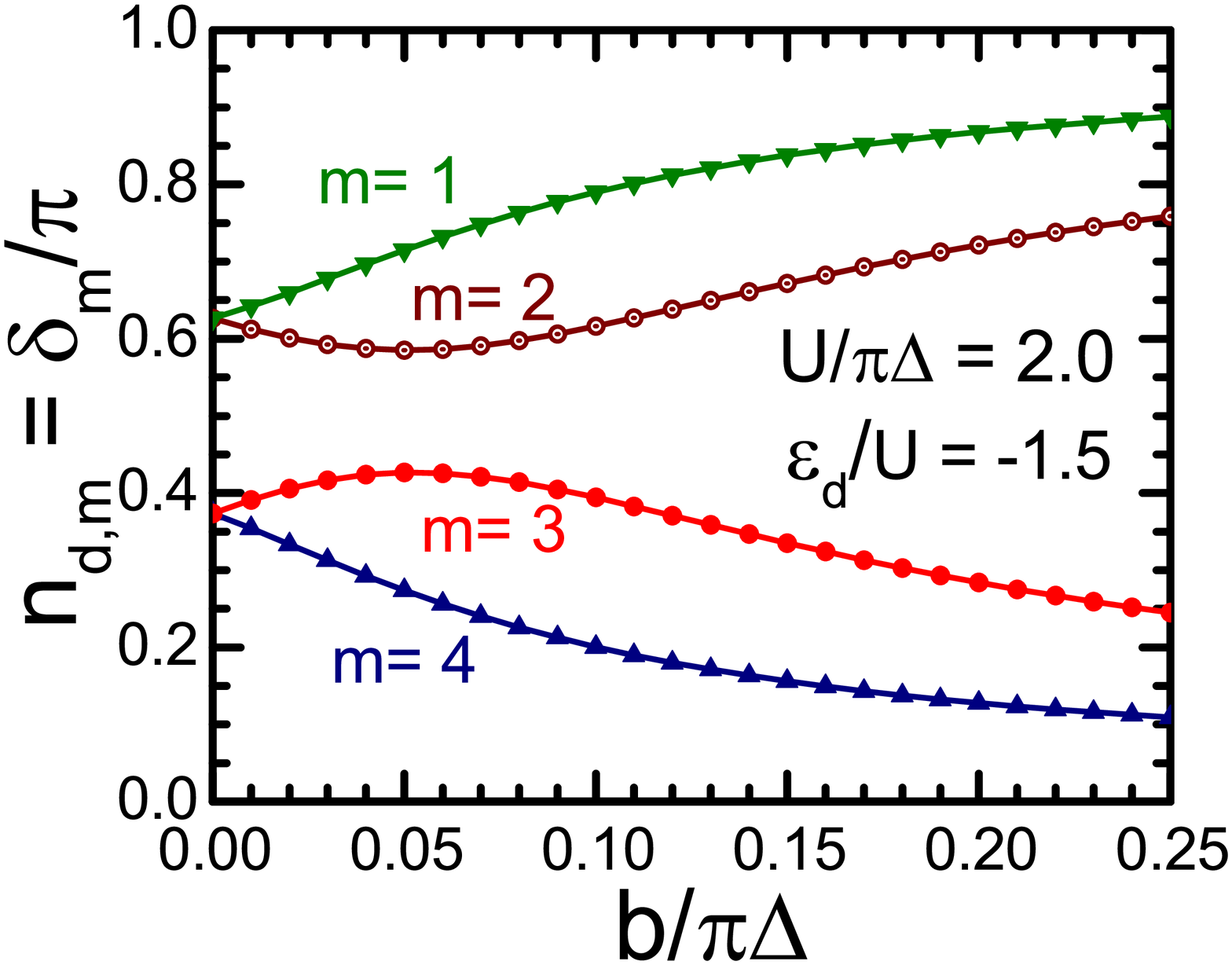}
\end{minipage}
\\
\begin{minipage}{0.7\linewidth}
\includegraphics[width=\linewidth]{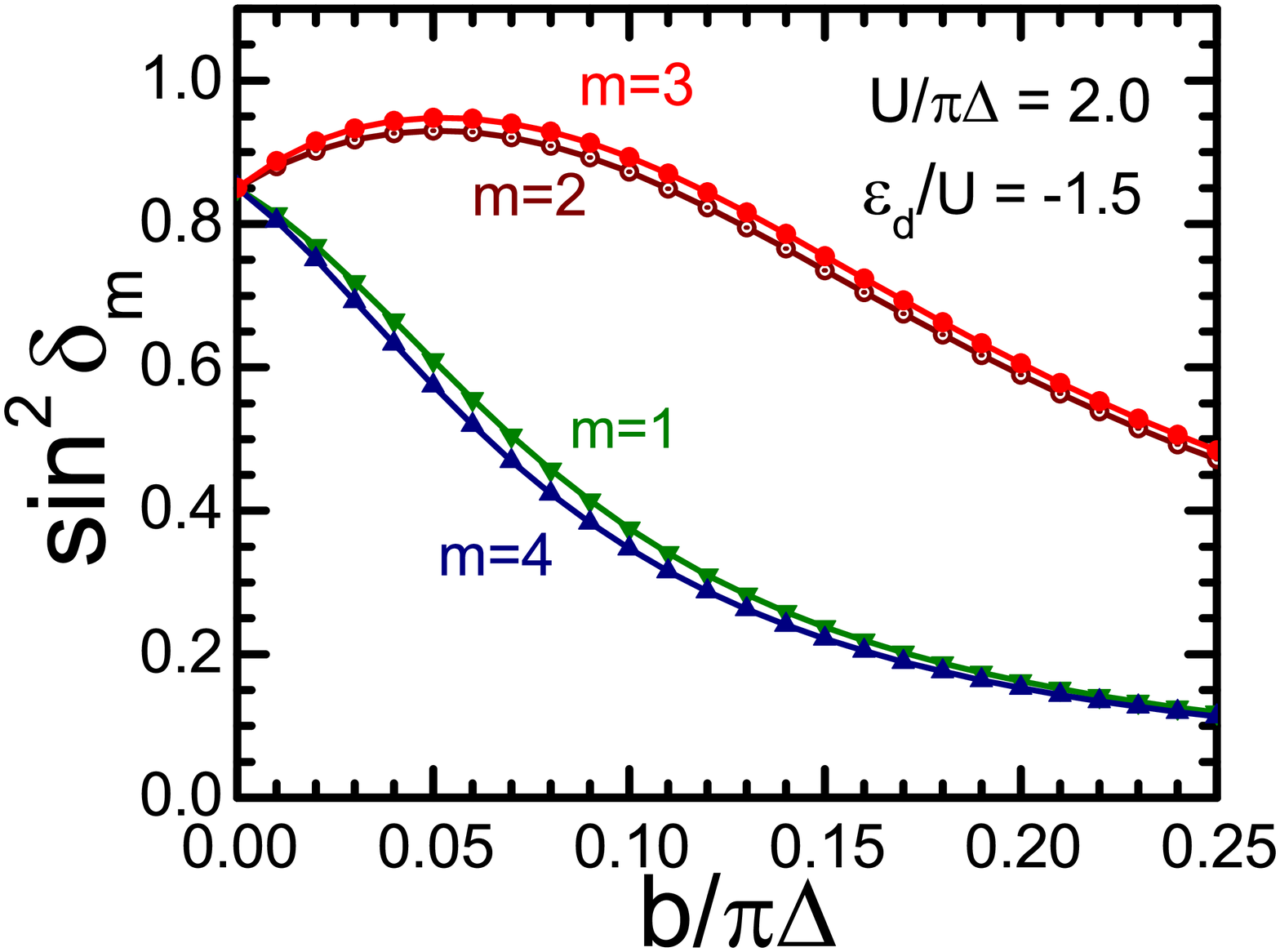}
\end{minipage}
\\
\rule{0,08cm}{0cm}
\begin{minipage}{0.68\linewidth}
\includegraphics[width=\linewidth]{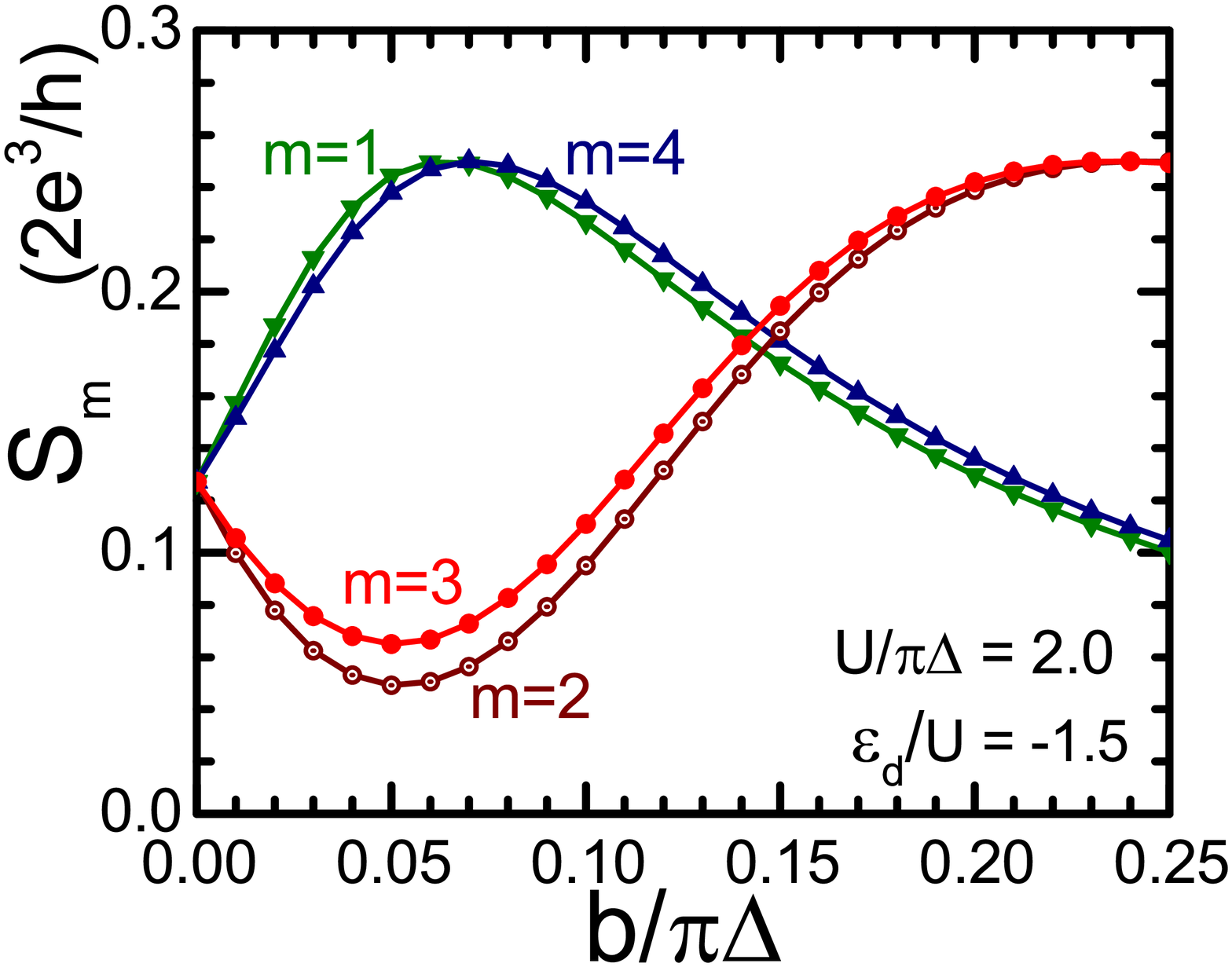}
\end{minipage}
\end{center}
 \caption{ 
(Color online) 
$\langle n_{d,m}^{} \rangle=\delta_m/\pi$, 
$\sin^2 \delta_m$,  and  current noise 
$\mathcal{S}_m$ are plotted vs $b$  
for  $\varepsilon_d =-1.5U$ and $U/(\pi\Delta)=2.0$,  
and $\epsilon_m$ deduced from Eq.\ \eqref{eq:caseB}.
}
 \label{fig:nd_cond_half_B}
\end{figure}

\begin{figure}[t]
\rule{0.2cm}{0cm}
\begin{minipage}{0.68\linewidth}
 \includegraphics[width=\linewidth]{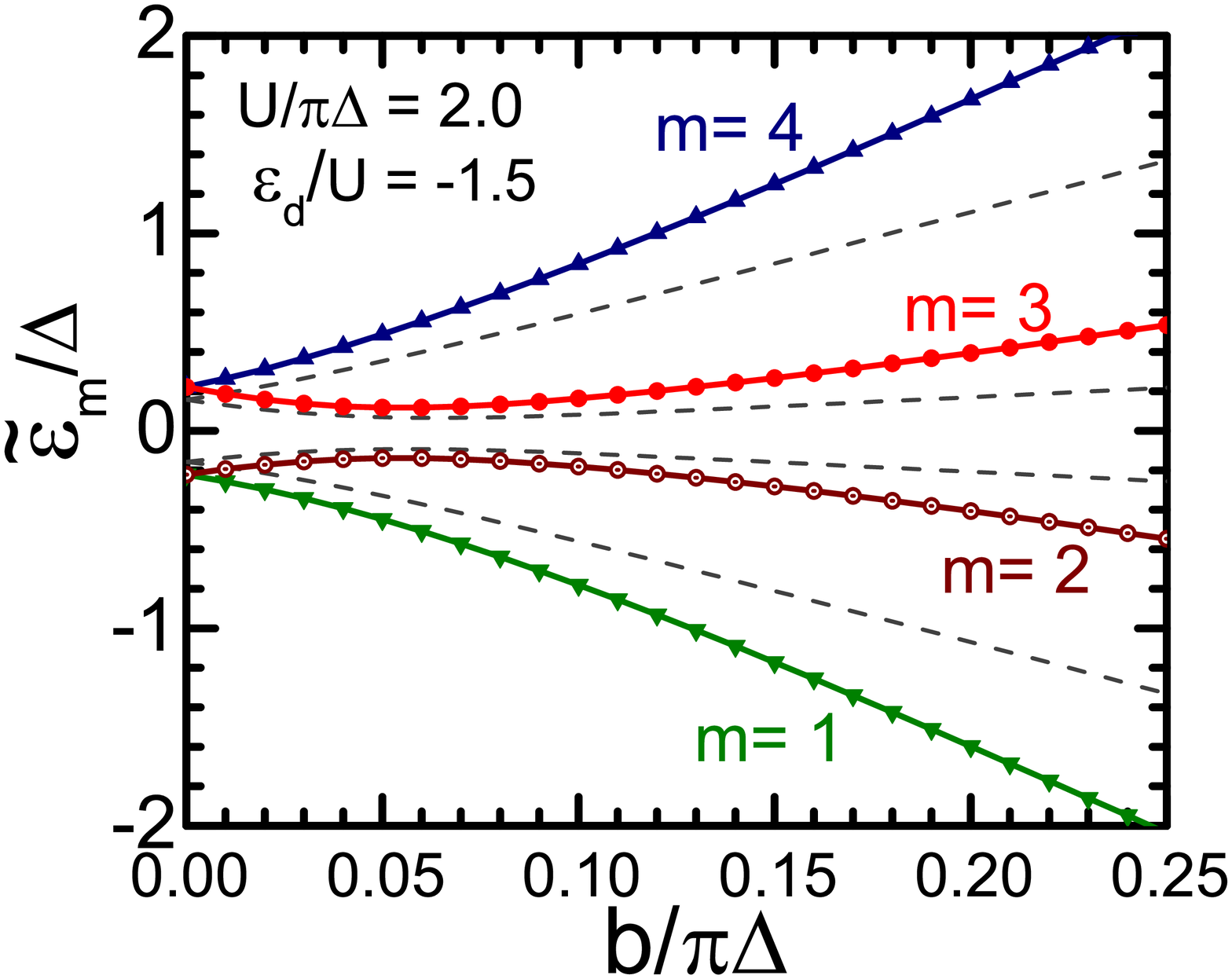}
\end{minipage}
\\
\begin{minipage}{0.7\linewidth}
\includegraphics[width=\linewidth]{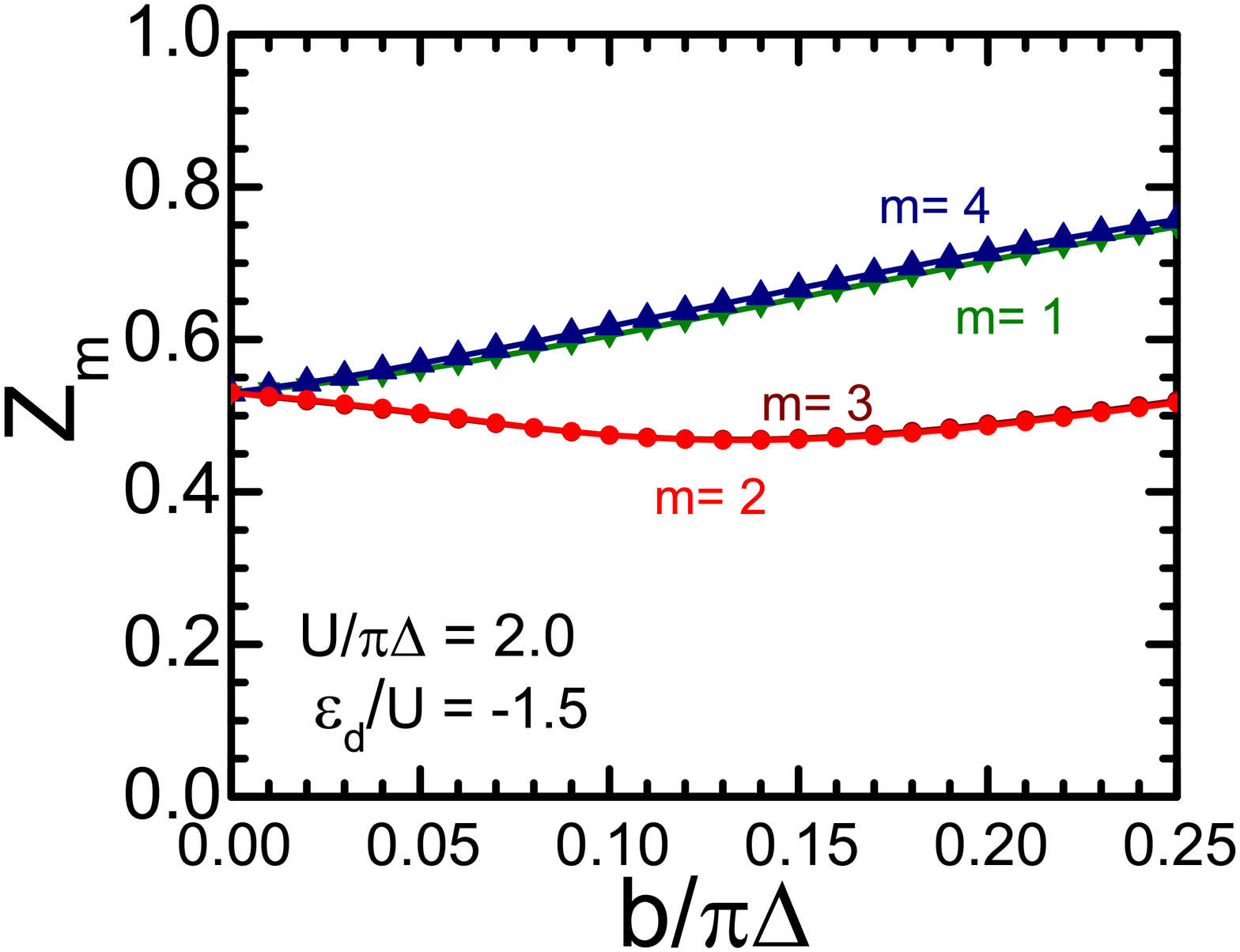}
\end{minipage}
\\
\begin{minipage}{0.7\linewidth}
\includegraphics[width=\linewidth]{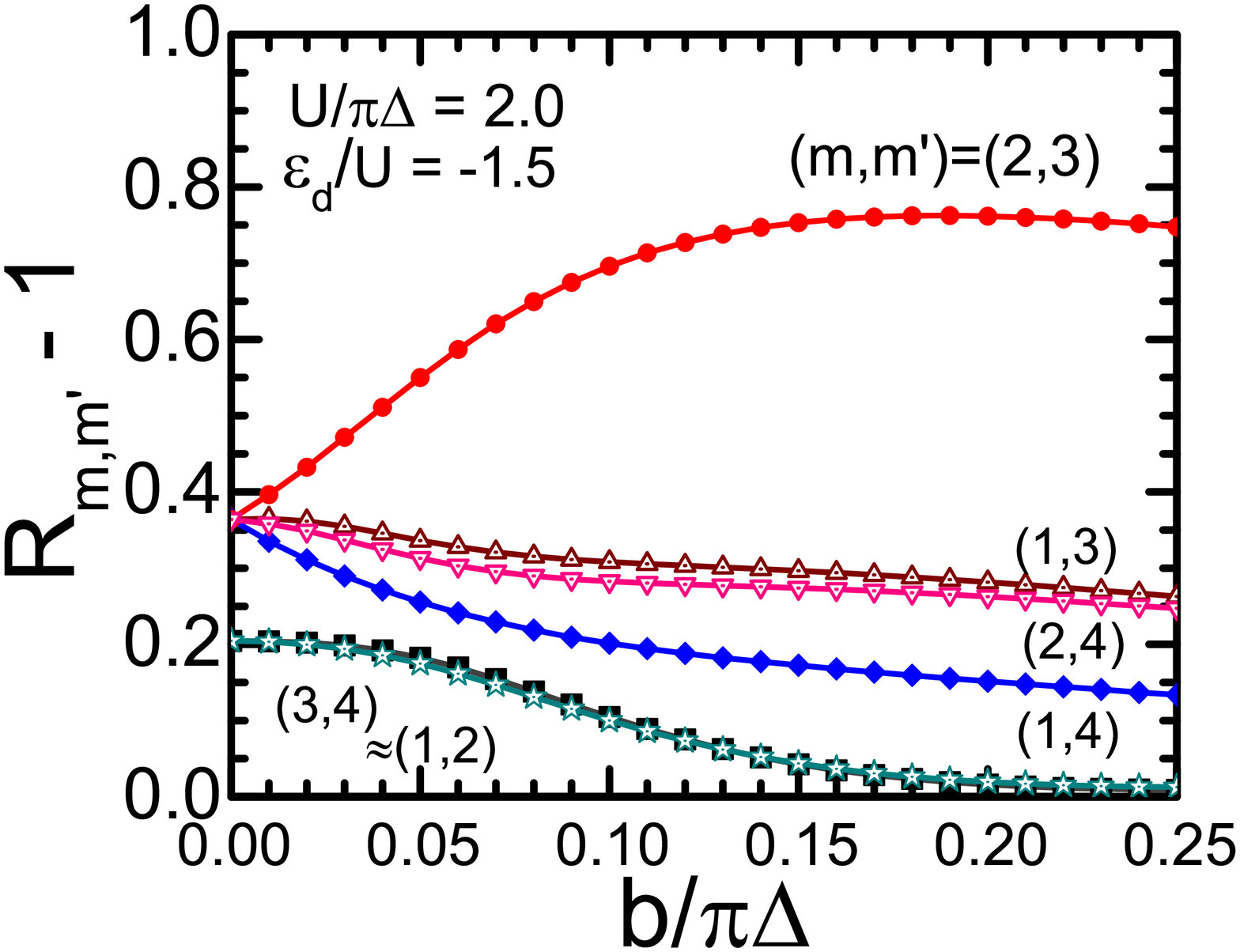}
\end{minipage}
 \caption{ 
(Color online) 
Resonance position  $\widetilde{\epsilon}_m$, 
 $Z_m$  and  $R_{mm'}-1$ 
 plotted vs $b$ 
for  $\varepsilon_d =-1.5U$ and $U/(\pi\Delta)=2.0$,  
and $\epsilon_m$ deduced from Eq.\ \eqref{eq:caseB}.
The dashed line in the top panel 
shows the noninteracting level position for $U=0$. 
}
 \label{fig:z_uren_half_caseB}
\end{figure}


Figure \ref{fig:Conductance_B} shows 
the NRG results  obtained for the parameter set given 
in  Eq.\ \eqref{eq:caseB}  
as a function of $\varepsilon_d$. 
We have chosen the same values 
for the Coulomb interaction $U$ and hybridization energy scale  
 $\Delta$ with $\Delta_L=\Delta_R$  
as those for the SU($2$)$_{m=2,3}$ symmetric case. 
In the upper panel, 
the linear conductance $\mathcal{G}$ is plotted 
for several values of the magnetic field $b$.
We see that a flat plateau emerges for 
$-2.0\lesssim \varepsilon_d/U \lesssim -1.0$ at zero field 
although the height  $\mathcal{G} \approx 3.4 \,e^2/h$ 
is smaller than the unitary limit value. 
This can be compared to the results 
shown in Fig.\ \ref{fig:Conductance_A}.  
The flat structure which is consistent with the recent measurements\cite{Note1}
 is still preserved for small fields $b/(\pi \Delta) \lesssim 0.1$, 
namely up to  $B\approx 5.0$ T in real scale of magnetic field.
For larger fields $b/(\pi\Delta) \gtrsim 0.1$,
the plateau deforms into 
two separate peaks,
 and also there emerge the other two outer sub-peaks.
We can also see in the lower panel of Fig.\  \ref{fig:Conductance_B}  
 how the four conductance peaks 
 are decomposed into  the contributions of 
each conducting channel $m$ at $b/(\pi \Delta)=0.2$.
The separation between the two peaks in the middle and 
their width determine a magnitude of the field,  
at which the plateau collapses.
Such a field  depends significantly on 
the Fermi-liquid corrections.\cite{HewsonBauerOguri} 
The peak structures in the lower panel also represent 
the density of states at the Fermi level, 
defined in Eq.\ \eqref{eq:DOS_int}. 
The density of states, $\rho_{d,m}(0)=\sin^2 \delta_m/(\pi\Delta)$,  
has a long tail in the off-resonance region 
with some steps, at which a resonance peak crosses the Fermi level  
and the occupation number $\langle n_{d,m} \rangle$ shows an abrupt change. 
Note that these results are 
{\it not\/} fully symmetric 
with respect to the point $\varepsilon_d =-1.5U$, except for $b=0$,  
as the particle-hole symmetry defined in 
Eq.\ \eqref{eq:p_h_symmetry_caseB_b0} does not hold for $b \neq 0$.

In Fig.\ \ref{fig:nd_cond_half_B},  
the NRG results that have been deduced from the phase shift $\delta_m$ 
are plotted as a function of the magnetic field $b$,  
for $\varepsilon_d = -1.5U$. 
These results can be compared with 
those  shown in Fig.\ \ref{fig:nd_cond_half_A} 
for the SU($2$)$_{m=2,3}$ symmetry case. 
We see in the top panel that      
the phase shift for $m=2$ and that for $m=3$ are {\it not\/} 
locked at $\pi/2$  in the preset case.
This is because 
that the gap due to the valley mixing and spin-orbit interaction 
lifts the degeneracy 
and the cancellation between the  spin and orbital 
Zeeman effects does {\it not\/} occur.
Nevertheless, 
these phase shifts, $\delta_2$ and $\delta_3$, 
take the values which 
are not far from $\pi/2$ 
for  $b/(\pi\Delta) \lesssim 0.1$, 
and these two  approach closest to each other at $b/(\pi\Delta) \approx 0.05$.
Correspondingly, the transmission probabilities  
through these two levels take the values around 
$\sin^2 \delta_2 \simeq \sin^2 \delta_3 \approx 0.9$.  
The current noise becomes finite 
in the presence of the perturbations  
but it is still not large $\mathcal{S}_2 \simeq \mathcal{S}_3 \approx 0.3$ 
for  $b/(\pi\Delta) \lesssim 0.1$. 
The behavior of the other two levels, $m=1$ and $m=4$, 
are  similar to those in the SU($2$)$_{m=2,3}$ symmetry case 
shown in Fig.\ \ref{fig:nd_cond_half_A},  
except for the region 
near zero field $b/(\pi\Delta) \approx 0.0$, 
where the energy gap due to $\Delta_\mathrm{KK'}$ and $\Delta_\mathrm{SO}$ 
dominate the spin and orbital Zeeman splittings for $m=1$ and $m=4$.

Figure \ref{fig:z_uren_half_caseB} shows 
the results of the renormalized local-Fermi-liquid parameters,  
obtained for the same parameter set.
At zero magnetic field, 
all the wavefunction renormalization factors, namely $Z_m$ for $m=1,2,3,4$, 
become identical. Furthermore, the Wilson ratio $R_{mm'}$,  
or the residual interaction $\widetilde{U}_{mm'}$, 
has two independent components at $b=0$: 
 $R_{12}=R_{34}$ between two different levels with the same energy,  
and  $R_{23}=R_{13}=R_{24}=R_{14}$ between two levels separated by the gap.
This is caused by the SU($2$) and extended particle-hole symmetries, 
described in Eqs.\ \eqref{eq:SU2_caseB_b0} 
and \eqref{eq:p_h_symmetry_caseB_b0}.
We see in the middle and bottom panels that  
the renormalization factors 
 $Z_2$ and $Z_3$, and residual interactions  $R_{23}-1$,  
for the two levels in the middle $m=2$ and $3$,
show a clear crossover behavior 
which is quite similar to those for the SU($2$)$_{m=2,3}$ symmetric case    
shown in Fig.\ \ref{fig:z_uren_half}.
Namely, at $b/(\pi\Delta) \lesssim 0.1$,  
$Z_2$ and $Z_3$ decrease and simultaneously $R_{23}-1$ increases 
as the other two outside levels 
 $\widetilde{\epsilon}_1$ and $\widetilde{\epsilon}_4$ 
 move away from the Fermi level.
It represents that electron correlations are enhanced 
as the fluctuations due to the levels $m=1$ and $4$ are suppressed. 
Note that the results for $Z_2$ and 
those for $Z_3$  almost overlap each other in the middle panel.

The top panel of Fig.\ \ref{fig:z_uren_half_caseB}
shows the position of 
renormalized resonance $\widetilde{\epsilon}_m$ (solid line) 
and that of the bare 
one $\epsilon_m$ (dashed line). 
For small fields, 
the two levels $\widetilde{\epsilon}_2$ and $\widetilde{\epsilon}_3$ 
near the Fermi level approach closer to each other 
until they reach the extreme points at  $b/(\pi\Delta)\approx 0.05$. 
Then, these two levels 
separate again for large fields.  
At the extreme point, the peak separation becomes 
 $\widetilde{\epsilon}_3-\widetilde{\epsilon}_2 \approx 0.25 \Delta$.
This is still smaller than the renormalized 
resonance width $\widetilde{\Delta}_m \equiv Z_m \Delta \sim T_K$ 
because $Z_2 \simeq Z_3 \approx 0.5$. 
The situation, 
 $\widetilde{\epsilon}_3-\widetilde{\epsilon}_2 \lesssim T_K$,
 does not change for $b\lesssim 0.1 \pi \Delta$,  
namely  up to $B\approx 5.0$ T. 
Thus,  although two resonance peaks at  
 $\widetilde{\epsilon}_2$ and $\widetilde{\epsilon}_3$  are separated  
in the realistic case of Eq.\ \eqref{eq:caseB}, 
the  superposition of these two form a single broad peak  
at the Fermi level for $b/(\pi\Delta) \lesssim 0.1$ 
and determines the low-energy behavior.

We can also see in the middle panel that 
 $Z_2$ and $Z_3 $ take a minimum at $b/(\pi \Delta) \approx 0.13$, 
which is larger than the extreme points  
 of $\widetilde{\epsilon}_2$ and $\widetilde{\epsilon}_3$, and 
also than those of the occupation number 
  $\langle n_{d,2}\rangle$ and  $\langle n_{d,3}\rangle$. 
This is caused by the fact that evolution of $Z_2$ and $Z_3$ 
also depends sensitively on the the occupation numbers,  
$\langle n_{d,1}\rangle$ and  $\langle n_{d,4}\rangle$, 
of the other two levels $m=1$ and $4$. 
We see in the top panel of Fig.\ \ref{fig:nd_cond_half_B} 
that  $\langle n_{d,1}\rangle$ and  $\langle n_{d,4}\rangle$ 
still show a linear dependence on $b$ at the extreme point 
of $\langle n_{d,2}\rangle$ and $\langle n_{d,3}\rangle$, 
near $b/(\pi \Delta) \approx 0.05$. 
Around this point of magnetic field $b$, 
the renormalization factors $Z_2$ and $Z_3$ still decrease 
as the variations of  $\langle n_{d,1}\rangle$ and  $\langle n_{d,4}\rangle$  
dominate those of  $\langle n_{d,2}\rangle$ and  $\langle n_{d,3}\rangle$.

 \section{Summary}
 \label{sec:summary}

We have shown 
that the SU($N$) Fermi-liquid fixed point 
that Nozi\`{e}res and Blandin suggested 
for general impurity-electrons filling $M$  
can be interpreted as a Perron-Frobenius eigenvector  
for the composite pairs, each of which  consists of 
one impurity-electron and one conduction-{\it hole} 
carrying the same {\it flavor} \lq\lq $m$''.
It is equivalent to the {\it totally antisymmetric representation\/} 
in the SU($N$) symmetric case. 
The description in terms of the bosonic Perron-Frobenius vector 
does {\it not\/} require the SU($N$) symmetry,  
 and this unique nodeless eigenvector can evolve in a certain region of 
the Hilbert space keeping its components {\it positive definite\/}. 
This is one significant advantage of the hard-core boson interpretation, 
and it also clarifies that the {\it hole picture\/},  
which is introduced only for conduction electrons, 
can naturally describe evolutions of 
the  Fermi-liquid fixed point for $M \leq N/2$.
As an example, we have 
 considered the ground-state wavefunction of  
an anisotropic Coqblin-Schrieffer model with $M$ impurity-electrons 
in the limit of strong exchange couplings.

One of the most interesting features of carbon nanotube quantum dot 
is that various kinds of Kondo effects occur 
in a tunable-parameter space. 
We have shown that the field-induced crossover from the SU($4$) to SU($2$) 
Fermi-liquid  behavior, 
which has been observed in recent experiments 
at two impurity-electrons filling,
can be explained 
as a result of a matching of the spin and orbital Zeeman splittings. 
It yields an emergent SU($2$) symmetry, 
which induces the Kondo effect that is not suppressed by magnetic fields.
Such a matching is expected to be {\it not\/} rare for nanotube dots,  
at least approximately, as the orbital 
magnetic moment $g_\mathrm{orb}$ can take a larger value 
than the spin magnetic moment $g_\mathrm{s}/2$.  
NRG calculations have been carried out 
$i$) for the case with this emergent SU($2$) symmetry,  
and $ii$) for the other case where realistic perturbations 
that break this symmetry are taken into account.
The results for the linear-conductance 
show the behavior that is consistent with the measurements, 
which observe that the height of the Kondo plateau decreases 
as the field increases keeping the flat structure.
This behavior can also be seen for a realistic parameter set 
 at magnetic fields of  $B \lesssim 5.0$ T 
where the level splitting,  
which is caused by the valley mixing, 
the spin-orbit coupling, and mismatching of 
the spin and orbital Zeeman effects,
 becomes smaller than the Kondo energy scale $T_K$.
Furthermore, the NRG results of 
the local-Fermi-liquid parameters for quasiparticles show that 
quantum fluctuations are enhanced 
as the number of active one-particle levels gradually decreases from $4$ to $2$.
These results will be compared with the experiments elsewhere.\cite{Note1}

\begin{acknowledgments}
This work was supported by JSPS KAKENHI Grant Numbers 
JP26220711, JP26400319, JP16K17723, and JP15K17680.
%
\end{acknowledgments}

\appendix


\section{SU($N$) Kondo model for general $M$}
\label{sec:SUN_group}


\subsection{Hubbard operators \& SU($N$) generators}
\label{susec:CS_to_SUnKondo}

The Coqblin-Schrieffer model can be written 
in the form of the SU($N$) Kondo model, 
using a relation between the Hubbard operators 
and the SU($N$) generators. 
The first term of the Coqblin-Schrieffer Hamiltonian,  
given in Eq.\ \eqref{eq:Coqblin-Schrieffer} 
can be expressed in the form
\begin{align}
& 
\!\!\!\!
\sum_{mm'} 
 a_{m}^{\dagger} a_{m'}^{} d_{m'}^{\dagger} d_m^{}  
=
\sum_{mm'} 
\left( \bm{a}^\dagger\bm{X}^{mm'} \bm{a} \right) 
\left( \bm{d}^\dagger \bm{X}^{m'm}\bm{d} \right) 
\nonumber \\
& \quad   
=  \frac{1}{N} 
\,\left( \bm{a}^\dagger \bm{1} \bm{a} \right) \,
\left(\bm{d}^\dagger \bm{1} \bm{d}\right) \,+\, 
2 \left(\bm{a}^\dagger \bm{T}^{\mu} \bm{a}\right) \,
\left(\bm{d}^\dagger \bm{T}^{\mu} \bm{d} \right) .
\label{eq:CF_Kondo_SU4}
\end{align}
Here, $\bm{X}^{mm'}$ is an $N\times N$ matrix 
version of the Hubbard operator   
corresponding to $|m\rangle \langle m'|$, namely  
it has a single non-zero element with 
the value $1$ at $(m,m')$ and all the other elements are zero.
The second line of Eq.\ \eqref{eq:CF_Kondo_SU4}    
follows from a matrix identity 
which corresponds 
to the completeness relation,\cite{GeorgiLieAlg}  
\begin{align}
\!\!
\sum_{m=1}^N
\sum_{m'=1}^N
\bm{X}^{mm'} \otimes \bm{X}^{m'm}
= 
\frac{1}{N} \bm{1} \otimes \bm{1} 
 + 2\, \bm{T}^{\mu}  \otimes \bm{T}^{\mu} .
\label{eq:Hubbard-vs-Gellmann}
\end{align}
It can be proved using the explicit expressions of $\bm{T}^{\mu}$, 
\begin{align}
\bm{T}^{(k+1)^2-1} \equiv& \  
\frac{1}{\sqrt{2k(k+1)}} \left(
\sum_{m=1}^{k} \bm{X}^{mm} -  k\,\bm{X}^{k+1,k+1} \right), 
\label{eq:SUn_Generator_explicit_1}
\\ 
\bm{T}^{k^2-2+2j}\, \equiv&  \  \frac{\bm{X}^{j,k+1} +\bm{X}^{k+1,j}}{2} , 
\label{eq:SUn_Generator_explicit_2}
\\ 
\bm{T}^{k^2-1+2j}\, \equiv& \  \frac{\bm{X}^{j,k+1} -\bm{X}^{k+1,j}}{2i} , 
\label{eq:SUn_Generator_explicit_3}
\end{align} 
where  $j=1,2, \ldots, k$, and $k= 1,2,\ldots, N-1$.  
The assignment of $\mu= 1,2,\ldots, N^2-1$ for $\bm{T}^{\mu}$,   
follows a conventional way of labeling the Gell-Mann 
 matrices  $\bm{\lambda}^{\mu}\equiv \bm{T}^{\mu}/2$  of the SU($N$). 
These matrices have the properties:
$\mathrm{Tr} \left[ \bm{T}^{\mu} \right] =0$, and  
\begin{align}
&  
& \mathrm{Tr} \left[ \bm{T}^{\mu}\,\bm{T}^{\nu} \right] 
\,=\, \frac{1}{2}\, \delta^{\mu\nu} , 
\qquad 
  \bm{T}^{\mu}\,\bm{T}^{\mu} = \frac{N^2-1}{2N} \, \bm{1}\;. 
 \label{eq:Casimir_fundamental} 
\end{align}
Here, $\bm{1}$ is the $N \times N$ unit matrix.  

\subsection{Fock space for $M$ impurity-electrons}
\label{subsec:M_electron_mat}


There are $\binom{N}{M}$ configurations to distribute $M$ electrons 
into $N$ impurity levels, 
\begin{align}
\left| \{\alpha\}\right\rangle_M^{}   
\,\equiv \,
d_{\alpha_1}^{\dagger}
d_{\alpha_2}^{\dagger} \cdots 
d_{\alpha_M}^{\dagger} \left| 0 \right\rangle \;.  
\label{eq:impurity_part_Fock}
\end{align}
Here, $\{\alpha\}=\{\alpha_1,\alpha_2,\ldots,\alpha_M\}$ 
represents a set of $M$ occupied impurity levels. 
This state can be regarded as  
an {\it antisymmetric representation\/} 
of the SU($N$).  With this basis set, the operator 
$\left(\bm{d}^\dagger \bm{T}^{\mu} \bm{d} \right)$ 
can be written in a matrix form, 
\begin{align}
\left\{\mathcal{S}_{r_M^{}}^{\mu}\right\}_{\alpha'\alpha} = \  
{_M^{}}\!\left \langle \{\alpha'\} \right| 
\bm{d}^\dagger \bm{T}^{\mu} \bm{d} 
\left| \{\alpha\} \right\rangle_M^{}  
\;.
\label{eq:impurity_part_matrix}
\end{align}
The Casimir operator for this representation is given by
\begin{align}
\mathcal{S}_{r_M^{}}^{\mu} \mathcal{S}_{r_M^{}}^{\mu}
= & \ C_2(r_M^{}) \, \mathcal{I}_{r_M^{}}, 
\\
C_2(r_M^{})   
  \equiv& \  \frac{M(N-M)(N+1)}{2N} \;.
 \label{eq:Casimir_general} 
\end{align}
Here,  $\mathcal{I}_{r_M^{}}$ is 
the $\binom{N}{M}$ dimensional unit matrix.


\subsection{Poor man's scaling for $M$ impurity-electrons}

\label{sec:Poorman_SUn}

The one-loop scaling equation for $M$ impurity-electrons  
can be obtained in a symmetric way,  using the exchange interaction given 
in  Eq.\ \eqref{eq:SUn_Kondo_spin}. 
Following the standard prescription,\cite{PoormansSC,HewsonBook} 
we introduce $\widetilde{D} = D-\delta D$ 
for the bonding components of conduction electrons $c_{\epsilon m} \equiv 
\! \sum_{\nu=L,R} v_{\nu}^{} c_{\nu, \epsilon m}/v$, 
and obtain the order $J_K^2$ corrections, 
\begin{align}
& 
\delta \mathcal{H}_K^{(M)} 
= \,  J_K^2 \, \rho_c \! 
\int_{-D}^D \!\! d\epsilon 
\, d\epsilon'
\Biggl[ \,
\nonumber \\
& \qquad \quad 
\int_{\widetilde{D}}^{D} \!\! d\xi
\,d\xi' \, 
\bm{c}_{\epsilon'}^{\dagger} 
\bm{T}^{\mu} 
\bm{c}_{\xi}^{}  
\mathcal{S}_{r_M^{}}^{\mu}
\frac{\rho_c}{\omega-D+\epsilon}\,
\bm{c}_{\xi'}^{\dagger} 
\bm{T}^{\nu} 
\bm{c}_{\epsilon}^{}\mathcal{S}_{r_M^{}}^{\nu}
\nonumber \\
& \qquad  
+ 
\int_{-D}^{-\widetilde{D}} \!\!\!\!  d\xi
\, d\xi'\, 
\bm{c}_{\xi'}^{\dagger} 
\bm{T}^{\nu} 
\bm{c}_{\epsilon}^{}  
\mathcal{S}_{r_M^{}}^{\nu}
\frac{\rho_c}{\omega-D-\epsilon'}\,
\bm{c}_{\epsilon'}^{\dagger} 
\bm{T}^{\mu} 
\bm{c}_{\xi}^{}\ 
\mathcal{S}_{r_M^{}}^{\mu}
\,\Biggr] 
\nonumber \\ 
& \quad \simeq  
\,   J_K^2 \, 
\frac{\rho_c\delta D}{\omega-D}\ 
\bm{a}^{\dagger} 
\bm{T}^{\mu}\bm{T}^{\nu} 
\bm{a}\,  
\left( 
\mathcal{S}_{r_M^{}}^{\mu}\mathcal{S}_{r_M^{}}^{\nu}
- 
\mathcal{S}_{r_M^{}}^{\nu}\mathcal{S}_{r_M^{}}^{\mu}
\right) 
\nonumber \\
& \quad  =\, 
 J_K^2 \, 
\frac{\rho_c\delta D}{\omega-D}\,
\left(-\frac{1}{2}\right) 
f^{\mu\nu\lambda}
f^{\mu\nu\lambda'}
\bm{a}^{\dagger} 
\bm{T}^{\lambda'} 
\bm{a}\,  
\mathcal{S}_{r_M^{}}^{\lambda}
\nonumber \\
& \quad = \,
 J_K^2 \, 
\frac{\rho_c\delta D}{\omega-D}\,
\left(-\frac{N}{2}\right) 
\bm{a}^{\dagger} 
\bm{T}^{\lambda} 
\bm{a}\,  
\mathcal{S}_{r_M^{}}^{\lambda}
\;.
\label{eq:SUN_scaling}
\end{align}
Here, the factor $N$ emerges from 
$f^{\mu\nu\lambda}f^{\mu\nu\lambda'} 
= N \delta^{\lambda\lambda'}$.\cite{PeskinSchroeder}  
From Eq.\ \eqref{eq:SUN_scaling}
the scaling equation, which obviously agrees with  
the results of Nozi\`{e}res and Blandin given in 
Eq.\ (14) of Ref.\ \onlinecite{NozieresBlandin},  
follows
\begin{align}
- \,\frac{d}{d \widetilde{D}} \left( \rho_c \widetilde{J}_K \right)
\,=\, \frac{N}{2\widetilde{D}} \left(\rho_c \widetilde{J}_K\right)^2 \;. 
\label{eq:scaling_SUn}
\end{align}
This gives $T_K = D \, e^{ - \frac{2}{N \rho_c J_{K}}}$, defined such that  
the effective coupling 
diverges  $\widetilde{J}_K\to \infty$ 
at $\widetilde{D} \searrow T_K$. 
Note that the perturbative scaling 
does not depend on $M$ in the one-loop order with respect to $J_K$. 

\section{Bosonic Perron-Frobenius vector without the SU($N$) symmetry}
\label{sec:anisotropic_exchange}

\subsection{Anisotropic exchange interaction}

The Hamiltonian $\mathcal{H}$ no longer has the SU($N$) symmetry 
in the case where the $N$-fold degeneracy of the one-particle impurity levels 
are lifted.
In this case, 
the effective Hamiltonian for the subspace 
with fixed $M$ impurity-electrons 
$\mathcal{H}_\mathrm{eff}^\mathrm{ais}$ can be obtained, 
extending Eq.\ \eqref{eq:Coqblin-Schrieffer_def} as follows.
Replacing $E_M$ in the energy denominator 
in the right-hand side of Eq.\ \eqref{eq:Coqblin-Schrieffer_def} 
by the lowest energy $E_{M}^\mathrm{min}$, and  inserting the complete set 
$\left| \{\alpha\} \right\rangle_M^{}$ for impurity states defined in 
Eq.\ \eqref{eq:impurity_part_Fock}, it takes the form 
%
\begin{align}
\mathcal{H}_\mathrm{eff}^\mathrm{ais}
\,= & \    
\sum_{\{\alpha\}}  
\mathcal{H}_{T}^{}\, 
\frac{1}{E_{M}^\mathrm{min}  -(\mathcal{H}_{d}^{}+\mathcal{H}_{c}^{} )}
\, \mathcal{H}_\mathrm{T}^{} 
\left| \{\alpha\} \right\rangle_M^{} \! 
{_M^{}}\!\left\langle \{\alpha\} \right| .
\label{eq:aniso_effective_Hamiltonian}
\end{align}
In this case, the energy of intermediate state 
depends on the initial impurity state $\{\alpha\}$.
We will use the notation $\mathcal{H}_d \left| \{\alpha\} \right\rangle_M^{} = 
E_{M}^{\{\alpha\}} \left| \{\alpha\} \right\rangle_M^{}$, and 
$E_{M}^\mathrm{min}$  ($E_{M}^\mathrm{max}$) 
is the lowest (highest)  energy 
in the $M$ impurity-electrons states.
%
Taking only into account the exchange-interaction part of 
 Eq.\ \eqref{eq:aniso_effective_Hamiltonian}, 
we consider a model defined by 
\begin{align}
&\mathcal{H}_\mathrm{K}^\mathrm{ais} \equiv   
\frac{1}{2}
\sum_{mm'} 
\sum_{\{\alpha\}}
\biggl(
J_{m'm}^{\{\alpha\}}
 a_{m}^{\dagger} a_{m'}^{} d_{m'}^{\dagger} d_m^{}
\nonumber \\ 
& \qquad \qquad \quad \ \ 
- 
\frac{\overline{J}}{N} 
a_{m'}^{\dagger} a_{m'}^{} d_{m}^{\dagger} d_{m}^{}
\biggr)
\left| \{\alpha\} \right\rangle_M^{} \! 
{_M^{}}\!\left\langle \{\alpha\} \right|  , 
\label{eq:J_aniso_appendix_def}
\\
&J_{m'm}^{\{\alpha\}} 
 \,\equiv\,  
2\biggl(
\frac{v^2}{
E_{M}^{\{\alpha\}}-E_{M}^\mathrm{min} +MU +\epsilon_{m'}
}\,
\nonumber 
\\
& \qquad \qquad \quad 
+
\,
\frac{v^2}{
E_{M}^{\{\alpha\}} 
-E_{M}^\mathrm{min}
 -(M-1)U -\epsilon_{m}  }
\biggr) ,
\label{eq:J_aniso_appendix}
\\
&
\overline{J}
 \,\equiv \,
\frac{1}{\binom{N}{M}}
\sum_{\{\alpha\}}
\frac{1}{M}
\sum_{m} 
J_{mm}^{\{\alpha\}}\  
{_M^{}}\!\left\langle \{\alpha\} \right| 
  d_{m}^{\dagger} d_m^{} \!
\left| \{\alpha\} \right\rangle_M^{}  
 .
\label{eq:J_aniso_mean_appendix}
\end{align}
Here, $\overline{J}$ is defined such that  
$\mathcal{H}_\mathrm{K}^\mathrm{ais}$ becomes traceless,
\begin{align}
&
\sum_{\{\alpha\}}
\sum_{\{\gamma\}}\, 
\left\langle \{\alpha\},\{\gamma\} \right| 
\mathcal{H}_\mathrm{K}^\mathrm{ais} 
\left| \{\gamma\}, \{\alpha\} \right\rangle 
= \, 0 , \\
&\left| \{\gamma\}, \{\alpha\} \right\rangle 
 \equiv   \,  
 \prod_{i=1}^{N_a} \,
 a^{\dagger}_{\gamma_i^{}}  \, 
 \prod_{j=1}^{M} \, 
 d^{\dagger}_{\alpha_j^{}}  \, |0 \rangle\; , 
%
\end{align}
where $\{\gamma_1^{},\gamma_2^{},\ldots,\gamma_{N_a}^{}\}$ represents 
a set of occupied conduction electron levels.

We assume that $M=1,2,\ldots N/2$ in the following, 
as the results for the cases $M>N/2$ can be deduced from 
those for $M<N/2$ through the particle-hole transform.
To be specific, we consider the case where each of $\epsilon_m$'s 
takes a certain value bounded in the range $\delta \epsilon$ 
near the middle of the $M$-electron region  Eq.\ \eqref{eq:M_electron_step},
%
\begin{align}
\epsilon_{m} =-\left( M - \frac{1}{2} \right) U + \delta \epsilon_m, 
\quad \    -\frac{\delta \epsilon}{2} < 
\delta \epsilon_m <\frac{\delta \epsilon}{2}.
\label{eq:model_aniso}
\end{align}
We assume that the range $\delta \epsilon$ to be $0\leq \delta \epsilon<U$.  
In this case, the exchange coupling is positive 
$J_{m'm}^{\{\alpha\}} >0$ for all $m$, $m'$, and $\{\alpha\}$. 
However,  this condition is still not sufficient 
 for setting up all the $M\pm 1$ impurity-electrons energies 
 to be much larger  than the $M$ impurity-electrons energies. 
The additional conditions, 
 $E_{M\pm 1}^\mathrm{min} - E_{M}^\mathrm{max} \gg  0$, are required:  
\begin{align}
E_{M+1}^\mathrm{min} - E_{M}^\mathrm{max} 
=& \  
E_{M}^\mathrm{min}   
+ \frac{U}{2} + \delta \epsilon_\mathrm{LU}^{} - E_{M}^\mathrm{max} \gg 0 ,
\label{eq:M+1_states}
\\
E_{M-1}^\mathrm{min} - E_{M}^\mathrm{max} 
=& \  
E_{M}^\mathrm{min} 
+ \frac{U}{2} - \delta \epsilon_\mathrm{HO}^{} - E_{M}^\mathrm{max}\gg 0 .
\label{eq:M-1_states}
\end{align}
Here, 
the lowest-unoccupied level $\delta\epsilon_\mathrm{LU}^{}$ 
and  the highest-occupied level  $\delta\epsilon_\mathrm{HO}^{}$ 
are defined with respect to the lowest $M$ impurity-electrons ground state,  
and are measured from $-(M-1/2)U$ as Eq.\ \eqref{eq:model_aniso}. 
These two conditions, Eqs.\ \eqref{eq:M+1_states} and \eqref{eq:M-1_states}, 
are sufficiently satisfied if the Coulomb repulsion is 
large much larger than the energy separation 
between $M$-electrons impurity states  such that 
\begin{align}
\frac{U}{2} \gg \left(M + \frac{1}{2}\right) \delta \epsilon  .
\label{eq:asym_conditions}
\end{align}
Note that the energy separation 
is bounded in the range 
 $ E_{M}^\mathrm{max} -  E_{M}^\mathrm{min}<M \delta \epsilon$ 
by definition given in Eq.\ \eqref{eq:model_aniso}.

The lower bound of the exchange interaction 
can be estimated from Eq.\ \eqref{eq:J_aniso_appendix}
through the matrix element for which 
 $E_{M}^{\{\alpha\}} = E_{M}^\mathrm{max}$,  
\begin{align}
J_{m'm}^{\{\alpha\}} 
 > & \   
\frac{2v^2}{
M \delta \epsilon +\frac{U}{2} +\delta \epsilon_{m'}
}\,
+
\,
\frac{2v^2}{
M \delta \epsilon 
 +\frac{U}{2} -\delta\epsilon_{m}  } 
\nonumber \\
>& \   
 \frac{4v^2}{
\left(M+\frac{1}{2}\right)  \delta \epsilon +\frac{U}{2}
} \ = \ \frac{J_K^0}{1+x} 
\;.
\label{eq:Jmm_Min}
\end{align}
Here,  $J_K^0 \equiv 8v^2/U$ 
and $x = \frac{2 \delta \epsilon}{U}(M+1/2)$. 
Note that  $x \ll 1$ from Eq.\ \eqref{eq:asym_conditions}.
Similarly, the upper bound of the exchange interaction is  
estimated through the matrix element for which 
 $E_{M}^{\{\alpha\}} = E_{M}^\mathrm{min}$,  
\begin{align}
J_{m'm}^{\{\alpha\}}  < & \ 
\frac{2 v^2}{ \frac{U}{2} +\delta \epsilon_{m'} }
+\frac{2 v^2}{ \frac{U}{2} -\delta \epsilon_{m} } 
\nonumber \\
< & \ 
 \frac{8 v^2}{ U - \delta \epsilon } \, 
= \frac{J_K^0}{1 -\frac{x}{2M+1} } \, .
\label{eq:Jmm_Max}
\end{align}
This also determines the upper bound of the average $\overline{J}$,
defined in Eq.\ \eqref{eq:J_aniso_mean_appendix}, as 
\begin{align}
 \frac{J_K^0}{1 -\frac{x}{2M+1} } > \overline{J} .
\label{eq:J_mean_upper_bound}
\end{align}

We next consider the following inequality 
\begin{align}
\frac{J_K^0}{1+x} > 
\frac{J_K^0}{1 -\frac{x}{2M+1} } \,\frac{M}{N},
\label{eq:J_temporal}
\end{align}
which is satisfied for  $0\leq x \ll 1$, 
namely in the case where Eq.\ \eqref{eq:asym_conditions} holds 
with $M/N \leq 1/2$. 
In this case, we obtain the following relation, 
using Eqs.\ \eqref{eq:Jmm_Min}, \eqref{eq:J_mean_upper_bound},
and \eqref{eq:J_temporal}, 
\begin{align}
J_{mm'}^{\{\alpha\}} \,>\,\frac{\overline{J}M}{N} .
\label{eq:inequality}
\end{align}


%

\subsection{Strong exchange coupling limit}
\label{subsec:ground-state_anisotropic}

The eigenstates of 
the Hamiltonian $\mathcal{H}_K^\mathrm{ais}+\mathcal{H}_d$, 
which is described in in Eq.\ \eqref{eq:CF_hard_core_boson_aniso_2_2site} 
with the {\it hole-picture\/},
can be expanded using the basis set  
Eqs.\ \eqref{eq:basis_aniso_1} and \eqref{eq:basis_aniso_2}. 

As mentioned in Sec.\ \ref{subsec:boson_general}, 
 the pair wavefunction  $\psi_{\mathrm{pair}}^{}(\{m_{\mathrm{p}}\})$   
is an eigenvector of 
a $\binom{N_\mathcal{UB}^{}}{N_\mathrm{pair}}$ dimensional 
Hamiltonian matrix, which has  
  $\left(N_\mathcal{UB}^{}-N_\mathrm{pair}\right)N_\mathrm{pair}$ 
 {\it negative\/} off-diagonal elements, $- J_{m'm}^{\{\alpha\}}/2$, 
in each column and their Hermitian-conjugate elements in each row. 
The gain of the {\it hopping\/} energy of the pairs is maximized for 
 $N_{\mathcal{O}_d} = N_{\mathcal{O}_h} = 0$, 
where the pairs are not blocked by the unpaired objects.
In this case, the dimension of the subspace becomes largest 
\begin{align}
& \binom{N_\mathcal{UB}^{}}{N_\mathrm{pair}}
= 
\binom{N-N_{\mathcal{O}_d}-N_{\mathcal{O}_h}}{M-N_{\mathcal{O}_d}} 
\leq \binom{N}{M} ,
\end{align}
as the numbers of unpaired objects take values in the range:  
 $N_{\mathcal{O}_d}=0,1,\ldots,M$ and 
$N_{\mathcal{O}_h}=0,1,\ldots,N-M$.  
Furthermore, the number of the states which are  
directly linked by the negative  off-diagonal elements is maximized   
\begin{align}
\left(N_\mathcal{UB}^{}-N_\mathrm{pair}\right)N_\mathrm{pair} 
=  & \  (N-M-N_{\mathcal{O}_h})(M-N_{\mathcal{O}_d}) 
\nonumber 
\\ 
\leq & \ (N-M) \, M .
\end{align}

We next examine the diagonal elements.  
The diagonal matrix elements of $\mathcal{H}_K^\mathrm{ais}+\mathcal{H}_d$ 
with respect to the basis set 
Eq.\ \eqref{eq:basis_aniso_1}, or \eqref{eq:basis_aniso_2}, 
are determined  
by the last three terms of Eq.\ \eqref{eq:CF_hard_core_boson_aniso_2_2site}, 
\begin{align}
&\sum_{m_d \in \mathcal{O}_d}^{N_{\mathcal{O}_d}} \frac{J_{m_d,m_d}^{\{\alpha\}}}{2} 
\,-\,
\frac{\overline{J}M\bigl[N-(M-N_{\mathcal{O}_d}+N_{\mathcal{O}_h})\bigr]}{2N} 
+ E_{M}^{\{\alpha\}} .
\label{eq:aniso_2site_diagonal}
\end{align}
Here, the impurity part $\left| \{\alpha\} \right\rangle_M^{}$ 
contains $N_{\mathcal{O}_d}$ unpaired electrons at ${\{m_d\}}$  
and $N_\mathrm{pair}$ electrons consisting pairs at ${\{m_\mathrm{p}\}}$. 
As $\overline{J}>0$, the diagonal element increases with $N_{\mathcal{O}_h}$,
and thus it is minimized at $N_{\mathcal{O}_h}=0$. 
In order to minimize the diagonal elements further 
varying $N_{\mathcal{O}_d}$, 
we rewrite Eq.\ \eqref{eq:aniso_2site_diagonal} in the following form 
taking  $N_{\mathcal{O}_h}=0$, 
\begin{align}
\frac{1}{2}\sum_{m_d \in \mathcal{O}_d}^{N_{\mathcal{O}_d}} 
\left( J_{m_d,m_d}^{\{\alpha\}} - \frac{\overline{J}M}{N} \right) 
-
\frac{\overline{J}M (N-M)}{2N} 
+ E_{M}^{\{\alpha\}} .
\label{eq:aniso_2site_diagonal_NOh0}
\end{align}
The first term has a lower bound which  
follows from Eqs.\ \eqref{eq:asym_conditions} and 
\eqref{eq:inequality}, 
\begin{align}
\frac{1}{2}\sum_{m_d \in \mathcal{O}_d}^{N_{\mathcal{O}_d}} 
\left( J_{m_d,m_d}^{\{\alpha\}} - \frac{\overline{J}M}{N} \right) 
\geq 0
\;,
\end{align}
where the  equality holds at $N_{\mathcal{O}_d}=0$.  
Thus, 
each of the diagonal elements for given $\{\alpha\}$ 
takes the smallest value for $N_{\mathcal{O}_d}=N_{\mathcal{O}_h}=0$: 
\begin{align}
 -\frac{\overline{J}M\,(N-M)}{2N} + E_{M}^{\{\alpha\}}
\;.
\end{align}

Due to these structures of the Hamiltonian matrix, 
the ground state of $\mathcal{H}_K^\mathrm{ais}+\mathcal{H}_d$  
is given by the Perron-Frobenius vector 
for $N_{\mathcal{O}_d}=N_{\mathcal{O}_h}=0$.

\section{Ward identities}
\label{sec:Ward_identities}

The occupation number of the impurity levels can be written, using 
the Friedel sum rule given in Eq.\ \eqref{eq:Friedel}, 
\begin{align}
\left\langle n_{d,m}\right\rangle \, =\, 
\frac{1}{\pi}\, \cot^{-1} 
\left[\frac{\epsilon_m +\Sigma_{m}(0)}{\Delta}\right] .
\end{align}
Taking a derivative with respect to $\epsilon_{m'}$,  we obtain,
\begin{align}
\frac{\partial \left\langle n_{d,m}\right\rangle }{\partial \epsilon_{m'}}  
\,= & \  
-\rho_{d,m}^{}(0)\, \widetilde{\chi}_{mm'}^{} \;,
\label{eq:dn_depsilon}
\\
\widetilde{\chi}_{mm'}^{} 
\, \equiv & \  
\delta_{mm'}^{} 
+ \frac{\partial\Sigma_{m}(0)}{\partial \epsilon_{m'}}  .
\label{eq:chi_tilde}
\end{align}
The Ward identities relate the derivative of the self-energy to the 
vertex corrections,\cite{Yoshimori} 
and can be expressed in the following form 
for the multi-orbital Anderson impurity $\mathcal{H}$
defined in Eqs.\ \eqref{eq:Hd}--\eqref{eq:HT},
\begin{align}
&
\delta_{mm'} 
+ 
\frac{\partial \Sigma_{m}(i\omega)}{\partial \epsilon_{m'}}
\,+\,
\,\Gamma_{mm':m'm}(i\omega, 0; 0, i\omega ) 
\,\rho_{d,m'}(0)
\nonumber 
\\
& 
\!\!
= \ 
\delta_{mm'} 
\left(
1- \frac{\partial \Sigma_{m'}(i \omega)}{\partial i \omega}
\right)
\;.
\label{eq:YYY}
\end{align}
At zero frequency $\omega=0$, this can be rewritten in terms of 
the enhancement factor $\widetilde{\chi}_{mm'}^{}$, 
defined by Eq.\ \eqref{eq:chi_tilde},  and the renormalization factor $Z_m$, 
as  
\begin{align}
\!
\widetilde{\chi}_{mm'}^{} = \,  
\frac{1}{Z_m} \,\delta_{mm'}
-\Gamma_{mm':m'm}(0, 0; 0,0) 
\,\rho_{d,m'}(0) . 
\label{eq:YYY_sp}
\end{align}
Note that  $\Gamma_{mm:mm}(0 , 0; 0, 0)=0$ for $m'=m$. 
The {\it Fermi-liquid relations\/} for the 
coefficients ${\partial \left\langle n_{d,m}\right\rangle }/{\partial \epsilon_{m'}}  $, 
given in  Eq.\ \eqref{eq:two_Fermi_liquid_relations}, follow from 
this identity and Eq.\ \eqref{eq:dn_depsilon}.

The Ward identity  Eq.\ \eqref{eq:YYY} can be proved following Yoshimori's 
Feynman-diagrammatic derivations.\cite{Yoshimori}
In our purpose, to calculate the derivative of the self-energy 
with resect to $\omega$, 
we can shift the frequencies of the propagators along 
the closed loops of two different groups: 
one group carrying the external label $m$ and 
the other group carrying $m'$ ($\neq m$)  
chosen from the rest of the orbital indecies. 
The identity in the form  Eq.\ \eqref{eq:YYY} 
can also be deduced from a current conservation law,  
which in the present case corresponds to the local charge conservation  
in each of the {\it flavors\/} \lq\lq$m$'',  
\begin{align}
\frac{\partial\, n_{d,m}^{}}{\partial t} +J_m^{} =0, 
\quad \ \ 
 J_m^{} \equiv 
 i v \left( 
 d_{m}^{\dagger} a_{m}^{} -  a_{m}^{\dagger} d_{m}^{} 
 \right) .
\label{eq:current_conserve}
\end{align}
The derivation  along this line uses three-point vertex functions of 
a Ward-Takahashi form\cite{SchriefferBook,aoJPSJ2001} 
and makes it clear the fact that Eq.\ \eqref{eq:YYY} is the relation 
which represents the local charge conservation 
 Eq.\ \eqref{eq:current_conserve}.


%

\end{document}